\documentclass[aps,prd,10pt,a4paper,twocolumn,preprintnumbers,nofootinbib,floatfix,superscriptaddress]{revtex4-1}
\pdfoutput=1
\usepackage[a4paper, hdivide={1.90cm,,1.13cm}, vdivide={1.83cm,,2.5cm}]{geometry}

\usepackage{amstext,amssymb}
\usepackage[intlimits]{amsmath}

\usepackage[ansinew]{inputenc}
\usepackage{graphicx,multirow}
\usepackage{longtable}
\usepackage{color}
\usepackage{slashed}
\usepackage{units}
\usepackage[hyperfootnotes=false]{hyperref}

\usepackage[autostyle=false, style=english]{csquotes}
\MakeOuterQuote{"}

\def \MSbar {$\overline{\text{MS}}$\ } 

\def \i {\mathrm{i}\mkern1mu} 

\def \L {\mathcal{L}} 

\def \epsilon {\varepsilon} 

\def \vec#1{{\boldsymbol{#1}}}

\newcommand{\dd}{\mathrm{d}}
\newcommand{\matrixx}[1]{\begin{pmatrix} #1 \end{pmatrix}} 
\newcommand{\tr}{\mathrm{tr}}
\newcommand{\hc}{\ensuremath{\text{h.c.}}}

\begin{document}

\preprint{ULB-TH/18-14, UCI-TR-2018-17, \href{https://arxiv.org/abs/1811.07910}{arXiv:1811.07910}}

\title{SO(10) paths to dark matter}

\author{Sacha Ferrari}
\email{sacha.ferrari@ulb.ac.be}
\affiliation{Service de Physique Th\'eorique, Universit\'e Libre de Bruxelles, Boulevard du Triomphe, CP225, 1050 Brussels, Belgium}

\author{Thomas Hambye}
\email{thambye@ulb.ac.be}
\affiliation{Service de Physique Th\'eorique, Universit\'e Libre de Bruxelles, Boulevard du Triomphe, CP225, 1050 Brussels, Belgium}

\author{Julian Heeck}
\email{julian.heeck@uci.edu}
\affiliation{Service de Physique Th\'eorique, Universit\'e Libre de Bruxelles, Boulevard du Triomphe, CP225, 1050 Brussels, Belgium}
\affiliation{Department of Physics and Astronomy, University of California, Irvine, CA 92697-4575, USA}

\author{Michel H.G. Tytgat}
\email{mtytgat@ulb.ac.be}
\affiliation{Service de Physique Th\'eorique, Universit\'e Libre de Bruxelles, Boulevard du Triomphe, CP225, 1050 Brussels, Belgium}

\hypersetup{
pdftitle={SO(10) paths to dark matter},   
pdfauthor={Sacha Ferrari, Thomas Hambye, Julian Heeck, Michel H.G. Tytgat}
}

\begin{abstract}
The grand-unification gauge group $SO(10)$ contains matter parity as a discrete subgroup. This symmetry could be at the origin of dark matter stability. The properties of the dark matter candidates depend on the path along which $SO(10)$ is broken, in particular through Pati--Salam or left--right symmetric subgroups. We systematically determine the non-supersymmetric dark matter scenarios that can be realized along the various paths. We emphasize that the dark matter candidates may have colored or electrically charged partners at low scale that belong to the same $SO(10)$ multiplet. These states, which in many cases are important for co-annihilation, could be observed more easily than the dark matter particle. We determine the structure of the tree-level and loop-induced mass splittings between the dark matter candidate and their partners and discuss the possible phenomenological implications.
\end{abstract}

\maketitle

\tableofcontents

\section{Introduction}

Among the various properties dark matter (DM) particles must have, the most intriguing one is their  stability on cosmological timescales. This property basically requires the existence of a new symmetry beyond the Standard Model (SM). 
This can be an \emph{ad hoc} symmetry imposed by hand or can be derived from a more fundamental principle, in particular from gauge invariance.
The way gauge symmetries can stabilize the DM particle(s) can be either \emph{direct}, if the stability results from the fact that a global or local subgroup of these gauge symmetries remains unbroken, or \emph{indirect}, if the gauge symmetries imply an accidental symmetry which is not a subgroup of these gauge symmetries, see e.g.~\cite{Hambye:2010zb}.
Beside the lightest neutrino which is stable by Lorentz invariance, all other stable particles in the SM are stable in such a direct (electron and photon) or indirect way (proton).

In the following, we will be interested in \emph{direct} stability for DM consisting of Weakly Interacting Massive Particles (WIMPs). Since direct stability requires an extra gauge group, natural candidate models are grand unified theories (GUTs), in particular the ones based on the group $SO(10)$~\cite{Georgi:1974my,Fritzsch:1974nn}. $SO(10)$ contains the $U(1)_{B-L}$ subgroup whose discrete subgroup $Z_2^{3(B-L)}$ can stabilize the DM particle~\cite{Kadastik:2009dj,Kadastik:2009cu,Frigerio:2009wf}.
This is the mechanism used to stabilize the neutralino in the Minimal Supersymmetric Standard Model (MSSM), as the $R$-symmetry assumed in the MSSM can be traded for $Z_2^{3(B-L)}$ \cite{Mohapatra:1986su,Martin:1992mq,Aulakh:1999cd,Aulakh:2000sn}. More recently this mechanism has been shown to also be operative for the non-supersymmetric case for a scalar~\cite{Kadastik:2009dj,Kadastik:2009cu} or fermion~\cite{Frigerio:2009wf} DM candidate. The various DM candidates that could emerge in the lowest dimensional $SO(10)$ representations have been determined~\cite{Frigerio:2009wf} and specific candidates have been considered in some details~\cite{Mambrini:2013iaa,Mambrini:2015vna,Nagata:2015dma} (see also Refs.~\cite{Boucenna:2015sdg,Arbelaez:2015ila,Bandyopadhyay:2017uwc}).
So far most of these $SO(10)$ DM scenarios have been discussed from the low-energy point of view, basically independently of the way $SO(10)$ is broken down to the SM, i.e.~disregarding the scalar content.

However, as we will show in this article, the way $SO(10)$ is broken has a clear impact on the low-energy phenomenology.
If the breaking path is such that one or several $SO(10)$ subgroups larger than the SM group are broken only around the TeV scale and/or at an intermediate scale, the DM phenomenology will drastically change. These symmetries can not only predict low-energy gauge bosons into or through which DM can (co)-annihilate but can also predict the low energy presence of some DM \emph{partners} belonging to the same $SO(10)$ multiplet. Depending on the breaking path, some of these partners may show up at low scale, with a pattern of mass splittings and decays between these particles which greatly affects the DM phenomenology and the viability of the DM scenario.
In some cases, we will show that there is no breaking path leading to a viable phenomenology for some otherwise good candidates. Similarly some candidates, a priori excluded from the start from the low energy perspective, turn out to be viable along specific $SO(10)$ breaking paths.
Moreover some of these partners could be produced and seen in a much easier way by colliders than the DM particle itself, because they are colored or charged.

In this work, adopting a list of simple minimality criteria that a model must fulfill, we determine in a systematic way the candidates that show up by explicitly considering the various possible $SO(10)$ breaking paths and discuss the phenomenology deriving from these paths.

The plan of this article is as follows. 
We first recap the possible $SO(10)$ breaking chains and subgroups in Sec.~\ref{sec:chains} and the $SO(10)$  representations with DM candidates in Sec.~\ref{sec:SO10candidates}. The potential DM partners for given $SO(10)$ multiplets and the rationale behind their  mass splitting with respect to the DM are discussed in Sec.~\ref{sec:DMsplittings}. In Sec.~\ref{sec:constraints} we  discuss the various constraints that we will impose on DM candidates. In Sec.~\ref{sec:lowscale} we list all the possible low-scale DM scenarios for representations up to  $\vec{210}'$. This constitutes the core of our work. Having listed all the candidates, we discuss in Sec.~\ref{sec:viable}  how they can concretely be realized through $SO(10)$ breaking. For completeness we discuss the possibility of accidental DM stability in Sec.~\ref{sec:indirect_way}. We summarize our main results and draw our conclusions in Sec.~\ref{sec:summary}. Appendix~\ref{app:rge} gives an introduction to renormalization group evolution that is relevant for radiative mass splittings within multiplets. In App.~\ref{app:chem_dec} we discuss the condition of chemical equilibrium relevant for co-annihilation processes. 
App.~\ref{app:tables} provides tables of tree-level mass splittings of relevant $SO(10)$ multiplets by scalars in representations $\vec{45}$, $\vec{54}$, and $\vec{210}$.

\section{\texorpdfstring{$SO(10)$}{SO(10)} breaking chains and subgroups}
\label{sec:chains}

\begin{figure}[b]
\includegraphics[width=0.5\textwidth]{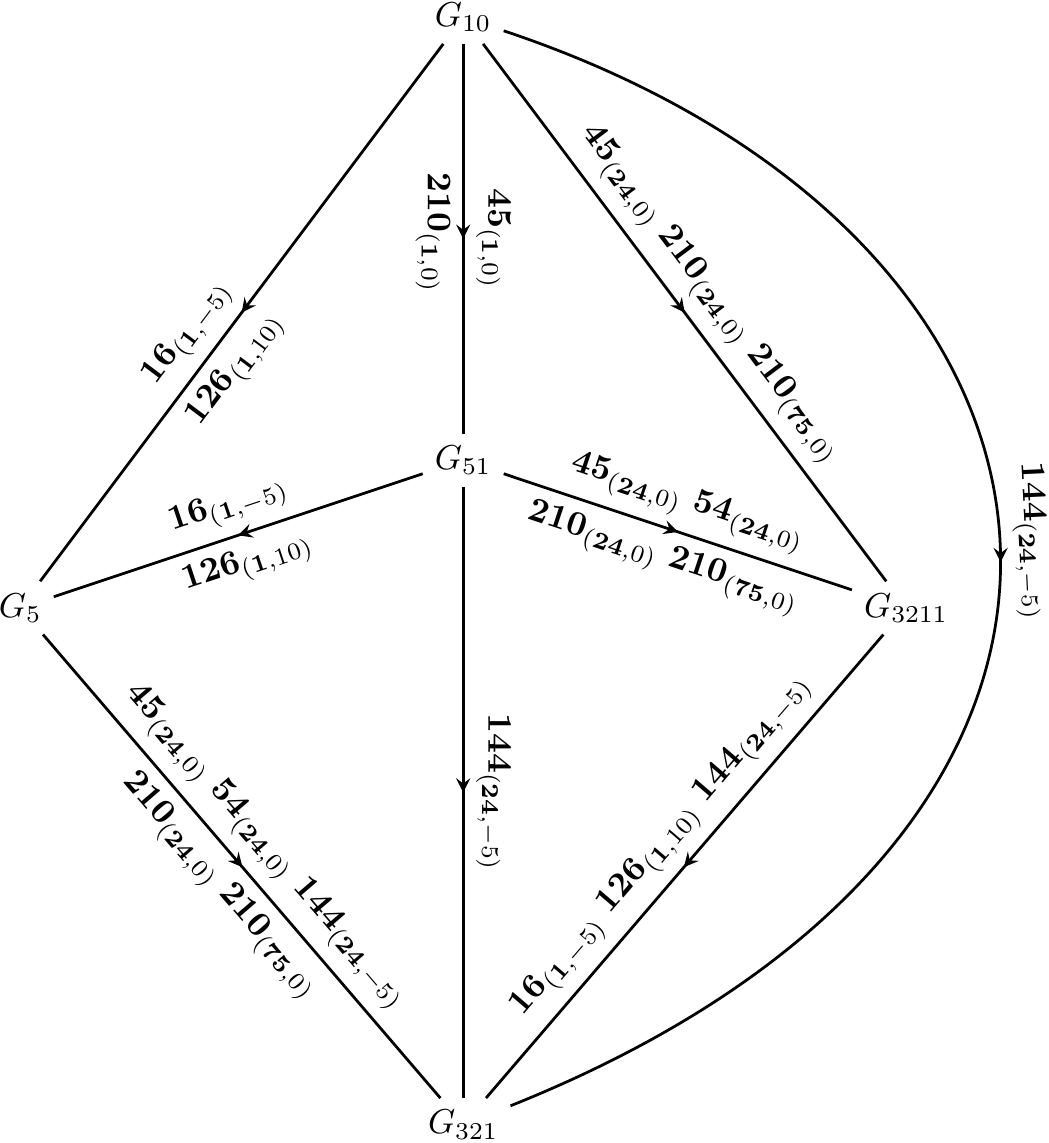}
\caption{
Breaking paths of $SO(10)$ along the Georgi--Glashow route. Each path corresponds to the VEV of an $SO(10)$ scalar, for example $\vec{144}_{(\vec{24},-5)}$ denotes the VEV of the $(\vec{24},-5)$ $SU(5)\times U(1)$ subcomponent of a scalar $SO(10)$ representation $\vec{144}$.  
}
\label{fig:so(10)_breaking_paths_GG}
\end{figure}

\begin{figure*}[t]
\includegraphics[width=0.80\textwidth]{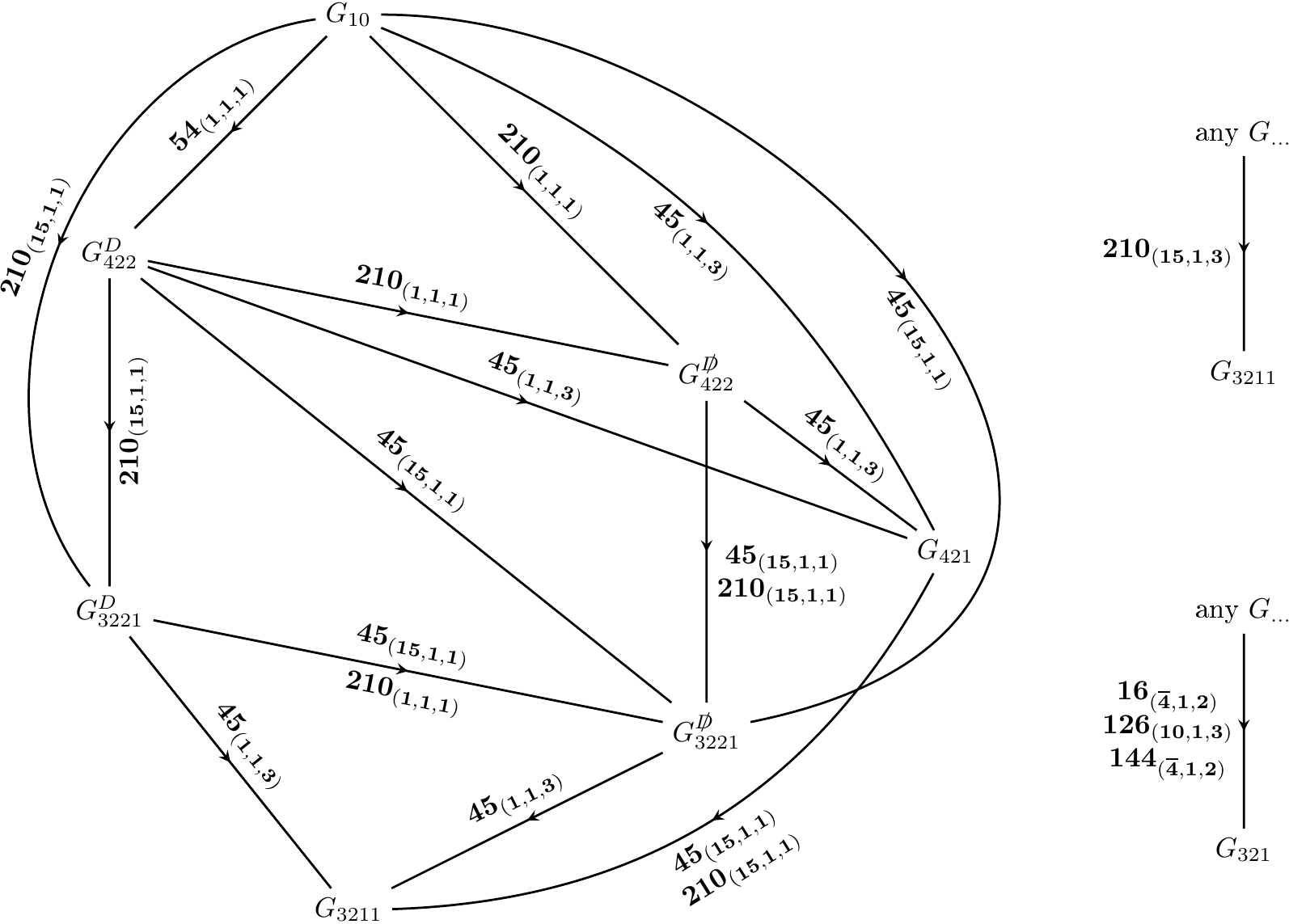}
\caption{
Breaking paths of $SO(10)$ along the Pati--Salam route. Each path corresponds to the VEV of an $SO(10)$ scalar, for example $\vec{210}_{(\vec{15},\vec{1},\vec{3})}$ denotes the VEV of the $(\vec{15},\vec{1},\vec{3})$ PS subcomponent of a scalar $SO(10)$ representation $\vec{210}$. The subgroups $G^D$ ($G^\slashed{D}$) conserve (violate) $D$ parity~\cite{Kuzmin:1980yp,Slansky:1980gc,Kibble:1982dd,Chang:1983fu,Chang:1984uy,Chang:1984qr}. 
The VEVs of $\vec{16}_{(\overline{\vec{4}},\vec{1},\vec{2})}$,  $\vec{126}_{(\vec{10},\vec{1},\vec{3})}$, and  $\vec{144}_{(\overline{\vec{4}},\vec{1},\vec{2})}$ can be used at any step to break to the SM and are omitted for illustration's sake, similar for the $\vec{210}_{(\vec{15},\vec{1},\vec{3})}$ which breaks any group to $G_{3211}$.}
\label{fig:so(10)_breaking_paths_PS}
\end{figure*}

As is well known, $G_{10}\equiv SO(10)$~\cite{Georgi:1974my,Fritzsch:1974nn} can be broken along various paths depending on the scalar representation content of the model and on the scalar potential of these representations.\footnote{We will be following the representation naming conventions (bars and primes) of \texttt{LieART}~\cite{Feger:2012bs} and \texttt{Susyno}~\cite{Fonseca:2011sy}, which differ slightly from e.g.~Slansky~\cite{Slansky:1981yr}, notably in the assignment $\vec{126}$ vs.~$\overline{\vec{126}}$.} These $SO(10)$ breaking paths are given in Figs.~\ref{fig:so(10)_breaking_paths_GG} and \ref{fig:so(10)_breaking_paths_PS}. There are two classes of paths, the ones proceeding through the maximal subgroup $G_{51}=SU(5)\times U(1)$ and/or its $G_5=SU(5)\times Z_2$ subgroup (to which we refer as Georgi--Glashow (GG)~\cite{Georgi:1974sy} paths) and the ones proceeding through the maximal Pati--Salam (PS) group $G_{422}=SU(4)_c\times SU(2)_L\times SU(2)_R$ and/or subgroups (Pati--Salam~\cite{Pati:1974yy} paths); see Refs.~\cite{Deshpande:1992au,Deshpande:1992em,Chakrabortty:2017mgi} for thorough discussions.
This is shown in Figs.~\ref{fig:so(10)_breaking_paths_GG} and \ref{fig:so(10)_breaking_paths_PS}, where the possible scalar representations up to $\vec{210}$ whose vacuum expectation values (VEVs) are at the origin of these breaking paths are also displayed.
The Pati--Salam paths can involve more intermediate subgroups than the Georgi--Glashow ones. On top of the PS group $G_{422}$, the paths may involve one of the three subgroups of $G_{422}$: 
\begin{align}
\hspace{-1ex} G_{421}  &=SU(4)_c\times SU(2)_L\times U(1)_{2 R} \,,\\
\hspace{-1ex} G_{3221} &=SU(3)_c\times SU(2)_L\times SU(2)_R\times U(1)_{3(B-L)} \,,\\
\hspace{-1ex} G_{3211} &=SU(3)_c\times SU(2)_L\times U(1)_{2 R}\times U(1)_{3(B-L)} \,.
\end{align}
Note that we have chosen a convenient normalization for the $U(1)$ generators in order to make the $U(1)$ charges integer.
For scalar representations up to $\vec{210}$, this leads to many possible paths, as shown in Fig.~\ref{fig:so(10)_breaking_paths_PS}.
Here we have also taken  $D$ parity~\cite{Kuzmin:1980yp,Slansky:1980gc,Kibble:1982dd,Chang:1983fu,Chang:1984uy,Chang:1984qr} into account, which corresponds to a discrete $Z_2$ left--right exchange symmetry with important implications when it comes to DM partners.
The last breaking step is of course to the SM gauge group, 
\begin{align}
G_\text{SM}\equiv G_{321} = SU(3)_c\times SU(2)_L\times U(1)_{Y}\,,
\end{align}
for which we chose the usual hypercharge normalization $Y = Q -T_3^L$, with electric charge $Q$ and diagonal $SU(2)_L$ generator $T_3^L$. Hypercharge can also be defined in terms of $SU(2)_R\times U(1)_{3(B-L)}$ or $U(1)_{2 R}\times U(1)_{3(B-L)}$ as
\begin{align}
Y = T_3^R +\frac16 \, [3(B-L)] = \frac12 \, [2 R] +\frac16 \, [3(B-L)]\,,
\label{eq:hypercharge}
\end{align}
with $T_3^R$ being the diagonal $SU(2)_R$ generator, which becomes the $U(1)_{R}$ generator $R$.

The massive gauge bosons in the coset $SO(10)/G_\text{SM}$ form representations under $G_\text{SM}$ (see Tab.~\ref{tableLowerLimits}) and have well-known couplings to SM fermions. 
We can calculate the masses that arise from the various VEVs along the PS path (Fig.~\ref{fig:so(10)_breaking_paths_PS}) with the help of \texttt{Susyno}~\cite{Fonseca:2011sy}:
\begin{widetext}
\begin{align}
\begin{split}
\frac{m_U^2}{g^2} &= \frac{v_{\vec{16}_{(\overline{\vec{4}},\vec{1},\vec{2})}}^2}{4} +\frac{1}{6}\left(v_{\vec{45}_{(\vec{15},\vec{1},\vec{1})}} +\sqrt{\frac{3}{2}} v_{\vec{45}_{(\vec{1},\vec{1},\vec{3})}}\right)^2 +\frac{5 v_{\vec{54}_{(\vec{1},\vec{1},\vec{1})}}^2}{12}+ \frac{v_{\vec{126}_{(\vec{10},\vec{1},\vec{3})}}^2}{2}+\frac{v_{\vec{144}_{(\overline{\vec{4}},\vec{1},\vec{2})}}^2}{4} \\
&\quad+\frac{v_{\vec{210}_{(\vec{1},\vec{1},\vec{1})}}^2}{2}+\frac{v_{\vec{210}_{(\vec{15},\vec{1},\vec{1})}}^2}{3}+\frac{v_{\vec{210}_{(\vec{15},\vec{1},\vec{3})}}^2}{4}-\frac{v_{\vec{210}_{(\vec{1},\vec{1},\vec{1})}} v_{\vec{210}_{(\vec{15},\vec{1},\vec{3})}}}{\sqrt{6}}+\frac{\sqrt{2}}{3}  v_{\vec{210}_{(\vec{15},\vec{1},\vec{1})}} v_{\vec{210}_{(\vec{15},\vec{1},\vec{3})}}\,,
\end{split}\\
\begin{split}
\frac{m_V^2}{g^2} &= \frac{1}{6}\left(v_{\vec{45}_{(\vec{15},\vec{1},\vec{1})}} -\sqrt{\frac{3}{2}} v_{\vec{45}_{(\vec{1},\vec{1},\vec{3})}}\right)^2 +\frac{5 v_{\vec{54}_{(\vec{1},\vec{1},\vec{1})}}^2}{12} + \frac{5 v_{\vec{144}_{(\overline{\vec{4}},\vec{1},\vec{2})}}^2}{12}\\
&\quad +\frac{v_{\vec{210}_{(\vec{1},\vec{1},\vec{1})}}^2}{2}+\frac{v_{\vec{210}_{(\vec{15},\vec{1},\vec{1})}}^2}{3}+\frac{v_{\vec{210}_{(\vec{15},\vec{1},\vec{3})}}^2}{4} +\frac{v_{\vec{210}_{(\vec{1},\vec{1},\vec{1})}} v_{\vec{210}_{(\vec{15},\vec{1},\vec{3})}}}{\sqrt{6}}-\frac{\sqrt{2}}{3}  v_{\vec{210}_{(\vec{15},\vec{1},\vec{1})}} v_{\vec{210}_{(\vec{15},\vec{1},\vec{3})}}\,,
\end{split}\\
\begin{split}
\frac{m_X^2}{g^2} &= \frac{v_{\vec{16}_{(\overline{\vec{4}},\vec{1},\vec{2})}}^2}{4} +\frac{2 v_{\vec{45}_{(\vec{15},\vec{1},\vec{1})}}^2}{3} + \frac{v_{\vec{126}_{(\vec{10},\vec{1},\vec{3})}}^2}{2} +\frac{v_{\vec{144}_{(\overline{\vec{4}},\vec{1},\vec{2})}}^2}{4}+\frac{2 v_{\vec{210}_{(\vec{15},\vec{1},\vec{1})}}^2}{3}+\frac{2 v_{\vec{210}_{(\vec{15},\vec{1},\vec{3})}}^2}{3}\,,
\end{split}\\
\begin{split}
\frac{m_{W_R}^2}{g^2} &= \frac{v_{\vec{16}_{(\overline{\vec{4}},\vec{1},\vec{2})}}^2}{4} +v_{\vec{45}_{(\vec{1},\vec{1},\vec{3})}}^2+ \frac{v_{\vec{126}_{(\vec{10},\vec{1},\vec{3})}}^2}{2} +\frac{v_{\vec{144}_{(\overline{\vec{4}},\vec{1},\vec{2})}}^2}{4}+v_{\vec{210}_{(\vec{15},\vec{1},\vec{3})}}^2\,,
\end{split}\\
\frac{m_{Z'}^2}{g^2} &= \frac{5 v_{\vec{16}_{(\overline{\vec{4}},\vec{1},\vec{2})}}^2}{8} +\frac{5 v_{\vec{126}_{(\vec{10},\vec{1},\vec{3})}}^2}{2}+\frac{5 v_{\vec{144}_{(\overline{\vec{4}},\vec{1},\vec{2})}}^2}{8} \,,
\label{eq:gauge_boson_masses}
\end{align}
\end{widetext}
using the lowest-order approximation of one common gauge coupling $g$. Since the SM gauge bosons are massless at this level, we do not show them. The mass contributions of the $\vec{16}$, $\vec{45}$, and $\vec{126}$ match those of Refs.~\cite{Bertolini:2009es,Bertolini:2012im} up to notational differences.
These explicit expressions can give additional insight into the breaking path and also illustrate the different viewpoints of PS and GG. For example, if only the $\vec{126}$ obtains a VEV one can see that $V$ remains massless, while all other new gauge bosons become heavy. This in fact shows that $\langle\vec{126}\rangle$ by itself breaks $SO(10)\to SU(5)$, as expected from Fig.~\ref{fig:so(10)_breaking_paths_GG}, with $V$ being precisely the gauge boson in $SU(5)/G_\text{SM}$ needed to complete $SU(5)$. The VEV $\langle\vec{126}\rangle$ (or $\langle\vec{16}\rangle$) can therefore not be used alone to break $SO(10)\to G_\text{SM}$ contrary to the statement of Fig.~\ref{fig:so(10)_breaking_paths_PS}, but rather requires an additional VEV that contributes to $m_V$.
As another interesting special case, one can observe that setting $v_{\vec{45}_{(\vec{15},\vec{1},\vec{1})}} =\sqrt{\frac{3}{2}} v_{\vec{45}_{(\vec{1},\vec{1},\vec{3})}}$ again keeps $V$ massless and thus actually breaks $SO(10)\to G_{51}$~\cite{Bertolini:2009es}, not visible from the graphical breaking path in Fig.~\ref{fig:so(10)_breaking_paths_PS}.
Indeed, the linear combination $\sqrt{\frac{2}{5}} v_{\vec{45}_{(\vec{15},\vec{1},\vec{1})}}-\sqrt{\frac{3}{5}} v_{\vec{45}_{(\vec{1},\vec{1},\vec{3})}}$ is actually nothing but $v_{\vec{45}_{(\vec{24},0)}}$ in Georgi--Glashow notation, which is precisely the $\vec{45}$ VEV to provide a mass to the $SU(5)$ gauge boson $V$ if non-zero.

\begin{table}[b]
\begin{center}
\small{
\begin{tabular}{llll}
\hline
Gauge boson & breaking step & lower limit & observable \\ 
\hline
$U\sim \left(\vec{3},\vec{2},\tfrac{1}{6}\right)$ & \multirow{2}{*}{$SO(10)\to G_{422}$} & \multirow{2}{*}{$\unit[10^{12}]{TeV}$} & \multirow{2}{*}{proton decay}\\
$V\sim \left(\vec{3},\vec{2},-\tfrac{5}{6}\right)$ & & & \\
$X\sim \left(\vec{3},\vec{1},\tfrac{2}{3}\right)$ & $G_{422}\to G_{3221}$ & $10^2$--$\unit[10^{3}]{TeV}$ & $K_L\to \mu^\pm e^\mp$\\
$W_R \sim (\vec{1},\vec{1},1)$ & $G_{3221}\to G_{3211}$ & $\unit[3]{TeV}$ & meson mixing\\
$Z' \sim (\vec{1},\vec{1},0)$ & $G_{3211}\to G_\text{SM}$ & $\mathcal{O}(\unit{TeV})$ & LHC dilepton\\
\hline
\end{tabular}}
\end{center}
\caption{Lower limits on the heavy $SO(10)$ gauge boson masses along one PS path, given in their SM representations. See text for details and references.
\label{tableLowerLimits}}
\end{table}

We expect $SO(10)$ breaking to happen at energy scales above $m_{U,V}\gtrsim\unit[10^{12}]{TeV}$ in order to avoid stringent bounds from proton decay~\cite{Nath:2006ut}, but the subgroups could be broken at lower scales. For the $SU(5)$ paths, there is little room to push the scale down, as proton decay (mediated by the $SU(5)$ gauge boson $V$) is equally dangerous here. The Pati--Salam paths on the other hand do not lead to dangerous gauge-boson induced proton decay and could be valid all the way down to $\unit[10^{3}]{TeV}$, where limits from rare meson decays such as $K_L\to \mu^\pm e^\mp$~\cite{Valencia:1994cj,Ambrose:1998us,Smirnov:2007hv} put constraints on the massive PS gauge boson $X$. This PS scale can be pushed down an order of magnitude further by playing with the quark and lepton mixing matrices~\cite{Kuznetsov:2012ai,Smirnov:2018ske}, but we will not make use of this for the most part. Similar lower bounds hold for the $G_{421}$ subgroup.
Finally, the left--right (LR) subgroup $G_{3221}$~\cite{Mohapatra:1974gc,Senjanovic:1975rk,Senjanovic:1978ev} can easily be at the TeV scale before running into problems with meson--anti-meson oscillations and direct searches~\cite{Beall:1981ze,Bertolini:2014sua}. This makes the low-scale left--right group an obvious candidate to stabilize DM and produce the right amount, as discussed in Refs.~\cite{Heeck:2015qra,Garcia-Cely:2015quu}.
$G_{3211}$ can in principle be even lower than the LR scale, although one still has lower limits of order TeV from dilepton searches at the LHC, depending on the details of the breaking~\cite{Patra:2015bga,Arcadi:2017atc}. For GUT-inspired couplings the lower limits are roughly between $4$ and $\unit[5]{TeV}$, ignoring all non-SM $Z'$ decay channels~\cite{Sirunyan:2018exx}.
We will pay special attention to possible low-scale $SO(10)$ subgroups as they can have a big effect on DM phenomenology.
The heavy \emph{colored} gauge bosons $U$, $V$, and $X$ induce interactions that are too weak to lead to viable freeze-out scenarios, but they can still have an impact on the phenomenology, in particular with regards to co-annihilation, see below.

\section{\texorpdfstring{$SO(10)$}{SO(10)} DM candidates}
\label{sec:SO10candidates}

\begin{table}[t]
\begin{center}
\small{
\begin{tabular}{|l|c|c|}
\hline
& {fermions} &{scalars} \\ \hline
\begin{tabular}{c} DM multiplet \\$n_Y$    \end{tabular} &   
\begin{tabular}{c} even  $SO(10)$ \\ multiplet \end{tabular}   & 
\begin{tabular}{c} odd  $SO(10)$ \\ multiplet \end{tabular}  \\ \hline
\hline
\quad$1_0$  & $\vec{45}$, $\vec{54}$, $\vec{126}$, $\vec{210}$ 
 & $\vec{16}$, $\vec{144}$ 
\\
\hline
\quad$2_{\pm1/2}$  &  $\vec{10}$, $\vec{120}$, $\vec{126}$, $\vec{210}$, $\vec{210}'$ 
&  $\vec{16}$, $\vec{144}$ 
 \\
\hline
\quad$3_0$ & $\vec{45}$, $\vec{54}$, $\vec{210}$ 
& $\vec{144}$  \\
\hline
\quad$3_{\pm1}$  & $\vec{54}$, $\vec{126}$ 
 & $\vec{144}$ \\
\hline 
\quad$4_{\pm1/2}\,\,\,$  &  $\vec{210}'$ & $\vec{560}$ \\
\hline
\quad$4_{\pm3/2}$  & $\vec{210}'$  & $\vec{720}$  \\
\hline 
\quad$5_0$  & $\vec{660}$    & $\vec{2640}$  \\
\hline 
\quad\dots &  \dots & \dots  \\
\hline 
\end{tabular}}
\end{center}
\caption{The first column gives $SU(2)_L\times U(1)_Y$ multiplets with a neutral component, $n$ being the $SU(2)_L$ dimension. 
The second (third) column shows the $P_M$-even ($P_M$-odd) $SO(10)$ representations
that contain a multiplet with the given electroweak charges and no color (for representation up to $\vec{210}$ or giving the smallest possible representation if this one is larger than $\vec{210}$). The triplet candidates with hypercharge $\pm 1$ are given for completeness but they will not be considered further because they are not viable, see text.
\label{tab:ListOfEWMultiplets}}
\end{table}

We define matter parity as $P_M=(-1)^{3(B-L)}$, which is a $Z_2$ subgroup of $SO(10)$~\cite{Martin:1992mq,Kadastik:2009dj,Kadastik:2009cu,Frigerio:2009wf}.\footnote{If we would break along the Georgi--Glashow route matter parity would be the $Z_2$ subgroup of $U(1)_\chi$ in $SU(5)\times U(1)_\chi$~\cite{Kadastik:2009dj,Ma:2018uss,Ma:2018zuj}.}
 If we limit ourselves to representations up to dimension 210, only the representations $\vec{16}$ and $\vec{144}$ are odd under matter parity, while all others are even.
Given the fact that the SM fermions of one generation are in an odd $\vec{16}$ and the SM scalar doublet in an even $\vec{10}$ representation, the lightest component of a newly introduced \emph{even fermion}  or  an \emph{odd scalar} representation is therefore exactly stable. (This stability is no longer guaranteed if we have scalar VEVs $\langle \vec{16} \rangle$ or $\langle \vec{144} \rangle$, but could still survive as an accidental symmetry, see Sec.~\ref{sec:indirect_way}.)
This leads to the list of possible DM candidates given in table~\ref{tab:ListOfEWMultiplets}~\cite{Frigerio:2009wf}. It is also useful to decompose the $SO(10)$ representations in terms of their Pati--Salam and left--right group representations, which allows us to identify the DM candidates in each $SO(10)$ representation and the quantum numbers they have under these subgroups. This is given in the tables~\ref{tab:DM_A} and~\ref{tab:DM_B} for representations up to $\vec{210}'$. For these lists of candidates we exclude from the start any multiplet which is colored; even though it was recently argued in Ref.~\cite{DeLuca:2018mzn} that there could in fact be colored DM particles, our candidates below do not come in the required representations.

Similarly we exclude candidates with a non-vanishing hypercharge because these are typically excluded by $Z$-mediated direct detection. An exception to this will be the left--right bi-doublet and bi-quadruplet, whose neutral Dirac fermion is in general split into two non-degenerate Majorana fermions at loop level, which can evade direct detection constraints~\cite{Garcia-Cely:2015quu}. 
Under these criteria tables~\ref{tab:DM_A} and~\ref{tab:DM_B} show that the $\vec{10}$, $\vec{45}$, $\vec{54}$, $\vec{120}$, $\vec{126}$, $\vec{210}$, and $\vec{210}'$ fermion representations contain 1, 3, 2, 2, 2, 4, and 3 DM candidates, respectively, for a total of 17 candidates. Note that the representations $\vec{45}$, $\vec{120}$, and $\vec{210}$ have several candidates with the same 
SM quantum numbers, namely two $1_0$, two $2_{1/2}$, and three $1_0$ candidates, respectively. Similarly for a scalar representation the $\vec{16}$ contains two DM candidates while the $\vec{144}$ has four candidates, for a total of six candidates (with two $1_0$ candidates).

\section{DM mass splittings}
\label{sec:DMsplittings}

Having determined $SO(10)$ representations that are stabilized by matter parity and contain an electrically neutral particle as a DM candidate, we have to worry about its multiplet partners. As long as $SO(10)$ is not broken, all the particles within an irreducible representation are necessarily degenerate, leading in all cases to colored and charged partners at the DM scale. These DM partners need nevertheless to be \emph{heavier} than DM, so that they can decay sufficiently fast to not leave an imprint in cosmological observables~\cite{Mambrini:2015vna}.
One must distinguish two kinds of partners:
\begin{itemize}

\item \underline{The high-scale DM partners}: 
These are the color partners that decay only by mediation of the very heavy $SO(10)$ gauge bosons $U$ or $V$, discussed in Tab.~\ref{tableLowerLimits}, on which we have strong \emph{lower} mass bounds from proton decay. The color DM partners then must have masses several orders of magnitude larger than the DM component in order to decay sufficiently fast. As a conservative limit we impose that the lifetime is shorter than $\sim \unit[0.1]{s}$ in order that the decay occurs before Big Bang nucleosynthesis (BBN).
For a typical decay width $\Gamma\propto m_\text{DM partner}^5/m_{U,V}^4$ and given the lower bound on $m_{U,V}$ from Tab.~\ref{tableLowerLimits} (assuming the GUT scale close to current proton-decay bounds), we obtain a typical lower bound $m_\text{DM partner} > \unit[\mathcal{O}(10^5)]{TeV}$. These DM partners are then clearly inaccessible experimentally and play no role for our further discussion. They are nevertheless crucial to the discussion of DM in $SO(10)$, as it is non-trivial to obtain the required mass splitting $m_\text{DM}\lll m_\text{DM partner}$. This large mass splitting can be induced only at tree level, proportional to the various breaking scales, which gives strong constraints on the scalar representations and $SO(10)$ breaking path.

{
\begin{table}[!tb]
\begin{tabular}{|c|c|c|c|}
\hline
$SO(10)$ & $G_{422}$ & $G_ {3221}$ & DM-$G_{321}$ 
  \\ \hline
\hline

\multirow{3}{*}{$\vec{10}$} 	& $(\vec{1},\vec{2},\vec{2})$ 					& $(\vec{1},\vec{2},\vec{2},0)$ 	& $2_{1/2}$\\ 
\cline{2-4} 					& \multirow{2}{*}{$(\vec{6},\vec{1},\vec{1})$} & $(\overline{\vec{3}},\vec{1},\vec{1},2)$ 	& / \\
\cline{3-4} 					& 												& $(\vec{3},\vec{1},\vec{1},-2)$ 	& / \\
\hline

\multirow{4}{*}{$\vec{16}$} & \multirow{2}{*}{$(\vec{4},\vec{2},\vec{1})$}
 & $(\vec{3},\vec{2},\vec{1},1)$ & /
\\ \cline{3-4} & &$(\vec{1},\vec{2},\vec{1},-3)$&$2_{1/2}$ \\
\cline{2-4}
& \multirow{2}{*}{$(\overline{\vec{4}},\vec{1},\vec{2})$} & 
 $(\overline{\vec{3}},\vec{1},\vec{2},-1)$ & / \\
\cline{3-4}
  &  
& $(\vec{1},\vec{1},\vec{2},3)$ & $1_0$\\
\hline

\multirow{8}{*}{$\vec{45}$}  	&  $(\vec{1},\vec{3},\vec{1})$ 	& $(\vec{1},\vec{3},\vec{1},0)$ 	& $3_0$ \\
\cline{2-4} 					&  $(\vec{1},\vec{1},\vec{3})$ 	& $(\vec{1},\vec{1},\vec{3},0)$ 	& $1_0$ \\
\cline{2-4} 					&  \multirow{2}{*}{$(\vec{6},\vec{2},\vec{2})$} 	& $(\overline{\vec{3}},\vec{2},\vec{2},2)$ 	& / \\
\cline{3-4} 					&   												& $(\vec{3},\vec{2},\vec{2},-2)$ 	& / \\
\cline{2-4} 					&  \multirow{4}{*}{$(\vec{15},\vec{1},\vec{1})$} 	& $(\vec{3},\vec{1},\vec{1}, 4)$ 	& / \\
\cline{3-4} 					&   												& $(\overline{\vec{3}},\vec{1},\vec{1},-4)$ 	& / \\
\cline{3-4} 					&  													& $(\vec{8},\vec{1},\vec{1},0)$ 	& / \\
\cline{3-4} 					&  													& $(\vec{1},\vec{1},\vec{1},0)$ 	& $1_0$

 \\
\hline
\multirow{7}{*}{$\vec{54}$} 	& $(\vec{1},\vec{1},\vec{1})$ 						& $(\vec{1},\vec{1},\vec{1},0)$ 	&$1_0$ \\
\cline{2-4} 					& $(\vec{1},\vec{3},\vec{3})$ 						& $(\vec{1},\vec{3},\vec{3},0)$ 	&$3_0$ \\
\cline{2-4} 					& \multirow{2}{*}{$(\vec{6},\vec{2},\vec{2})$} 		& $(\overline{\vec{3}},\vec{2},\vec{2},2)$ 	&/ \\
\cline{3-4} 					&  													& $(\vec{3},\vec{2},\vec{2},-2)$ 	&/ \\
\cline{2-4} 					& \multirow{3}{*}{$(\vec{20}',\vec{1},\vec{1})$} 	& $(\vec{6},\vec{1},\vec{1},4)$ 	&/ \\
\cline{3-4} 					& 												 	& $(\overline{\vec{6}},\vec{1},\vec{1},-4)$ 	&/ \\
\cline{3-4} 					& 													& $(\vec{8},\vec{1},\vec{1},0)$ 	&/ \\
\hline

\multirow{15}{*}{$\vec{120}$}  	& $(\vec{1},\vec{2},\vec{2})$ 						& $(\vec{1},\vec{2},\vec{2},0)$ 	&$2_{1/2}$\\
\cline{2-4} 					& \multirow{2}{*}{$(\vec{6},\vec{1},\vec{3})$} 		& $(\overline{\vec{3}},\vec{1},\vec{3},2)$ &/\\
\cline{3-4} 					&  													& $(\vec{3},\vec{1},\vec{3},-2)$ &/\\
\cline{2-4} 					& \multirow{2}{*}{$(\vec{6},\vec{3},\vec{1})$} 		& $(\overline{\vec{3}},\vec{3},\vec{1},2)$ &/\\
\cline{3-4} 					&  													& $(\vec{3},\vec{3},\vec{1},-2)$ &/\\
\cline{2-4}  					& \multirow{3}{*}{$(\vec{10},\vec{1},\vec{1})$} & $(\vec{1},\vec{1},\vec{1},-6)$ &/\\
\cline{3-4}  					&  													& $(\vec{3},\vec{1},\vec{1},-2)$ &/\\
\cline{3-4}  					&  													& $(\overline{\vec{6}},\vec{1},\vec{1},2)$ &/\\
\cline{2-4}  					& \multirow{3}{*}{$(\overline{\vec{10}},\vec{1},\vec{1})$} & $(\vec{1},\vec{1},\vec{1},6)$ &/\\
\cline{3-4}  					&  													& $(\overline{\vec{3}},\vec{1},\vec{1}, 2)$ &/\\
\cline{3-4}  					&  													& $(\vec{6},\vec{1},\vec{1},-2)$ &/\\
\cline{2-4}  					& \multirow{4}{*}{$(\vec{15},\vec{2},\vec{2})$} 	& $(\vec{1},\vec{2},\vec{2},0)$ &$2_{1/2}$\\
\cline{3-4} 					& 													& $(\vec{8},\vec{2},\vec{2},0)$ &/\\
\cline{3-4}  					&  													& $(\vec{3},\vec{2},\vec{2},4)$ &/\\
\cline{3-4}  					&  													& $(\overline{\vec{3}},\vec{2},\vec{2},-4)$ &/\\
\hline

\multirow{12}{*}{$\vec{126}$}  	& \multirow{2}{*}{$(\vec{6},\vec{1},\vec{1})$}  & $(\overline{\vec{3}},\vec{1},\vec{1},2)$ 	& / \\
\cline{3-4}  					& 												 & $(\vec{3},\vec{1},\vec{1},-2)$ 	& / \\
\cline{2-4}   					& \multirow{4}{*}{$(\vec{15},\vec{2},\vec{2})$} & $(\vec{1},\vec{2},\vec{2},0)$	&$2_{1/2}$ \\
\cline{3-4}   					&  												 & $(\vec{8},\vec{2},\vec{2},0)$	& /\\
\cline{3-4}   					& 												 & $(\vec{3},\vec{2},\vec{2},4)$	& /\\
\cline{3-4}   					& 												 & $(\overline{\vec{3}},\vec{2},\vec{2},-4)$	& /\\
\cline{2-4}   					& \multirow{3}{*}{$(\vec{10},\vec{1},\vec{3})$}	 & $(\vec{1},\vec{1},\vec{3},-6)$ 	& $1_{0}$\\
\cline{3-4}   					& 												 & $(\vec{3},\vec{1},\vec{3},-2)$	& /\\
\cline{3-4}  					& 												 & $(\overline{\vec{6}},\vec{1},\vec{3},2)$& /\\
\cline{2-4}   					& \multirow{3}{*}{$(\overline{\vec{10}},\vec{3},\vec{1})$} & $(\overline{\vec{3}},\vec{3},\vec{1},2)$& /\\
\cline{3-4}  					&												 & $(\vec{1},\vec{3},\vec{1},6)$	& /\\
\cline{3-4}  					& 												 & $(\vec{6},\vec{3},\vec{1},-2)$	& /\\

\hline
\end{tabular}
\caption{
Decomposition of the $SO(10)$ representations between $\vec{10}$ and $\vec{126}$ under the PS (LR) subgroup $G_{422}$ ($G_ {3221}$) in column 2 (3). In the last column we identify possible DM components in their SM notation, see Tab.~\ref{tab:ListOfEWMultiplets}.
\label{tab:DM_A}}
\end{table}
}

{
\begin{table}[!tb]
\begin{tabular}{|c|c|c|c|}
\hline
$SO(10)$ & $G_{422}$ & $G_ {3221}$ & DM-$G_{321}$ 
  \\ \hline
\hline

\multirow{16}{*}{$\vec{144}$} & \multirow{2}{*}{$(\vec{4},\vec{2},\vec{1})$}
 & $(\vec{3},\vec{2},\vec{1},1)$ & /
\\ \cline{3-4} & &$(\vec{1},\vec{2},\vec{1},-3)$& $2_{1/2}$\\
\cline{2-4}
& \multirow{2}{*}{$(\overline{\vec{4}},\vec{1},\vec{2})$} & 
 $(\overline{\vec{3}},\vec{1},\vec{2},-1)$ & / \\
\cline{3-4}
  &  
& $(\vec{1},\vec{1},\vec{2},3)$ &$1_0$ \\
\cline{2-4}
& \multirow{2}{*}{$(\vec{4},\vec{2},\vec{3})$}
 & $(\vec{3},\vec{2},\vec{3},1)$ & /
\\ \cline{3-4} & &$(\vec{1},\vec{2},\vec{3},-3)$& $2_{1/2}$\\
\cline{2-4}
& \multirow{2}{*}{$(\overline{\vec{4}},\vec{3},\vec{2})$} & 
 $(\overline{\vec{3}},\vec{3},\vec{2},-1)$ & / \\
\cline{3-4}
  &  
& $(\vec{1},\vec{3},\vec{2},3)$ & $3_0$\\
\cline{2-4}
  &   \multirow{4}{*}{$(\vec{20},\vec{2},\vec{1})$}
& $(\vec{3},\vec{2},\vec{1},1)$ &/  \\
\cline{3-4}
&
 & $(\overline{\vec{3}},\vec{2},\vec{1},5)$  & /
\\
\cline{3-4}
&
 & $(\vec{6},\vec{2},\vec{1},1)$ & /
\\ \cline{3-4} & &$(\vec{8},\vec{2},\vec{1},-3)$& /\\
\cline{2-4}
  &   \multirow{4}{*}{$(\overline{\vec{20}},\vec{1},\vec{2})$}
& $(\overline{\vec{3}},\vec{1},\vec{2},-1)$ & /\\
\cline{3-4}
&
 & $(\vec{3},\vec{1},\vec{2},-5)$ & /
\\
\cline{3-4}
&
 & $(\overline{\vec{6}},\vec{1},\vec{2},-1)$ & /
\\ \cline{3-4} & &$(\vec{8},\vec{1},\vec{2},3)$& /\\
\hline 
	
\multirow{21}{*}{$\vec{210}$}  	& $(\vec{1},\vec{1},\vec{1})$ 						& $(\vec{1},\vec{1},\vec{1},0)$  	& $1_0$\\
\cline{2-4} 					& \multirow{2}{*}{$(\vec{6},\vec{2},\vec{2})$} 		& $(\overline{\vec{3}},\vec{2},\vec{2},2)$  	& /\\
\cline{3-4} 					&  												 	& $(\vec{3},\vec{2},\vec{2},-2)$  	& /\\
\cline{2-4} 					& \multirow{3}{*}{$(\vec{10},\vec{2},\vec{2})$} 	& $(\vec{1},\vec{2},\vec{2},-6)$  	&/\\
\cline{3-4} 					&   												& $(\overline{\vec{6}},\vec{2},\vec{2},2)$  &/\\
\cline{3-4} 					&  													& $(\vec{3},\vec{2},\vec{2},-2)$  	&/\\
\cline{2-4} 					& \multirow{3}{*}{$(\overline{\vec{10}},\vec{2},\vec{2})$} 	& $(\vec{1},\vec{2},\vec{2},6)$  &/\\
\cline{3-4} 					&   												& $(\vec{6},\vec{2},\vec{2},-2)$  	&/\\
\cline{3-4} 					&  													& $(\overline{\vec{3}},\vec{2},\vec{2},2)$  &/\\
\cline{2-4} 					&  \multirow{4}{*}{$(\vec{15},\vec{1},\vec{1})$} 	& $(\vec{3},\vec{1},\vec{1},4)$  	&/\\
\cline{3-4} 					&   												& $(\overline{\vec{3}},\vec{1},\vec{1},-4)$  	&/\\
\cline{3-4} 					&   												& $(\vec{1},\vec{1},\vec{1},0)$  	&1$_0$\\
\cline{3-4} 					&  													& $(\vec{8},\vec{1},\vec{1},0)$  	&/\\
\cline{2-4} 					&  \multirow{4}{*}{$(\vec{15},\vec{1},\vec{3})$} 	& $(\vec{1},\vec{1},\vec{3},0)$  	& $1_0$\\
\cline{3-4} 					&   												& $(\vec{8},\vec{1},\vec{3},0)$  	&/\\
\cline{3-4} 					&   												& $(\vec{3},\vec{1},\vec{3},4)$		&/\\
\cline{3-4} 					&   												& $(\overline{\vec{3}},\vec{1},\vec{3},-4)$	&/\\
\cline{2-4} 					&  \multirow{4}{*}{$(\vec{15},\vec{3},\vec{1})$} 	& $(\vec{1},\vec{3},\vec{1},0)$  	& $3_0$\\
\cline{3-4} 					&  													& $(\vec{8},\vec{3},\vec{1},0)$  	&/\\
\cline{3-4} 					&  													& $(\vec{3},\vec{3},\vec{1},4)$		&/\\
\cline{3-4} 					&  													& $(\overline{\vec{3}},\vec{3},\vec{1},-4)$	&/\\

\hline 
\multirow{13}{*}{$\vec{210}'$}	& $(\vec{1},\vec{2},\vec{2})$  						& $(\vec{1},\vec{2},\vec{2},0)$		&$2_{1/2}$ \\
\cline{2-4}  					& \multirow{2}{*}{$(\vec{6},\vec{3},\vec{3})$}  	& $(\overline{\vec{3}},\vec{3},\vec{3},2)$		&/ \\
\cline{3-4}  					& 												  	& $(\vec{3},\vec{3},\vec{3},-2)$	&/ \\
\cline{2-4}  					& $(\vec{1},\vec{4},\vec{4})$  						& $(\vec{1},\vec{4},\vec{4},0)$		&$4_{1/2}, \,4_{3/2}$ \\
\cline{2-4}  					& \multirow{3}{*}{$(\vec{20}',\vec{2},\vec{2})$}	& $(\vec{8},\vec{2},\vec{2},0)$		& / \\
\cline{3-4} 					&  													& $(\vec{6},\vec{2},\vec{2},4)$		& / \\
\cline{3-4} 					&  													& $(\overline{\vec{6}},\vec{2},\vec{2},-4)$	& / \\
\cline{2-4} 					& \multirow{2}{*}{$(\vec{6},\vec{1},\vec{1})$}		& $(\overline{\vec{3}},\vec{1},\vec{1},2)$		&  /\\
\cline{3-4} 					&													& $(\vec{3},\vec{1},\vec{1},-2)$	&  /\\
\cline{2-4} 					& \multirow{4}{*}{$(\vec{50},\vec{1},\vec{1})$}  	& $(\overline{\vec{15}},\vec{1},\vec{1},2)$	& /\\
\cline{3-4} 					& 													& $(\vec{15},\vec{1},\vec{1},-2)$	& /\\
\cline{3-4} 					& 													& $(\overline{\vec{10}},\vec{1},\vec{1},6)$	& /\\
\cline{3-4} 					& 													& $(\vec{10},\vec{1},\vec{1},-6)$	& /\\
\hline 
\end{tabular}
\caption{
Decomposition of $SO(10)$ representations between $\vec{144}$ and $\vec{210}'$ under PS and LR. In the last column we identify possible DM components in their SM notation.
\label{tab:DM_B}}
\end{table}
}

\item \underline{The low-scale DM partners}: Even if colored or charged, some of the partners may be present at the low DM scale.
These are the partners belonging to the same irreducible $SO(10)$-subgroup representation as the DM component, i.e.~those particles whose decay into DM proceeds through comparably light gauge bosons, namely the SM gauge bosons, the LR gauge bosons $W_R$ and $Z'$, or even the Pati--Salam boson $X$. This requires these partners to be heavier than the DM component by an amount which depends on the mass of these gauge bosons. As we will see, the necessary splittings of the DM component with the low-scale partners can in some cases be generated radiatively, unlike the splittings between DM and the high-scale partners. 

\begin{figure}[t]
\includegraphics[width=0.4\textwidth]{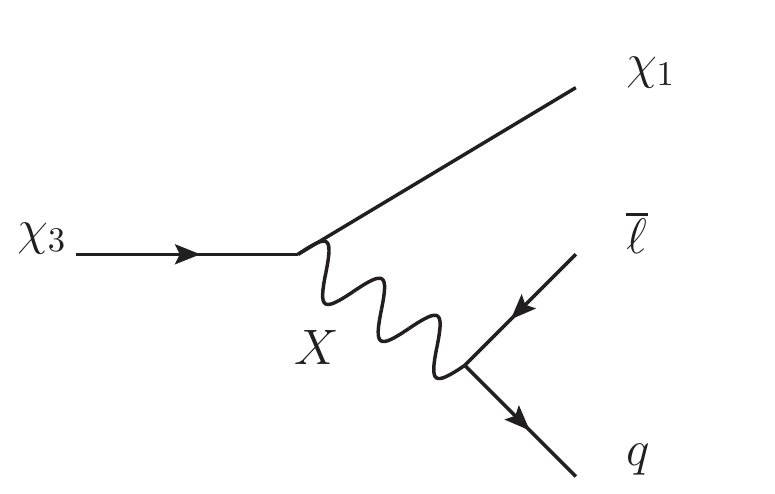}
\caption{Example for the decay of a DM partner into the DM particle. Here, the color triplet $\chi_3$ and Majorana singlet $\chi_1$ come from a $(\vec{15},\vec{1},\vec{1})$ under PS and thus decay via the PS gauge boson $X$ into each other plus SM fermions.}
\label{fig:decay_via_X}
\end{figure}

To be more explicit and quantitative on the decay of the low-scale partners, let us consider as an example a chiral $(\vec{15},\vec{1},\vec{1})$ multiplet of the Pati--Salam group $G_{422}$. This multiplet contains a color singlet Majorana $\chi_1$ (the DM candidate), two color triplets (which form one Dirac fermion $\chi_3$), and a Majorana color octet $\chi_8$. Obviously the triplet and octet must have decayed by today or even by the time of BBN, about $\unit[0.1]{s}$ after the Big Bang~\cite{Kawasaki:2004qu}.
The dominant decay channel for the triplet is into $\chi_1$ and a lepton-quark fermion pair (through a Pati--Salam gauge boson $X$), as shown in Fig.~\ref{fig:decay_via_X}, with relevant interactions
\begin{align}
\qquad g_4 X_\mu &\left[ \frac{1}{\sqrt{2}} \overline{Q}\gamma^\mu P_L L + \frac{1}{\sqrt{2}} \overline{d}\gamma^\mu P_R \ell + \frac{1}{\sqrt{2}} \overline{u}\gamma^\mu P_R N \right. \nonumber\\
& \quad \left. + \sqrt{\frac{2}{3}} \overline{\chi}_3 \gamma^\mu \chi_1 + \frac{2^{1/4}}{\sqrt{3}} \overline{\chi}_3 \gamma^\mu \chi_8 \right] +\hc ,
\label{eq:Xinteractions}
\end{align}
where we also give the $X$ couplings to the SM fermions $Q$, $\ell$, $u$, $d$, as well as the right-handed neutrinos $N$.
Summing over all SM fermion channels and assuming these SM fermions to be massless (including for definiteness the right-handed neutrinos)
we find the $\chi_3$ decay rate
\begin{align}
\Gamma_{\chi_3} = \frac{m_3^5}{192\pi^3} \frac{g_4^4}{m_X^4} h \left( \frac{m_1}{m_3}\right).
\label{eq:decay_of_15}
\end{align}
$m_1$ and $m_3$ are the masses of the singlet and triplet, respectively, and we have the phase-space function
\begin{align}
\hspace{4ex} h(x) &= 1 - 2 x - 8 x^2 - 18 x^3 + 18 x^5 + 8 x^6 \nonumber \\
&\quad + 2 x^7 - x^8- 24 x^3 (1 + x + x^2) \log (x)\,,
\end{align}
which goes to $h(x)\simeq \frac{32}{5} (1 - x)^5$ if $x\simeq1$ (quasi-degenerate triplet--singlet case).
Note that the decay rate goes down by a factor of $3/4$ if the right-handed neutrinos are too heavy to be accessible.
Demanding that the triplet decays before BBN, roughly $\Gamma_{\chi_3}^{-1} <\unit[0.1]{s}$, gives the \emph{upper} bound on the PS scale
\begin{align}
m_X/g_4 \lesssim \unit[10^{6}]{TeV} \,\left(\frac{m_3}{\unit[2]{TeV}}\right)^{5/4} \,,
\end{align}
assuming conservatively $m_1 \ll m_3$, otherwise the decay will be further suppressed by phase space and $m_X/g_4$ needs to be even lower. This will set in many scenarios an upper bound on the PS scale.

The discussion for the LR scale for decays mediated by a charged to neutral component of a $SU(2)_R$ multiplet via $W_R$ is analogous. 
Note that decays via the neutral $Z'$ would be between two neutral particles, so the DM partner would be neutral and hardly a problem for BBN. Even in this case one could find constraints from e.g.~CMB on the energy injection from a long-lived $\chi\to\text{DM}+\text{SM}$, but this will not be the focus here. As for decays mediated by SM gauge bosons, it is well known that mass splittings of only a few MeV are enough to make the decay of the charged partner of DM in an SM multiplet fast enough, as discussed at length for the case of Minimal DM~\cite{Cirelli:2005uq}.

As we will see in the following, the low-scale mass splittings are not only crucial for allowing a sufficiently fast decay of the partners but also in some cases to get the right amount of DM relic density via co-annihilation. 

\end{itemize}

Given that mass splittings are crucial for the phenomenology of $SO(10)$ DM, let us discuss explicitly how they can be generated. Here we will discuss the tree-level and radiative splittings separately:
\begin{itemize}
\item \underline{Tree-level mass splittings}: To know what are the various possible structures of tree-level splittings we can get, 
one needs to know what are the various scalar fields that can contribute to their masses. For example, chiral multiplets $R$ can be split at tree level by introducing scalars in representations found in the product $(R\otimes R)$ and letting them acquire VEVs.
One also needs to know the Clebsch--Gordan coefficients which weight the various VEVs for the various components. These can be determined efficiently with the program \texttt{Susyno}~\cite{Fonseca:2011sy} or (partly) found in Ref.~\cite{Fukuyama:2004ps}. In App.~\ref{app:tables} we give all the relevant Clebsch--Gordan coefficients that appear in the product $(R\otimes R)$ for $R$ up to $\vec{210}'$. Note that along the various possible breaking paths of $SO(10)$ there are in most cases breaking scales that do not contribute to the mass formula of the DM multiplet, either because they do not show up in $(R\otimes R)$ or because they contribute only to the anti-symmetric combination, which vanishes if we have only one generation of $R$. Thus, to see what are the possible mass spectra of DM and its multiplet partners, one will not need to consider explicitly each possible breaking path.
It is sufficient to consider the various possible hierarchies one could have between the mass contributions of the various scalar representations entering in the mass formula, no matter that along an explicit given $SO(10)$ breaking path there could be other scalar representations not contributing to the mass formula. This largely simplifies the discussion for all DM representations except for the $\vec{210}$ where all possible scalar representations contribute to the mass (except for $\vec{126}$), which makes the discussion more involved.

In particular, for the Pati--Salam ways and scalar representations up to $\vec{210}$, there are seven scalar fields entering in the symmetric product $(R\otimes R)_S$. The mass for each component of the DM multiplet is thus in full generality a combination of the universal mass $m_\vec{1}$ all $SO(10)$ multiplet components receive (i.e.~the GUT symmetry \emph{conserving} contribution) and of the VEVs $v$ of these seven scalar fields, each one multiplied by the corresponding Clebsch--Gordan coefficient $c$. For any fermion $f$ one thus has the general tree-level "master" mass formula
\begin{align}
m_f = m_\vec{1} &+c^f_{\vec{45}_{(\vec{15},\vec{1},\vec{1})}} v_{\vec{45}_{(\vec{15},\vec{1},\vec{1})}}+c^f_{\vec{45}_{(\vec{1},\vec{1},\vec{3})}} v_{\vec{45}_{(\vec{1},\vec{1},\vec{3})}}\nonumber \\
&+c^f_{\vec{54}_{(\vec{1},\vec{1},\vec{1})}}v_{\vec{54}_{(\vec{1},\vec{1},\vec{1})}} +c^f_{\vec{210}_{(\vec{1},\vec{1},\vec{1})}} v_{\vec{210}_{(\vec{1},\vec{1},\vec{1})}} \\
&+c^f_{\vec{210}_{(\vec{15},\vec{1},\vec{1})}} v_{\vec{210}_{(\vec{15},\vec{1},\vec{1})}}+c^f_{\vec{210}_{(\vec{15},\vec{1},\vec{3})}} v_{\vec{210}_{(\vec{15},\vec{1},\vec{3})}} \,, \nonumber
\end{align}
where the various scalar fields are defined according to their Pati--Salam  $G_{422}$ quantum numbers.
We omitted the VEV of a $\vec{126}$ scalar even though it can be used to break $SO(10)$ because it does not couple to any of the DM multiplets under discussion here. There are hence only \emph{six} scalar VEVs relevant for the DM splitting.
Computing the relevant $SO(10)$ Clebsch--Gordan coefficients, given in App.~\ref{app:tables}, one can determine the particles in the DM $SO(10)$ multiplet which can arise at the same scale as the DM particle. 

Similarly for the Georgi--Glashow paths (Fig.~\ref{fig:so(10)_breaking_paths_GG}) there are also six relevant scalar fields and we have
\begin{align}
m_f = m_\vec{1} &+c^f_{\vec{45}_{(\vec{1},0)}} v_{\vec{45}_{(\vec{1},0)}}+c^f_{\vec{45}_{(\vec{24},0)}} v_{\vec{45}_{(\vec{24},0)}}\nonumber \\
&+c^f_{\vec{54}_{(\vec{24},0)}}v_{\vec{54}_{(\vec{24},0)}}
+c^f_{\vec{210}_{(\vec{1},0)}} v_{\vec{210}_{(\vec{1},0)}} \\
&+c^f_{\vec{210}_{(\vec{24},0)}} v_{\vec{210}_{(\vec{24},0)}}+c^f_{\vec{210}_{(\vec{75},0)}} v_{\vec{210}_{(\vec{75},0)}} \,,\nonumber
\end{align}
where the various scalar fields are defined according to their Georgi--Glashow $G_{51}$ quantum numbers.
Note that we will mostly focus on the PS paths and not the Georgi--Glashow paths, because the former allow for more low-energy subgroups. This stems from the fact that proton decay constrains the $SU(5)$  group to be broken at a very high scale, above $\sim \unit[10^{13}]{TeV}$. The only Georgi--Glashow subgroup allowed at lower scales is $G_{3211}$ from $SU(5)\times U(1)_\chi \to G_\text{SM}\times U(1)_\chi$~\cite{Masiero:1980dd}. This has qualitatively the same phenomenology as the $G_{3211}$ obtained from PS, differing only in the $Z'$ couplings. We will therefore not give the Clebsch--Gordan coefficients along the Georgi--Glashow paths in this work.

Non-zero tree-level mass splittings require Yukawa couplings of the DM multiplet to scalars. In the following we will neglect these interactions when it comes to DM phenomenology, e.g.~in the freeze-out process. This is a valid approximation if all the scalars are sufficiently heavier than the DM particle or the Yukawa couplings are very small. While the former is a rather natural outcome of a multi-scale theory, it is also highly unwelcome, seeing as at least the SM scalar doublet has to be light. It is hence perfectly possible that other scalars could also be light and thus relevant for our discussion of DM. We have to neglect these scenarios here due to the sheer number of additional free parameters and possibilities this would introduce.

\item \underline{Radiative mass splittings}: Given a set of scalar representations contributing to the masses at tree level, in general a set of sum rules among the masses will be obtained, i.e.~certain linear combinations of masses vanish, $\sum_j d_j m_j =0$, with $m_j$ the masses of the various DM multiplet components. 
However, even if the gauge group is further broken only by scalars that do not couple directly to $R$, these sum rules will in general be broken at loop level, as can be verified by calculating the fermion self-energies with gauge bosons in the loop~\cite{Weinberg:1972ws,Weinberg:1973ua}. $\sum_j d_j m_j$ then becomes a calculable observable, seeing as  the lack of counterterms for this quantity implies \emph{finite} loop corrections. Two popular examples here are wino and minimal DM~\cite{Cirelli:2005uq}, which transform as $\vec{3}$ and $\vec{5}$ under $SU(2)_L$, respectively, and do not couple to any scalars. At one-loop level the degenerate electric-charge $Q$ eigenstate components split,
\begin{align}
\hspace{-1.01ex} m_Q - m_0 \simeq Q^2 
\begin{cases}
\alpha_2 m_W \sin^2\left(\frac{\theta_W}{2}\right) & \hspace{-1ex}\text{for } m_W \ll m_0 \;,\\
\frac{3 \alpha }{2 \pi } m_0 \log \left(\frac{m_W}{m_0}\right) & \hspace{-1ex} \text{for } m_W \gg m_0 \;,
\end{cases}
\label{eq:wino_splitting}
\end{align}
where the masses are physical pole masses and the couplings are running $\overline{\text{MS}}$ parameters, with a renormalization scale dependence that is canceled by higher loop orders~\cite{Ibe:2012sx}. One sum rule remains unbroken at this level for the quintuplet, $(m_2 -m_0) -4 (m_1 - m_0) =0$, which, surprisingly, seems to be even satisfied at the two-loop level~\cite{Yamada:2009ve,McKay:2017xlc}.

The approximate mass splitting for the hierarchy $m_W \gg m_0$ can equivalently be calculated using effective field theory. For this we convert the pole masses $m_Q$ to \MSbar masses $M_Q (\mu)$, which we run up to the high scale $m_W$ using standard QED formulas and impose the common-multiplet boundary conditions $M_Q (m_W ) = M_0 (m_W )$ $\forall Q$. We can easily do this more generally for any group $G$ that is broken to a subgroup $H$ at scale $m_G$: fermions with degenerate mass $m \ll m_G$  that used to form an irreducible representation of $G$ are now split into representations $R_j$ of $H$, with pole-mass splitting
\begin{align}
\hspace{4ex}\frac{m(R_i) - m(R_j)}{m} &\simeq \frac{3 \alpha (m) }{2 \pi } (\Delta C_2)_{ij} \log \left(\frac{m_G}{m}\right) ,
\label{eq:mass_splitting}
\end{align}
$(\Delta C_2)_{ij}  \equiv C_2 (R_i)-C_2 (R_j)$ being the difference of quadratic Casimir invariants (a selection is given in Tab.~\ref{tab:casimirs} in App.~\ref{app:rge}) and $\alpha$ the fine structure constant of $H$. If $H$ is a direct product of groups the right-hand side of Eq.~\eqref{eq:mass_splitting} becomes a sum over the corresponding couplings and Casimirs.
The above approximation breaks down for large representations under strongly coupled groups and should be replaced by  a more careful treatment that sums up the large logs, similar to the procedure for gluino masses~\cite{Martin:2006ub}. We refer to App.~\ref{app:rge} for a discussion.

Despite the potentially large logarithm $\log (m_G/m)$ in Eq.~\eqref{eq:mass_splitting}, it is clear that purely radiative mass splitting cannot generate huge hierarchies among the components of a GUT multiplet, explaining why we said above that the splittings between the DM component and its heavy partners must be necessarily induced at tree level.

\end{itemize}

\section{List of DM viability constraints}
\label{sec:constraints}

In this section we give the full list of criteria we apply to retain a DM scenario. This list contains the criteria already discussed above as well as other simple ones, based on viability and minimality. They focus on the DM pattern, without trying to solve many other issues that arise in $SO(10)$ GUTs, as these issues could easily depend on many extra ingredients that 
are to a large extent independent of the DM pattern. These criteria and assumptions are as follows:
\begin{itemize}
\item \underline{DM representation}: We limit ourselves to a single $SO(10)$ representation with dimension up to 210. Adding several candidates that mix with each other can lead to interesting phenomenology~\cite{Arbelaez:2015ila,Boucenna:2015sdg} but would lead us too far.

\item \underline{Color-singlet DM}: As already mentioned above, it is easy to show that all the colored DM candidates which show up are excluded, so that they will not appear anywhere below. 

 \item \underline{Relic density}: We check that the DM scenario can lead to the observed relic density in a thermal way, i.e.~from (co-)annihilation freeze-out. In App.~\ref{app:chem_dec} we explain how DM can be obtained through co-annihilation, as relevant for the scenarios we will consider. This requires that the co-annihilating particles are in chemical equilibrium, a condition that is also derived in this appendix.
 
\item \underline{Low-scale DM partners}: As already mentioned above, given the strong constraints which exist on any stable colored or charged particle relic density today, one must make sure that any colored or charged DM partner decays.  This depends on the mass and mass splittings involved, see the discussion above.

\item \underline{Direct detection}: We will make sure that the DM--nucleon cross section  induced in the various scenarios is not already excluded by current direct detection experiments \cite{Aprile:2017iyp,Aprile:2018dbl}. As mentioned above, direct detection constraints exclude all the candidates with a non-vanishing hypercharge which could show up otherwise below, except the left--right bi-doublet and bi-quadruplet.
We will not look at indirect detection constraints, except when these constraints have already been studied in the literature. 

\item \underline{Fine-tuning of the DM mass}: In all scenarios below we need to have some DM partners at a much higher scale than the low-scale fermion DM candidate, the above called "high-scale DM partners."
This disparity of scales
turns out to imply in all cases a fine-tuning, i.e.~a cancellation of at least two DM mass contributions which are both larger than the DM mass.
This might not look like much of a surprise given the fact that the framework we consider is non-supersymmetric, but it is actually a rather subtle point (see the discussion of the DM representation $\vec{120}$ in Sec.~\ref{sec:120} below). In the following we will allow for one DM mass contribution cancellation and not more. Of course, if one allows for one cancellation, nothing prevents one from having more cancellations, but already with one we cover a wide range of possibilities. We will discuss briefly  in Sec.~\ref{sec:Otherscenarios} what could change when more cancellations are assumed.
Similarly, we will not try to solve other fine-tuning issues that may arise in non-supersymmetric GUTs, especially in the scalar sector, including the doublet--triplet splitting problem. We will also not consider any explicit scalar potential to see how the patterns of symmetry breaking scales assumed in the various scenarios could actually be achieved.
 
\item \underline{Yukawa interactions}: DM couplings to scalars are unavoidable in order to obtain the tree-level mass splittings necessary for viable mass spectra. If these couplings are large and the scalars not too heavy they could have a major impact on the DM phenomenology, say the relic abundance. We will neglect these interactions in the following and only study gauge interactions. 
 
\item \underline{Gauge unification}: We will not demand our setup to lead to gauge unification, because this does not only depend on the general breaking pattern and DM multiplet considered here, but also (see e.g.~Refs.~\cite{Altarelli:2013aqa,Dueck:2013gca,Chakrabortty:2017mgi,Ohlsson:2018qpt}) on the exact values of the breaking scales, the values of the Yukawa couplings multiplying these scales, on the boundary conditions assumed at the GUT scale, on renormalization group running effects, and on the existence of other possible $SO(10)$ multiplets whose masses could show up basically anywhere without much affecting our DM discussion. Note nevertheless that in some cases these extra multiplets could affect the masses of the DM multiplet components. Thus, whenever we use running parameters in the following, they are to be understood as benchmark values within the most minimal models and hence could be subject to change in full models.
Of course, the cases below that have many low-energy states could have a very large effect on the running of the gauge couplings, which would typically require other states at intermediate scales in order to have gauge unification, but again we will not consider this.

\item \underline{Fermion masses}: For the same reasons as for gauge unification, we will not look at the way SM fermion masses and neutrino masses can be accounted for.

\end{itemize}

\section{List of low-scale DM scenarios}
\label{sec:lowscale}

This section displays the practical low-energy  output of the analysis performed in this work: here we list the various low-scale scenarios obtained from considering stable fermionic DM  representations up to $\vec{210}'$ and briefly discuss the phenomenology of each of these scenarios. On the basis of the constraints described in the previous section, this list is obtained in a top-down way from the detailed and more involved discussion given in Sec.~\ref{sec:viable}. We will present first the scenarios which do not lead to any colored partner at low scale and subsequently the ones which do predict low-energy colored partners.
This order corresponds mostly to the order which would show up if we were presenting the scenarios in the order they appear starting from a  $\vec{10}$ DM representation all the way up to the $\vec{210}'$.  Note that in the following the number indicated in the name of each scenario refers to the number of tree-level fermionic degrees of freedom this scenario involves. 
For example, the "octet--singlet bi-doublet 32+4" scenario of Sec.~\ref{sec:Octet-singlet_bi-doublet_32+4} contains at low energy one color octet that is also a bi-doublet under $SU(2)_L\times SU(2)_R$ for a total number of $8\times 2\times 2=32$ degenerate states plus one color-singlet bi-doublet with four degenerate states, but with a different mass from the octet.
To our knowledge, the majority of scenarios listed below are basically new (see scenarios in subsections D,G,H,J-R).

\subsection{LR bi-doublet 4}
\label{sec:LR_bi-doublet_4}

\noindent
\hspace*{2.5mm}- \underline{\it Low-scale content}: A tree-level degenerate chiral bi-doublet of $SU(2)_L\times SU(2)_R$ coming from the $(\vec{1},\vec{2},\vec{2})_\text{PS}$ in $\vec{10}$ or $\vec{120}$ or the $(\vec{15},\vec{2},\vec{2})_\text{PS}$ in $\vec{120}$. DM is the lightest neutral component of it.
\vspace*{1mm}\\
\hspace*{2.5mm}- \underline{\it Mass splittings}: The bi-doublet forms one Dirac doublet $\left(\vec{1},\vec{2},\tfrac{1}{2}\right)$ under the SM group; this Higgsino-like doublet with mass $m$ is ultimately split by electroweak loops, which make the charged component $\chi^+$ approximately $\alpha \,m_{Z}/2\simeq \unit[360]{MeV}$ heavier than the neutral one~\cite{Thomas:1998wy,Cirelli:2005uq}. Still, the neutral component of the doublet is naively excluded as a DM candidate due to its hypercharge, which implies a coupling to the $Z$ boson and thus large direct-detection cross sections. However, once the LR symmetry is broken, this neutral Dirac component is actually split into two Majorana fermions $\chi_{1,2}$ at the one-loop level via $W_L$--$W_R$ mixing (see Ref.~\cite{Garcia-Cely:2015quu}) leading to the mass splitting within the Dirac doublet $\left(\vec{1},\vec{2},\tfrac{1}{2}\right)$
\begin{align}
m_{\chi^+} - m_{\chi_{1,2}} &\simeq \frac{\alpha}{2} m_{Z_1} \label{eq:bidoublet_splitting} \\
&\quad\pm \frac{\alpha_2}{8\pi} \frac{g_R}{g_L} m \sin (2\xi) \left[ f(r_{W_2})-f(r_{W_1})\right] , \nonumber
\end{align}
with $W_L$--$W_R$ mixing angle $\xi$, mass ratio $r_V \equiv m_V/m$, and loop function
\begin{align}
f(r)&\equiv 2\int_0^1 \dd x \, (1+x) \log \left[ x^2 + (1-x) r^2\right]\\
&= -5-r^2+r^4 \log r \nonumber\\
&\quad +\frac{r}{2}  \sqrt{r^2-4} \left(2+r^2\right) \log \left[ \frac{r^2-2-r \sqrt{r^2-4}}{2}\right] .\nonumber
\end{align}
For LR scales below $\sim \unit[75]{TeV}$, the induced splitting between the neutral components can be above $\unit[200]{keV}$, enough to kinematically forbid the now inelastic direct-detection scattering~\cite{TuckerSmith:2001hy,Nagata:2014aoa}. 
Thus one can get a viable DM candidate only if the LR group is broken below $\sim \unit[75]{TeV}$. This hence requires a symmetry breaking path proceeding through a low-scale LR group. 
The final mass spectrum of the $\vec{10}$ is illustrated in Fig.~\ref{fig:spectrum_10}.
\vspace*{1mm}\\
\hspace*{2.5mm}- \underline{\it Relic density}: If the LR gauge bosons $W_R$ and $Z'$ are much heavier than the DM, the relic abundance is mainly set by DM annihilation into SM gauge bosons, which fixes the Higgsino-like DM mass to be $\unit[1.2]{TeV}$~\cite{Profumo:2004at,Cirelli:2007xd}. However, much larger masses up to $\unit[30]{TeV}$ become viable if DM annihilates into SM particles via the $W_R$ or $Z'$ resonances, i.e.~for $m_\text{DM} \simeq m_{W_R}/2$ or $m_{Z'}/2$~\cite{Garcia-Cely:2015quu}. 
The exact way the relic density is obtained depends on whether the LR group is broken directly to the SM or goes through the $G_{3211}$ intermediate step, as this changes the ratio $m_{W_R}/m_{Z'}$.
Without the intermediate step $G_{3211}$ it has been shown in Ref.~\cite{Garcia-Cely:2015quu} that the bi-doublet of the LR group leads to a good DM candidate with mass between TeV and $\unit[30]{TeV}$ and a low LR scale.
\emph{With} the intermediate $G_{3211}$ step one can achieve a gauge-boson hierarchy $m_{Z'}\ll m_{W_R}$ which can potentially change the DM phenomenology. We still need $m_{W_R}\lesssim \unit[75]{TeV}$ for a large enough DM mass splitting, so $Z'$ and $W_R$ cannot actually be too far apart considering $m_{Z'}$ has to be multi-TeV to evade existing constraints (Tab.~\ref{tableLowerLimits}). For $m_\text{DM} \ll m_{Z'}\ll m_{W_R}$, the DM abundance is once again set by annihilation into SM gauge bosons, so one requires again $\unit[1.2]{TeV}$ for the correct relic abundance. Considering $m_\text{DM} \sim m_{Z'}\ll m_{W_R}$ allows one to go to higher DM masses as long as one sits on the $Z'$ resonance. The last possible hierarchy, $m_{Z'}\ll m_\text{DM}$, in which the additional annihilation channel into $Z' Z'$ would become relevant is excluded by the lower bounds on $m_{Z'}$ (Tab.~\ref{tableLowerLimits}). The phenomenology with or without $G_{3211}$ is hence qualitatively similar. 
\vspace*{1mm}\\
\hspace*{2.5mm}- \underline{\it LHC}: Both direct and indirect detection are heavily suppressed~\cite{Garcia-Cely:2015quu}, so the best signature of this model is the existence of relatively light gauge bosons $W_R$ and $Z'$ as well as the charged DM partners at the DM mass scale modulo the small $\unit[360]{MeV}$ mass splitting.
The LHC phenomenology of the bi-doublet is very similar to (pseudo-Dirac) Higgsinos, in particular the decay rate of $\chi^+$ into DM and pions~\cite{Thomas:1998wy,Garcia-Cely:2015quu}. As such, the LHC will not be able to probe the relevant parameter space, but a $\unit[100]{TeV}$ collider could probe the TeV mass region~\cite{Mahbubani:2017gjh,Krall:2017xij,Han:2018wus,Chigusa:2018vxz}, i.e.~the non-resonant regime.

\subsection{$SU(2)_L$ triplet 3}
\label{sec:SU(2)_L_triplet_3}

\noindent
\hspace*{2.5mm}- \underline{\it Low-scale content}: One $SU(2)_L$  triplet
coming from the $(\vec{1},\vec{3},\vec{1})_\text{PS}$ in $\vec{45}$ or $(\vec{15},\vec{3},\vec{1})_\text{PS}$ in $\vec{210}$.
This is the usual \emph{wino} DM scenario that is well known from the MSSM:
\begin{align}
m_{(\vec{1},\vec{3},0)} \equiv m_\text{DM}\,.
\end{align}
\hspace*{2.5mm}- \underline{\it Mass splittings}: $m_+ - m_0 \simeq \unit[166]{MeV}$ [Eq.~\eqref{eq:wino_splitting}].
\vspace*{1mm}\\
\hspace*{2.5mm}- \underline{\it Relic density}: 
The annihilation proceeds into SM gauge bosons in the usual wino way, which gives~\cite{Hisano:2006nn,Mitridate:2017izz}
\begin{align}
\Omega_L h^2 \simeq 0.12 \left(\frac{m_\text{DM}}{\unit[2.7]{TeV}}	\right)^2\,.
\label{eq:wino_abundance}
\end{align}
The above formula misses the Sommerfeld resonances, e.g.~at $m_\text{DM}\sim\unit[2.4]{TeV}$, but is a good approximation overall.
If all DM is in the form of this triplet, we need $m_\text{DM}\simeq\unit[2.7]{TeV}$~\cite{Mitridate:2017izz}, which leads to a Sommerfeld-enhanced production of monochromatic photons and is basically excluded by observation for the most standard  galactic DM density profiles Navarro--Frenk--White and Einasto; it is, however, still allowed for an isothermal profile~\cite{Cohen:2013ama,Baumgart:2018yed}.
\vspace*{1mm}\\
\hspace*{2.5mm}- \underline{\it LHC}: The current lower limit on the wino mass is $\unit[460]{GeV}$~\citep{Aaboud:2017mpt}, and future limits can be found in Refs.~\cite{Cirelli:2014dsa,Han:2018wus}. The LHC will not be able to probe the relic-density motivated region $m_\text{DM}\sim\unit[2.7]{TeV}$, but a $\unit[100]{TeV}$ collider might.

Note that this is the only $SO(10)$ DM scenario that does not give any restrictions on $SO(10)$ subgroups and scales, thanks to the fact that it can annihilate into SM gauge bosons and that electroweak loops provide the necessary mass splittings.

\subsection{$SU(2)_R$ triplet 3}
\label{sec:SU(2)_R_triplet_3}

\begin{figure}[t]
\includegraphics[width=0.5\textwidth]{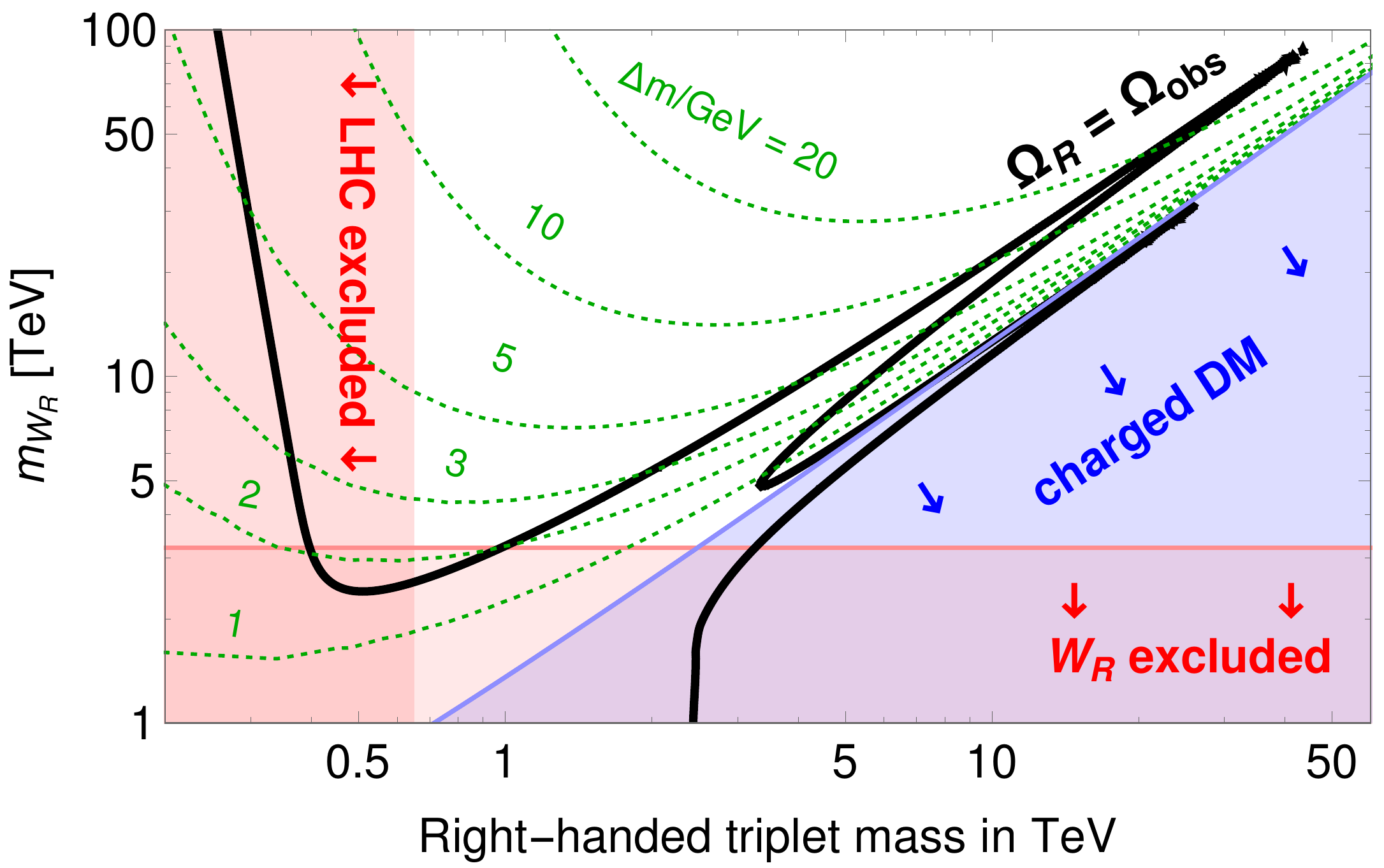}
\caption{
The correct relic density of a right-handed triplet (black), assuming $g_R=g_L$, which fixes $m_{Z'}/m_{W_R}$. The dotted green lines show radiative-mass-splitting contours as determined by Eq.~\eqref{eq:RH_mass_splitting}. In the blue region the radiative mass splitting within the triplet is negative, rendering the charged component stable~\cite{Heeck:2015qra,Garcia-Cely:2015quu}.
The orange region is excluded by meson mixing constraints and searches for singly-charged collider-stable particles.}
\label{fig:right-handed_wino}
\end{figure}

\noindent
\hspace*{2.5mm}- \underline{\it Low-scale content}: A tree-level degenerate $SU(2)_R$ triplet from the $(\vec{1},\vec{1},\vec{3})_\text{PS}$ in $\vec{45}$ or $(\vec{15},\vec{1},\vec{3})_\text{PS}$ in $\vec{210}$:
\begin{align}
m_{(\vec{1},\vec{1},1)}=m_{(\vec{1},\vec{1},0)} \equiv m_\text{DM}\,.
\end{align}
Note that a scenario with only this triplet at low scale [and not also the $SU(2)_L$ triplet at low scale as in Refs.~\cite{Heeck:2015qra,Garcia-Cely:2015quu}] has to our knowledge been previously considered only in Ref.~\cite{Bandyopadhyay:2017uwc}.
\vspace*{1mm}\\
\hspace*{2.5mm}- \underline{\it Mass splittings}: The  mass splitting of the $SU(2)_R$ triplet induced by the  gauge interactions is given by
\begin{align}
\begin{split}
m_{Q}^R-m_{0}^R &\simeq \frac{\alpha_2}{4\pi} \frac{g_R^2}{g_L^2} m_\text{DM} Q^2 \left[ f(r_{W_2}) - c_M^2 f(r_{Z_2})\right.\\
&\quad\left.-s_W^2 s_M^2 f(r_{Z_1})- c_W^2 s_M^2 f(r_\gamma)\right] ,
\end{split}
\label{eq:RH_mass_splitting}
\end{align}
where $s_M = \sin (\theta_M) = \tan (\theta_W) g_L/g_R$ and $c_M =\cos (\theta_M)$. 
Here $Z_1$ and $W_1$ ($Z_2$ and $W_2$) denote the mass eigenstates which are essentially the SM gauge bosons [the $SU(2)_R$ gauge bosons~\cite{Heeck:2015qra,Garcia-Cely:2015quu}].
Some region of parameter space is excluded because the charged component $\chi^+_R$ is lighter than the neutral one (see Fig.~\ref{fig:right-handed_wino}), at least for $g_R = g_L$.
\vspace*{1mm}\\
\hspace*{2.5mm}- \underline{\it Relic density}: 
Since the neutral Majorana component $\chi^0_R$ of the $SU(2)_R$ triplet does not couple to the $Z'$, the relic density is necessarily
set by processes which couple $\chi^0_R$ to the charged component $\chi^+_R$ and a $W_R$.
There are two types of relevant processes: direct co-annihilation of the neutral state into light SM states $\chi^+_R \chi^0_R \to W_R \to \text{SM} \,\text{SM}$ and two-step co-annihilation, i.e.~the conversion $\chi^0_R \,\text{SM} \to \chi^+_R\, \text{SM}$ followed by $\chi^+_R \chi^-_R$ annihilation into SM particles mediated by $\gamma$ and $Z$. The values of the DM and $W_R$ masses needed to reproduce the observed relic density, as calculated using {\sc MicrOMEGAs}~\cite{Belanger:2006is} and taking into account the radiative mass splitting, are given in Fig.~\ref{fig:right-handed_wino}.

From the direct process, one can reproduce the observed relic density for DM masses up to $\unit[50]{TeV}$ around the $W_R$ resonance $m_\text{DM} \sim m_{W_R}/2$.
For larger $m_{W_R}$ to $m_\text{DM}$ ratio, the relic density can be achieved through the $\chi^0_R \,\to \chi^+_R$ conversion driven co-annihilation process, with subsequent annihilation of these charged states into photons and $Z$ bosons.
This leads to the correct relic density for a DM mass around $300$--$\unit[500]{GeV}$~\cite{Heeck:2015qra,Garcia-Cely:2015quu}, as shown in Fig.~\ref{fig:right-handed_wino}. This calculation is similar to the one performed in Refs.~\cite{Heeck:2015qra,Garcia-Cely:2015quu} but takes into account the finite value of the radiative mass splitting. Co-annihilation requires
the transition rate $\chi^0_R\to \chi^+_R$ to be in thermal equilibrium around the time of freeze-out, which is easily achievable for $m_{W_R}/g_R \lesssim \unit[\mathcal{O}(10^2)]{TeV}$.\footnote{More precisely, using the results from App.~\ref{app:chem_dec}, we obtain that chemical equilibrium requires $m_{W_R}\lesssim \unit[170]{TeV}$ for $m_0^R =  \unit[300]{GeV}$ and a mass splitting of $\unit[5]{GeV}$, assuming $g_R=g_L$ and heavy right-handed neutrinos. For the case of $m_0^R = \unit[500]{GeV}$ and zero splitting, relevant for the scenario of Sec.~\ref{sec:SU(2)_R_triplet_2+1}, we get instead $m_{W_R} \lesssim \unit[260]{TeV}$. }
Finally, the charged states could also annihilate into SM particles via a $Z'$ in the $s$ channel. If one lies at the resonance of this process, $m_{Z'}\sim 2 m_{\chi_R^+}$, one can also reproduce the observed relic density, which leads 
to the lower resonance peak in Fig.~\ref{fig:right-handed_wino}. This region is, however, invalid because the radiative mass splitting turns negative in that region, at least for the LR symmetric case, $g_R = g_L$, i.e.~using the usual $m_{Z'}/m_{W_R} \simeq \sqrt{1+\sec 2\theta_W}$ mass relation.
If $g_R>g_L$, the allowed parameter space shrinks because this leads to a more extended excluded region where the charged state is lighter than the neutral one and because in this case the LHC lower limit on $m_{W_R}$ increases.
The case $g_R<g_L$ leads to a pattern similar to the symmetric limit; it requires a slightly smaller value of $m_{W_R}$ for a given $m_\text{DM}$, and puts the $Z'$ resonance peak in the allowed region.

\hspace*{2.5mm}- \underline{\it LHC}: LHC searches for singly-charged \emph{stable} particles give strong constraints around $\unit[650]{GeV}$~\cite{Chatrchyan:2013oca,Khachatryan:2016sfv}. This excludes the entire co-annihilation region around $m_\text{DM}\simeq 300$--$\unit[500]{GeV}$, seeing as the small radiative mass splitting gives a very suppressed $\chi^+_R$ decay rate. Specifically, the decay  $\chi^+_R\to \chi^0_R e^+ N$ with massless right-handed neutrino $N$ gives
\begin{align}
\Gamma (\chi^+_R\to \chi^0_R e^+ N) &\simeq \frac{\Delta m^5}{240\pi^3}  \left(\frac{g_R}{m_{W_R}}\right)^4 \label{eq:decay_of_RH_triplet}\\
& \simeq \frac{1}{\unit[20]{m}} \left(\frac{\Delta m}{\unit[2]{GeV}}\right)^5 \left(\frac{\unit[3]{TeV}}{m_{W_R}}\right)^4 ,\nonumber
\end{align}
whereas the rate is a color factor $N_c =3$ larger for the light-quark modes $\chi^+_R\to \chi^0_R \overline{d} u$. This makes $\chi^+_R$ stable on collider scales and thus excludes $m_\text{DM}\lesssim \unit[650]{GeV}$ (considering here only the quark channels).
The only viable region of parameter space left is then near the $W_R$ resonance with $m_\text{DM}\gtrsim \unit[1]{TeV}$, as shown in Fig.~\ref{fig:right-handed_wino}, which given the fact that one can get the observed relic density for $m_\text{DM}$ up to $\sim \unit[50]{TeV}$ implicitly also gives an upper bound on the LR scale, $m_{W_R}\lesssim \unit[100]{TeV}$.

\subsection{$SU(2)_R$ triplet 2+1}
\label{sec:SU(2)_R_triplet_2+1}

\noindent
\hspace*{2.5mm}- \underline{\it Low-scale content}: One $SU(2)_R$ triplet with tree-level mass splitting
\begin{align}
 m_{(\vec{1},\vec{1},1)}\neq m_{(\vec{1},\vec{1},0)} \equiv m_{\text{DM}}\,,
\end{align}
coming from the  $(\vec{1},\vec{1},\vec{3})_\text{PS}$ in $\vec{45}$ or $(\vec{15},\vec{1},\vec{3})_\text{PS}$ in $\vec{210}$.
\vspace*{1mm}\\
\hspace*{2.5mm}- \underline{\it Mass splittings}: Compared to Sec.~\ref{sec:SU(2)_R_triplet_3} we have a tree-level mass splitting $\Delta m \equiv m_{(\vec{1},\vec{1},1)}-m_{\text{DM}}$ between the charged and neutral components.
\vspace*{1mm}\\
\hspace*{2.5mm}- \underline{\it Relic density \& LHC}: With respect to Sec.~\ref{sec:SU(2)_R_triplet_3} the tree-level mass splitting allows us to decouple the value of the splitting from the values of the other parameters. 

\begin{figure}[t]
\includegraphics[width=0.5\textwidth]{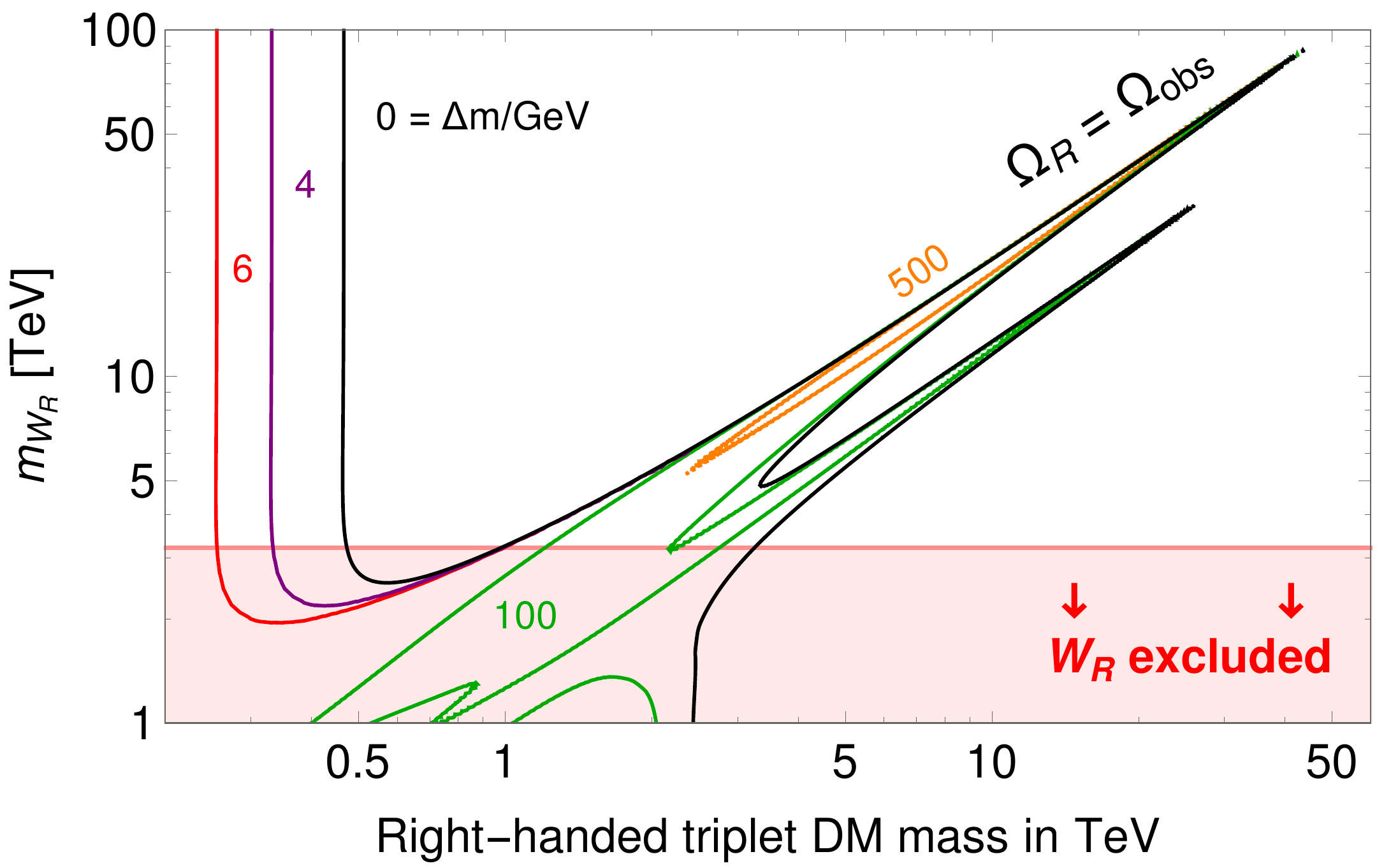}
\caption{
The correct relic density of a right-handed triplet, assuming $g_R=g_L$, for various fixed mass splittings.
}
\label{fig:SU2R_triplet_2plus1}
\end{figure}

For what concerns the $\chi^0_R \,\to \chi^+_R$ conversion driven co-annihilation regime, i.e.~the  $m_{\text{DM}}\sim 300$--$\unit[500]{GeV}$
region, to consider a larger mass splitting shortens the $\chi_R^+$ lifetime, which is welcome to soften the  $m_\text{DM}\gtrsim \unit[650]{GeV}$ LHC constraint (see Fig.~\ref{fig:right-handed_wino}). However, a larger mass splitting makes the co-annihilation process less efficient due to the Boltzmann-suppressed conversion rate $\langle \Gamma\rangle_{\chi^0_R \,\to \chi^+_R}\propto e^{-{\Delta m}/{T}}$ (App.~\ref{app:chem_dec}), which requires us to \emph{lower} the overall mass of $\chi_R^+$ and $\chi_R^0$ from $\unit[500]{GeV}$, potentially down to the LEP bound of $\unit[100]{GeV}$~\cite{Barate:2000tu,Heister:2002mn,Egana-Ugrinovic:2018roi}, as shown in Fig.~\ref{fig:SU2R_triplet_2plus1}. There is indeed a small region of parameter space where we can evade the LHC bound.	
Using for example $m_{\text{DM}}= \unit[250]{GeV}$ with a mass splitting $\Delta m =\unit[6]{GeV}$, we can obtain the correct relic density in the co-annihilation region (Fig.~\ref{fig:SU2R_triplet_2plus1}). For $m_{W_R}$ close to its current bound, the $\chi_R^+$ lifetime can be as low as $c \tau \sim \unit[1]{cm}$, short enough to evade the LHC constraints on stable charged particles. In fact, these decays could appear \emph{displaced} in ATLAS and CMS, a potentially advantageous search feature. As stated above, the dominant decay will be into light quark modes plus missing energy via $\chi^+_R\to \chi^0_R \overline{d} u$ and $\chi^0_R \overline{s} c$, with $\chi_R^+$ itself being pair-produced via Drell--Yan. Note that increasing $g_R > g_L$ will shorten the $\chi^+_R$ even further and render its decays prompt, without affecting much the co-annihilation calculation. The DM mass in this co-annihilation region is below $\unit[400]{GeV}$ and the charged partner less than $\unit[8]{GeV}$ heavier. This together with the necessarily as-light-as-possible $W_R$ make this region completely testable at the LHC. We strongly encourage a dedicated search for these states.

Increasing the mass splitting above $\sim\unit[8]{GeV}$ renders the co-annihilation region infeasible and thus requires us to go back to the resonant regions above $m_\text{DM}>\unit{TeV}$ (Fig.~\ref{fig:SU2R_triplet_2plus1}). Here the effects of $\Delta m$ are rather small until we increase it above $10\%$. A larger $\Delta m$ always makes (co-)annihilation less efficient, so one has to be closer and closer to the $W_R$ or $Z'$ resonances in order to end up with the desired abundance. For $\Delta m>\unit[500]{GeV}$ only the $\chi^+_R \chi^0_R \to W_R \to \text{SM} \,\text{SM}$ driven co-annihilation region is left, which furthermore requires $m_\text{DM}>\unit[2]{TeV}$.
The overall upper bound $m_{W_R} \lesssim \unit[100]{TeV}$ is still valid in all cases because the larger $m_{W_R}$ is the less efficient are the $W_R$ mediated processes.
The resonant region is clearly more difficult to probe, especially if the mass splitting is large, making the $\chi_R^+$ extremely short lived.

\subsection{LR triplet 6}
\label{sec:LR_triplet_6}

\noindent
\hspace*{2.5mm}- \underline{\it Low-scale content}: One triplet of $SU(2)_L$ and one of $SU(2)_R$ which are degenerate at tree level due to $D$~parity and can come from the $(\vec{1},\vec{3},\vec{1})_\text{PS}\oplus (\vec{1},\vec{1},\vec{3})_\text{PS}$ in $\vec{45}$ or $(\vec{15},\vec{3},\vec{1})_\text{PS}\oplus (\vec{15},\vec{1},\vec{3})_\text{PS}$ in $\vec{210}$. 
The tree-level masses of the fields, decomposed under $G_\text{SM}$, are
\begin{align}
m_{(\vec{1},\vec{3},0)}=m_{(\vec{1},\vec{1},1)}=m_{(\vec{1},\vec{1},0)} \equiv m_\text{DM}\,.
\end{align}
This is a two-component DM scenario, since the neutral Majorana components of both triplets are stable.
The phenomenology of this low-scale particle content has been studied in detail in Refs.~\cite{Heeck:2015qra,Garcia-Cely:2015quu}.
\vspace*{1mm}\\
\hspace*{2.5mm}- \underline{\it Mass splittings}: Radiative mass splittings as in Secs.~\ref{sec:SU(2)_L_triplet_3} and~\ref{sec:SU(2)_R_triplet_3}.
\vspace*{1mm}\\
\hspace*{2.5mm}- \underline{\it Relic density}: For the $SU(2)_L$ triplet the abundance is given by Eq.~\eqref{eq:wino_abundance}.
The $SU(2)_R$ triplet has an abundance $\Omega_R h^2$ that depends on the $W_R$ mass (Fig.~\ref{fig:right-handed_wino}) and can be used to obtain $\Omega_L h^2 +\Omega_R h^2\simeq 0.12$ for $\unit[1]{TeV}\leq m_\text{DM}\leq \unit[2.7]{TeV}$ (see Sec.~\ref{sec:SU(2)_R_triplet_3}). 
As already said above, the upper end of this mass range, where DM is mainly from the $SU(2)_L$ triplet, is in tension with indirect-detection constraints~\cite{Cohen:2013ama} (see Sec.~\ref{sec:SU(2)_L_triplet_3}). The $SU(2)_R$ triplet relic density is subdominant in this case, which can be obtained if one lies within the region bounded by both resonance peaks in Fig.~\ref{fig:right-handed_wino}.
This requires a very low LR scale of $m_{W_R}\lesssim \unit[7]{TeV}$.

The $m_\text{DM}\sim\unit[1]{TeV}$ lower end of this mass range, where DM is essentially from the $SU(2)_R$ triplet, corresponds to the lower bound we got in Fig.~\ref{fig:right-handed_wino}. It requires $m_{W_R}\lesssim \unit[4]{TeV}$ and is not without tension from indirect detection too, but at a lower level \cite{Mitridate:2017izz}.
The entire LR triplet 6 scenario can thus be excluded in the future using indirect detection data and slightly improved limits on $W_R$.
\vspace*{1mm}\\
\hspace*{2.5mm}- \underline{\it LHC}: With DM masses above TeV the LHC will not be able to see these particles.

\subsection{LR triplet 3+3}
\label{sec:LR_triplet_3+3}

\noindent
\hspace*{2.5mm}- \underline{\it Low-scale content}: One $SU(2)_L$ triplet  and one $SU(2)_R$ triplet which are split by a $D$-parity breaking tree-level mass contribution:
\begin{align}
\hspace{-2ex} m_{\text{DM},L} \equiv  m_{(\vec{1},\vec{3},0)} \neq m_{(\vec{1},\vec{1},1)}=m_{(\vec{1},\vec{1},0)} \equiv m_{\text{DM},R}\,.
\end{align}
The triplets can
come from the $(\vec{1},\vec{3},\vec{1})_\text{PS}\oplus (\vec{1},\vec{1},\vec{3})_\text{PS}$ in $\vec{45}$ or the $(\vec{15},\vec{3},\vec{1})_\text{PS}\oplus (\vec{15},\vec{1},\vec{3})_\text{PS}$ in $\vec{210}$. We have two-component DM since both triplets are stable.
\vspace*{1mm}\\
\hspace*{2.5mm}- \underline{\it Mass splittings}: We have a free tree-level mass splitting $m_{\text{DM},L}-m_{\text{DM},R}$ between both triplets, but within each triplet the radiative splittings are again given by Eq.~\eqref{eq:wino_splitting} and Eq.~\eqref{eq:RH_mass_splitting}.
\vspace*{1mm}\\
\hspace*{2.5mm}- \underline{\it Relic density}: With respect to the case in Sec.~\ref{sec:LR_triplet_6} where both triplets are degenerate at tree level, this leads to more freedom for the parameter space and allows one to alleviate largely the strong indirect detection constraints which hold for a pure wino or for the case where both triplets are degenerate.
We can have a wino-like subcomponent of DM with mass $m_{\text{DM},L}<\unit[2.7]{TeV}$ and fill the rest with right-handed triplet DM with mass $m_{\text{DM},R}$ up to $\unit[50]{TeV}$ (see Fig.~\ref{fig:right-handed_wino}).
We need  $m_{W_R}< \unit[100]{TeV}$ as in the \ref{sec:SU(2)_R_triplet_3} scenario.
\vspace*{1mm}\\
\hspace*{2.5mm}- \underline{\it LHC}: Since the wino is only a subcomponent of DM, it could lie around the corner, just beyond the current LHC bound of $\unit[460]{GeV}$~\citep{Aaboud:2017mpt}. The right-handed triplet with mass above TeV will, on the other hand, be elusive.

\subsection{LR triplet 5+1}
\label{sec:LR_triplet_5+1}

\noindent
\hspace*{2.5mm}- \underline{\it Low-scale content}: One $SU(2)_L$ triplet and one $SU(2)_R$ triplet whose neutral component is split by a tree-level mass contribution,
\begin{align}
\hspace{-2ex} m_{\text{DM},L} \equiv  m_{(\vec{1},\vec{3},0)} = m_{(\vec{1},\vec{1},1)}\neq m_{(\vec{1},\vec{1},0)} \equiv m_{\text{DM},R}\,,
\end{align}
coming from the $(\vec{1},\vec{3},\vec{1})_\text{PS}\oplus (\vec{1},\vec{1},\vec{3})_\text{PS}$ in $\vec{45}$.
\vspace*{1mm}\\
\hspace*{2.5mm}- \underline{\it Mass splittings}: Radiative mass splittings are the same as in previous cases. We need $m_{\text{DM},R} < m_{\text{DM},L}< \unit[2.7]{TeV}$ to get rid of the charged components and to not overclose the Universe [Eq.~\eqref{eq:wino_abundance}].
\vspace*{1mm}\\
\hspace*{2.5mm}- \underline{\it Relic density}: 
With respect to the \ref{sec:LR_triplet_6} scenario this scenario has more freedom in a way similar to  the \ref{sec:SU(2)_R_triplet_2+1} scenario with respect to the 
\ref{sec:SU(2)_R_triplet_3} scenario.
We still need approximately $m_{W_R}\lesssim \unit[7]{TeV}$ as in the \ref{sec:LR_triplet_6}  scenario if we live in the resonant region $m_{\text{DM},R}+m_{\text{DM},L}\sim m_{W_R}$.
Given the lower limit $\unit[460]{GeV}< m_{\text{DM},L}$ from LHC searches for winos, we cannot use the co-annihilation region of Sec.~\ref{sec:SU(2)_R_triplet_2+1}.
\vspace*{1mm}\\
\hspace*{2.5mm}- \underline{\it LHC}: Again, with DM masses above TeV, there is little to see at the LHC, but future colliders should be able to probe this scenario conclusively, e.g.~by looking for the wino.

\subsection{LR triplet 3+2+1}
\label{sec:LR_triplet_3+2+1}

\noindent
\hspace*{2.5mm}- \underline{\it Low-scale content}: One $SU(2)_L$ triplet and one $SU(2)_R$ triplet with all possible tree-level splittings,
\begin{align}
\hspace{-2ex} m_{\text{DM},L} \equiv  m_{(\vec{1},\vec{3},0)} \neq m_{(\vec{1},\vec{1},1)}\neq m_{(\vec{1},\vec{1},0)} \equiv m_{\text{DM},R}\,.
\end{align}
coming from the $(\vec{1},\vec{3},\vec{1})_\text{PS}\oplus (\vec{1},\vec{1},\vec{3})_\text{PS}$ in $\vec{45}$ or $(\vec{15},\vec{3},\vec{1})_\text{PS}\oplus (\vec{15},\vec{1},\vec{3})_\text{PS}$ in $\vec{210}$.
\vspace*{1mm}\\
\hspace*{2.5mm}- \underline{\it Mass splittings}: Radiative splitting between the wino components as usual.
\vspace*{1mm}\\
\hspace*{2.5mm}- \underline{\it Relic density}: This case gives the most freedom, it corresponds to the \ref{sec:SU(2)_R_triplet_2+1} scenario with an additional wino of arbitrary mass. Since indirect detection constraints prefer this wino component to be under-abundant, which translates into $m_{\text{DM},L} < \unit[2.7]{TeV}$, the phenomenology is similar basically to the one of Sec.~\ref{sec:SU(2)_R_triplet_2+1}. In particular, the $m_{W_R}< \unit[100]{TeV}$ bound remains valid.
\vspace*{1mm}\\
\hspace*{2.5mm}- \underline{\it LHC}: This scenario has the richest LHC phenomenology. In the region of under-abundant wino, it could be around the corner of the current LHC limit. In addition, the LHC could probe the sub-TeV region of the right-handed triplet, see the discussion in Sec.~\ref{sec:SU(2)_R_triplet_2+1}.

\subsection{LR bi-triplet 9}
\label{sec:LR_bi-triplet_9}

\noindent
\hspace*{2.5mm}- \underline{\it Low-scale content}: DM is the lightest neutral component of nine tree-level degenerate components within an $SU(2)_L\times SU(2)_R$ bi-triplet, coming from the $(\vec{1},\vec{3},\vec{3})_\text{PS}$ in $\vec{54}$. This bi-triplet contains one chiral wino and one triplet with hypercharge $Y=1$, which contains an electrically neutral Dirac fermion. This was discussed in detail in Ref.~\cite{Garcia-Cely:2015quu}.
\vspace*{1mm}\\
\hspace*{2.5mm}- \underline{\it Mass splittings}: Direct detection requires the wino to be the lightest of the two $SU(2)_L$ triplets, which in turn restricts the $m_\text{DM}$--$m_{W_R}$ parameter space~\cite{Garcia-Cely:2015quu}.
\vspace*{1mm}\\
\hspace*{2.5mm}- \underline{\it Relic density}:   For a low LR scale, co-annihilation is very efficient and makes it possible to obtain the correct relic abundance for DM masses between $\unit[1.8]{TeV}$ and $\unit[40]{TeV}$. However, it still behaves exactly as a wino, so indirect detection puts very strong constraints on most of the parameter space, leaving only some narrow regions near the $W_R$ resonance, which in particular requires $m_{W_R}< \unit[80]{TeV}$~\cite{Garcia-Cely:2015quu}.
\vspace*{1mm}\\
\hspace*{2.5mm}- \underline{\it LHC}: The DM masses above $\unit[1.8]{TeV}$ make it difficult to see the wino triplet at the LHC. The partner triplet \emph{with hypercharge}, on the other hand, has much larger production cross sections and could be probed even at these high masses at the HL-LHC~\cite{Cirelli:2005uq,Garcia-Cely:2015quu}, although a thorough analysis has yet to be performed.

\subsection{LR bi-quadruplet 16}
\label{sec:LR_bi-quadruplet_16}

\noindent
\hspace*{2.5mm}- \underline{\it Low-scale content}: An $SU(2)_L\times SU(2)_R$ bi-quadruplet from the $\vec{210}'$ representation. DM is the lightest
of the four neutral components.
\vspace*{1mm}\\
\hspace*{2.5mm}- \underline{\it Mass splittings}:  The bi-quadruplet can be written as a self-conjugate particle 
\begin{align}
\Psi = \tilde{\Psi} = \matrixx{
\Psi^0_B & \Psi^+_B & \Psi^{++}_B & \Psi^{+++} \\
\Psi^-_C & \Psi^0_A & \Psi^{+}_A & -\Psi^{++}_A \\
\Psi^{--}_A & \Psi^-_A & -(\Psi^{0}_A)^c & \Psi^{+}_C \\
\Psi^{---} & -\Psi^{--}_B & \Psi^{-}_B & -(\Psi^{0}_B)^C},
\end{align}
which contains two neutral Dirac particles: $\Psi^0_A$ and $\Psi^0_B$. As can be seen from the Lagrangian interactions, $\Psi^0_A$ is split into two Majorana fermions $\chi_{1,2}$ at the one-loop level, in complete analogy to the bi-doublet case and with the same mass formula with respect to $\Psi^{+}_A$ [Eq.~\eqref{eq:bidoublet_splitting}]. If the mass splitting $\Delta m$ between $\chi_{1,2}$ is above $\unit[200]{keV}$, we can evade direct-detection bounds. $\Psi^0_B$, on the other hand, remains Dirac even at the one-loop level and can therefore not be a good DM candidate due to the coupling to the $Z$ boson. We thus have to make sure that one of $\chi_{1,2}$  is the lightest of the bi-quadruplet particles, which requires us to calculate all the radiative mass splittings.
\noindent
After $SU(2)_R$ breaking, the bi-quadruplet can be described as two Dirac multiplets which transform under $G_\text{SM}$ according to
\begin{align}
(\Psi^0_B, \Psi^-_C,\Psi^{--}_A,\Psi^{---})^T &\sim (\vec{1},\vec{4},-3/2)\,,\\
(\Psi^+_B, \Psi^0_A,\Psi^{-}_A,-\Psi^{--}_B)^T &\sim (\vec{1},\vec{4},-1/2)\,.
\end{align} 
The mass splitting between these two multiplets is determined by $m_{W_R}$ and can be suitably chosen so that the one with $\Psi^0_A$ is lightest via $m_\text{DM}<m_{W_R}$. Within each $G_\text{SM}$ multiplet, the mass splittings then simply depend on the $W$ and $Z$ masses, with the mass formula well known from Minimal DM~\cite{Cirelli:2005uq}. In particular, one finds for the $(\vec{1},\vec{4},-1/2)$ multiplet,
\begin{align}
\hspace{-1ex} m_{\Psi^+_B}-m_{\Psi^0_A} &\simeq -\frac{\tan^2 (\theta_W/2)}{2} \alpha \, m_Z  \simeq -\unit[23]{MeV}\,,\\
\hspace{-1ex} m_{\Psi^-_A}-m_{\Psi^0_A} &\simeq \frac{1}{2} \alpha \, m_Z \simeq \unit[350]{MeV}\,,\\
\hspace{-1ex} m_{\Psi^{--}_B}-m_{\Psi^0_A} &\simeq \frac{1+2\cos(\theta_W)}{1+\cos (\theta_W)} \alpha \,  m_Z \simeq \unit[1]{GeV}\,.
\end{align}
The neutral component of $(\vec{1},\vec{4},-1/2)$ is hence \emph{not} the lightest particle, rendering it excluded at first sight. However, since our $\Psi^0_A$ splits further into two Majoranas via LR mixing, $m_{\Psi^0_A} \to m_{\Psi^0_A} \pm \Delta m/2$, and we can conceivably make this splitting $\Delta m$ large enough to push one of the Majoranas, say $\chi_1$, below the $\Psi^+_B$ mass. The $\chi_1$--$\chi_2$ mass splitting is hence no longer required to be just above $\unit[200]{keV}$, but rather above $\unit[46]{MeV}$! This requires a very low LR scale, say $m_{W_R}<\unit[5]{TeV}$, probably even lower, with large $W_L$--$W_R$ mixing. This scenario might already be excluded, but will definitely be conclusively probed by the LHC.
Assuming the scenario to still be viable from the $W_R$ perspective, it is clear that the mass splitting $m_{\Psi^+_B}-m_{\chi_1}$ will be tiny, at best $\unit[\mathcal{O}(10)]{MeV}$. Such a small mass splitting can render the charged partner fairly long lived, potentially wreaking havoc with BBN. 
\vspace*{1mm}\\
\hspace*{2.5mm}- \underline{\it Relic density}: 
The right relic density for a $(\vec{1},\vec{4},-1/2)$ multiplet can be obtained for a mass around $\unit[2.4]{TeV}$~\cite{Cirelli:2005uq}, neglecting Sommerfeld effects. Since we necessarily have a light $W_R$, one can have co-annihilations with the $(\vec{1},\vec{4},-3/2)$ multiplet and in particular access to the $W_R$ $s$-channel resonance, allowing one to increase the DM mass significantly.

Overall, this scenario is extremely constrained and will be excluded if we do not find a $W_R$ at the LHC.

\subsection{Octet-triplet-singlet 8+6+1}
\label{sec:Octet-triplet-singlet_8+6+1}

\noindent
\hspace*{2.5mm}- \underline{\it Low-scale content}: One Majorana color octet, one Dirac color triplet, and one Majorana singlet, split by a tree-level mass contribution,
\begin{align}
m_{(\vec{8},\vec{1},0)} \neq m_{(\vec{3},\vec{1},2/3)} \neq m_{(\vec{1},\vec{1},0)} \equiv m_\text{DM}\,,
\end{align}
coming from the $(\vec{15},\vec{1},\vec{1})_\text{PS}$ in $\vec{45}$ or $(\vec{15},\vec{1},\vec{3})_\text{PS}$ in $\vec{210}$.
\vspace*{1mm}\\
\hspace*{2.5mm}- \underline{\it Mass splittings}: Even if there were no tree-level splitting between the octet, triplet, and singlet, the degenerate $\vec{15}_\text{PS}$ would still be split at the one-loop level once the PS group is broken, see App.~\ref{app:rge}.
The singlet ends up lighter than its partners, making it a good DM candidate; the color octet is the heaviest component. For a low  PS scale $\sim \unit[10^{3}]{TeV}$ and the $\vec{15}_\text{PS}$ at the TeV scale, we use the one-loop renormalization group equations to run up to the PS scale and impose the degeneracy condition on the components of the  $\vec{15}$ (App.~\ref{app:rge}).
For a $\unit[1]{TeV}$ singlet mass, this gives triplet and octet pole masses at around $\unit[1.7]{TeV}$ and $\unit[2.5]{TeV}$, respectively, due to the large-log enhancement. These colored components then decay sufficiently fast into the DM candidate.  Larger PS scales lead to larger mass splittings.
\vspace*{1mm}\\
\hspace*{2.5mm}- \underline{\it Relic density}: From only the radiative splitting above the relic density cannot be obtained from a standard freeze-out because an annihilation of the singlet DM candidate can proceed only through the exchange of a PS scale gauge boson $X$ which suppresses the annihilation rate too much. Due to the presence of the color partners one could nevertheless think about co-annihilation between the DM singlet and these color partners. In more details, as explained in App.~\ref{app:chem_dec} this requires the PS scale gauge boson mediated processes involving the DM color singlet and the color triplet to be fast enough to keep the singlet in thermal equilibrium, such that it transforms into these color states, the latter states undergoing 
a thermal freeze-out into gluons, see Refs.~\cite{deSimone:2014pda,Mitridate:2017izz} for a general scenario of this kind. 
This condition of thermal equilibrium gives the approximate constraint $m_X/g_4 \lesssim \unit[900]{TeV}\,(m_1/\unit{TeV})^{3/4}$. The  experimental lower bound on $m_X \gtrsim \unit[2000]{TeV}$ of~\cite{Valencia:1994cj,Ambrose:1998us,Smirnov:2007hv} then requires considering DM masses $m_1 \gtrsim \unit[3]{TeV}$, as illustrated in Fig.~\ref{fig:chem_eq}.

However, in order that the DM singlet abundance is sufficiently depleted in this way, this would require a mass splitting between the singlet and one of the colored states to be at most $\unit[100]{GeV}$~\cite{deSimone:2014pda,Mitridate:2017izz}. The radiative splittings are hence too big for co-annihilation to work, which is the reason we did not retain an "octet-triplet-singlet 15" scenario in our list. However, with additional tree-level mass splittings to 8+6+1 states, one can have the required small mass splitting to get the observed relic density via co-annihilation.

Note that, as explained in Sec.~\ref{sec:viable}, one could also have only the singlet and octet components at low energy, but this scenario does not work because the gauge boson $X$ only couples the singlet to the triplet and the triplet to the octet. Conversion of a singlet to an octet thus requires a mediator triplet, which becomes inefficient if the triplet is much heavier than the singlet and octet. Thus one could think instead about a scenario with the singlet and triplet components as the only low-energy components, but this scenario does not show up from any of the representations we consider in Sec.~\ref{sec:viable} if one assumes at most one large fine-tuning between the various contributions in the mass formula. It could show up nevertheless with more than one of these tunings, see Sec.~\ref{sec:Otherscenarios}.
\vspace*{1mm}\\
\hspace*{2.5mm}- \underline{\it LHC}: The relevant interactions are given in Eq.~\eqref{eq:Xinteractions}. On top of a rather small mass splitting between the singlet and triplet, successful co-annihilation requires a DM mass above $\unit[3]{TeV}$ (see App.~\ref{app:chem_dec}, especially Fig.~\ref{fig:chem_eq}). 
The Dirac color triplet $\chi_3$ with electric charge $2/3$ will decay flavor-universal according to $\chi_3\to \chi_0 q \overline{\ell}$, which should be a good signature, especially since the decay length is typically large in the parameter region of interest, potentially even stable on collider scales. It will then form $R$-hadron-like bound states with SM particles~\cite{Farrar:1978xj} which can lead to specific energy loss signatures~\cite{Aaboud:2016uth}.

The color octet DM partner within the $(\vec{15},\vec{1},\vec{1})$ will behave similar to a gluino, albeit with different decay modes. With gluino limits currently around $\unit[1.5]{TeV}$~\cite{Khachatryan:2016xdt} our scenario should still be viable, although we urge the experimental collaborations to investigate the PS DM decay chain $pp\to \chi_8\chi_8$ with $\chi_8\to \chi_3 \ell \bar{q} \to \chi_0 \overline{q}q' \overline{\ell}'\ell$. Our gluino can easily be long lived on collider scales, which gives rise to different signatures~\cite{Aad:2011yf,Sirunyan:2017sbs}.

\subsection{Octet-triplet-singlet 14+1}
\label{sec:Octet-triplet-singlet_14+1}

\noindent
\hspace*{2.5mm}- \underline{\it Low-scale content}: One Majorana color octet, one Dirac color triplet and one Majorana DM singlet, the latter split by a tree-level mass contribution from the other degenerate states. 
\begin{align}
m_{(\vec{8},\vec{1},0)} = m_{(\vec{3},\vec{1},2/3)} \neq m_{(\vec{1},\vec{1},0)} \equiv m_\text{DM}\,.
\end{align}
This can arise from a low-scale $(\vec{15},\vec{1},\vec{1})_\text{PS}$ multiplet in the $\vec{45}$ representation.
\vspace*{1mm}\\
\hspace*{2.5mm}- \underline{\it Mass splittings}:  
The tree-level degenerate octet and triplet split at loop level by rather large amounts, as derived in App.~\ref{app:rge}. The tree-level mass splitting allows us to keep the triplet close to the singlet, as relevant for co-annihilation.
\vspace*{1mm}\\
\hspace*{2.5mm}- \underline{\it Relic density}: Thanks to the tree-level splitting between the singlet and the triplet, co-annihilations are possible for a low $\sim \unit[1000]{TeV}$ PS scale, see the discussion in Sec.~\ref{sec:Octet-triplet-singlet_8+6+1} and in App.~\ref{app:chem_dec}.
\vspace*{1mm}\\
\hspace*{2.5mm}- \underline{\it LHC}: DM and partners are between $3$ and $\unit[5]{TeV}$ (see Fig.~\ref{fig:chem_eq}), with signatures depending strongly on the mass splitting and PS scale.

\subsection{Octet-triplet-singlet bi-doublet 60}
\label{sec:Octet-triplet-singlet_bi-doublet_60}

\noindent
\hspace*{2.5mm}- \underline{\it Low-scale content}: A degenerate $(\vec{15},\vec{2},\vec{2})_\text{PS}$ multiplet in $\vec{120}$,  for a total of 60 states at low scale. DM is the lightest neutral component of the $(\vec{1},\vec{2},\vec{2},0)_\text{LR}$ LR bi-doublet within this PS multiplet.
\vspace*{1mm}\\
\hspace*{2.5mm}- \underline{\it Mass splittings}: Taking a PS scale above $\sim  \unit[10^3]{TeV}$, the biggest splitting within $(\vec{15},\vec{2},\vec{2})_\text{PS}$ comes from the RGE running as in App.~\ref{app:rge}:
\begin{align}
\frac{m_{(\vec{8},\vec{2},\vec{2},0)_\text{LR}}}{m_{(\vec{1},\vec{2},\vec{2},0)_\text{LR}}} \gtrsim 2.5\,, && \frac{m_{(\vec{3},\vec{2},\vec{2},4)_\text{LR}}}{m_{(\vec{1},\vec{2},\vec{2},0)_\text{LR}}} \gtrsim 1.7\,.
\end{align}
The lightest bi-doublet $(\vec{1},\vec{2},\vec{2},0)_\text{LR} \to (\vec{1},\vec{2},1/2)_\text{SM}$ is then split as in scenario \ref{sec:LR_bi-doublet_4}, which, to evade direct-detection constraints, requires a low LR scale $\lesssim \unit[75]{TeV}$.
\vspace*{1mm}\\
\hspace*{2.5mm}- \underline{\it Relic density}: The phenomenology is similar to the bi-doublet scenario of \ref{sec:LR_bi-doublet_4}, since the colored partners are too heavy to lead to co-annihilation. Away from the $W_R$ and $Z'$ resonances, this then requires a bi-doublet DM mass of $\unit[1.2]{TeV}$~\cite{Garcia-Cely:2015quu}. Using the RGE mass splitting, and the decay formula from Eq.~\eqref{eq:decay_of_15}, we find an upper bound of $\unit[10^6]{TeV}$ on the PS breaking scale in order to let the color triplet decay before BBN.
\vspace*{1mm}\\
\hspace*{2.5mm}- \underline{\it LHC}:
The color triplets $(\vec{3},\vec{2},1/6)_\text{SM} \oplus (\vec{3},\vec{2},7/6)_\text{SM}$ and octet $(\vec{8},\vec{2},1/2)_\text{SM}$ DM partners can be as light as $\unit[2]{TeV}$ and $\unit[3]{TeV}$, respectively, very much in reach of the LHC. Pushing the PS scale to $\unit[10^6]{TeV}$ increases these masses to $\unit[3]{TeV}$ and $\unit[5]{TeV}$, still rather low. If these are excluded by LHC searches, one has to put the DM bi-doublet close to the $W_R$ or $Z'$ resonance in order to push its mass to the multi-TeV range, which correspondingly increases the colored partner masses.

\subsection{Octet-triplet-singlet bi-doublet 32+24+4}
\label{sec:Octet-triplet-singlet_bi-doublet_32+24+4}

\noindent
\hspace*{2.5mm}- \underline{\it Low-scale content}: Same as the previous scenario, but with additional tree-level splittings of the $(\vec{15},\vec{2},\vec{2})_\text{PS}$ within $\vec{120}$, leading to a $32+24+4$ mass spectrum,
\begin{align}
m_{(\vec{8},\vec{2},\vec{2},0)_\text{LR}} \neq m_{(\vec{3},\vec{2},\vec{2},4)_\text{LR}} \neq m_{(\vec{1},\vec{2},\vec{2},0)_\text{LR}} \equiv m_\text{DM}\,.
\end{align}
\hspace*{2.5mm}- \underline{\it Mass splittings}: The bi-doublet $(\vec{1},\vec{2},\vec{2},0)_\text{LR} \to (\vec{1},\vec{2},1/2)_\text{SM}$ splits as in scenario \ref{sec:LR_bi-doublet_4}, which here too requires a low LR scale $\lesssim \unit[75]{TeV}$.
\vspace*{1mm}\\
\hspace*{2.5mm}- \underline{\it Relic density}: Either as scenario \ref{sec:LR_bi-doublet_4}  or via co-annihilation with the colored partners for small mass splittings. An efficient co-annihilation would require a PS scale similar to the one given in Eq.~(\ref{eq:bound_mx_approx}), i.e.~close to the current bound, and 
$<\mathcal{O}(\unit[100]{GeV})$ mass splittings. The DM mass scale in both cases is going to be above TeV.

Note that the $(\vec{8},\vec{2},1/2)_\text{SM}$ is not a good DM candidate because it either has hypercharge or splits into two Majorana octets, both options being disfavored~\cite{DeLuca:2018mzn}.
\vspace*{1mm}\\
\hspace*{2.5mm}- \underline{\it LHC}: The colored DM partners here can easily be around TeV, which should lead to interesting signatures at the LHC. Due to the decay via the heavy $X$ boson, these partners are typically long lived and should thus form $R$~hadrons~\cite{Farrar:1978xj}, which can lead to specific energy loss signatures~\cite{Aaboud:2016uth}. A detailed analysis will be performed elsewhere.

\subsection{Octet-singlet bi-doublet 32+4}
\label{sec:Octet-singlet_bi-doublet_32+4}

\noindent
\hspace*{2.5mm}- \underline{\it Low-scale content}: Same as the last two scenarios but with an additional tree-level contribution that
sends the color triplets, within the $(\vec{15},\vec{2},\vec{2})$ in the $\vec{120}$ representation, to a high scale, leaving a total of 32 states at low scale:
\begin{align}
m_{(\vec{8},\vec{2},\vec{2},0)_\text{LR}} \neq m_{(\vec{1},\vec{2},\vec{2},0)_\text{LR}} \equiv m_\text{DM}\,.
\end{align}
\hspace*{2.5mm}- \underline{\it Mass splittings}: As in scenario \ref{sec:LR_bi-doublet_4}  for the components of the bi-doublet. The decay of the octet must necessarily proceed through a virtual triplet going to the singlet, estimated as $\Gamma \propto \Delta m^{11}/(m_X^8 m_3^2)$. Given the $\sim \unit[1000]{TeV}$ lower bound on the mass of the $X$ boson this requires the triplet to be not too heavy, depending on the octet-singlet mass splitting. 
\vspace*{1mm}\\
\hspace*{2.5mm}- \underline{\it Relic density}: As in scenario \ref{sec:LR_bi-doublet_4} for large mass splitting.
 Here co-annihilation with the color octet would not work since the PS boson $X$ does not couple the color singlet with the color octet. As a result, this is essentially scenario~\ref{sec:LR_bi-doublet_4} with an additional color octet partner that is irrelevant for the relic abundance and has an arbitrarily (short) lifetime, at least if the color triplet is not too heavy.

\subsection{Octet-triplet-singlet $SU(2)_L$ triplet 45}
\label{sec:Octet-triplet-singlet_SU(2)_L_triplet_45}

\noindent
\hspace*{2.5mm}- \underline{\it Low-scale content}: One $(\vec{15},\vec{3},\vec{1})_\text{PS}$ from a $\vec{210}$
representation is present at low scale for a total of 45 tree-level degenerate states. DM is the neutral component of the $SU(2)_L$ triplet.
\vspace*{1mm}\\
\hspace*{2.5mm}- \underline{\it Mass splittings}: The biggest splitting comes again from the high PS breaking scale (App.~\ref{app:rge}),
\begin{align}
\frac{m_{(\vec{8},\vec{x},\vec{y},0)_\text{LR}}}{m_{(\vec{1},\vec{x},\vec{y},0)_\text{LR}}} \gtrsim 2.5\,, && \frac{m_{(\vec{3},\vec{x},\vec{y},4)_\text{LR}}}{m_{(\vec{1},\vec{x},\vec{y},0)_\text{LR}}} \gtrsim 1.7\,,
\end{align}
and afterwards the wino $(\vec{1},\vec{3},\vec{1},0)_\text{LR}$ splits radiatively as usual.
 \vspace*{1mm}\\
\hspace*{2.5mm}- \underline{\it Relic density}: The radiative mass splitting between the color octet/triplet and the color singlet are too large to allow successful co-annihilation. Thus for the relic density this scenario is like the wino \ref{sec:SU(2)_L_triplet_3} scenario above, which fixes the wino mass to $\unit[2.7]{TeV}$. This implies a color triplet (octet) mass above $\unit[4.6]{TeV}$ ($\unit[6.8]{TeV}$). Letting the triplet decay before BBN puts an upper bound of $\sim\unit[10^6]{TeV}$ on the PS scale, which then translates into upper limits on the triplet and octet masses of $\unit[6.2]{TeV}$ and $\unit[11.9]{TeV}$, respectively.
We note again that this wino scenario is disfavored by indirect detection.
\vspace*{1mm}\\
\hspace*{2.5mm}- \underline{\it LHC}: The colored partners in this scenario are rather heavy and thus possibly out of reach of the LHC, but as always could be probed at a future $\unit[100]{TeV}$ collider.

\subsection{SM singlet charged under extra $U(1)$}
\label{sec:ZprimeDM}

\begin{figure}[t]
\includegraphics[width=0.5\textwidth]{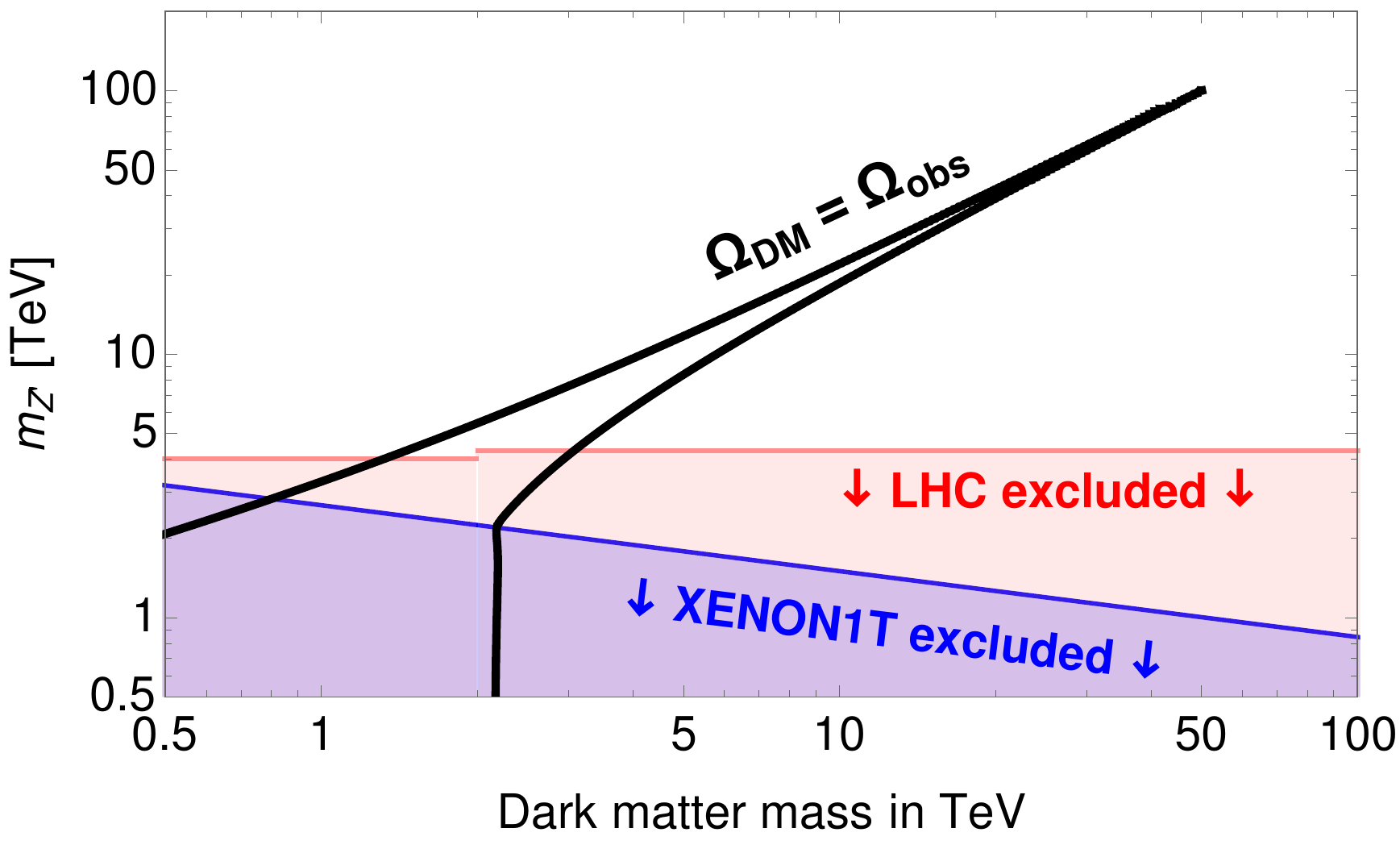}
\caption{
The correct relic density (black) of a Dirac singlet from $\vec{126}$ coupled to a low-scale $Z'$ from $G_{3211}$, as well as constraints from XENON1T (blue) and CMS (red).
}
\label{fig:126plot}
\end{figure}

\hspace*{2.5mm}- \underline{\it Low-scale content}: In none of the scenarios above, which are all based on assuming a single real DM representation, do we have only a singlet DM candidate with no partner from the same representation at low scale. This shows that a DM singlet scenario without partners does not result at all generically from a GUT theory, at least for a fermion candidate. However, it turns out that such a scenario can result from $SO(10)$ if one assumes a complex DM representation and its conjugate, for instance the  $\vec{126}$ representation and its $\overline{\vec{126}}$ conjugate, with the singlet coming from the $(\vec{10},\vec{1},\vec{3})_\text{PS}$ in these representations, see Sec.~\ref{sec:126}.
\vspace*{1mm}\\
\hspace*{2.5mm}- \underline{\it Mass splittings}: The Dirac singlet is the only state at low scale, i.e.~no low-scale mass splittings here.
\vspace*{1mm}\\
\hspace*{2.5mm}- \underline{\it Relic density}: The singlet is charged under the extra $U(1)$ in the intermediate $G_{3211}$ subgroup~\cite{Arcadi:2017atc}.
The relic density constraint requires this subgroup to be broken at low energy, so that DM can annihilate into a pair of $Z'$ or through an $s$-channel $Z'$-mediated transition into SM fermions. 
In Fig.~\ref{fig:126plot} we show the {\sc MicrOMEGAs}~\cite{Belanger:2006is} result for the relic density as well as constraints from direct detection (XENON1T~\cite{Aprile:2018dbl}). 
\vspace*{1mm}\\
\hspace*{2.5mm}- \underline{\it LHC}:  While the DM candidate itself is difficult to see at the LHC, we still have dilepton $Z'$ searches~\cite{Sirunyan:2018exx}. As shown in Fig.~\ref{fig:126plot} these searches exclude $Z'$ masses below $4$--$\unit[4.3]{TeV}$, depending on the DM mass (which can lower the $Z'\to \ell^-\ell^+$ branching ratio). As with most other $Z'$-mediated DM models one is forced to sit close to the resonance $m_\Psi \sim m_{Z'}/2$ in order to evade constraints. The DM mass is then necessarily between $\unit[1.3]{TeV}$ and $\sim\unit[50]{TeV}$.

\subsection{Other scenarios}
\label{sec:Otherscenarios}

As noted in the (next) Sec.~\ref{sec:viable}, a $\vec{210}$ DM representation can also lead to scenarios with more states and/or mass splittings, 
such as the ``\underline{octet-triplet-singlet $SU(2)_L$ triplet 24+18+3 scenario}'', where the entire $\left(\vec{15},\vec{3},\vec{1}\right)$ is present at low scale with tree-level splitting between the color octets, triplets, and singlets, and similarly the ``\underline{octet-triplet-singlet $SU(2)_R$ triplet 24+18+3 scenario}'' from the $\left(\vec{15},\vec{1},\vec{3}\right)$.
Combinations of the last two scenarios are also possible, leading to the ``\underline{octet-triplet-singlet $SU(2)_{L+R}$ triplet $n$ scenarios}'', with $n$ equal to $48+36+6$ or $3+3+18+18+24+24$ or $3+3+42+42$ (with in the latter case degenerate color triplet and octet). 
As stated in Sec.~\ref{sec:viable} we will only mention these complicated possibilities but not go into details.
Further scenarios for the $\vec{210}$ DM representation (which we do not list in Sec.~\ref{sec:viable}) arise if one considers the paths going through
an additional intermediate $G_{3211}$ subgroup. 
A further contribution from a vacuum expectation value leading to this subgroup can split states within the same LR multiplet, in particular the charged and neutral component of a right-handed triplet, so that the "3" above in the $(\vec{15},\vec{1},\vec{3})_\text{PS}$ in $\vec{210}$ becomes a $2+1$.

As emphasized also in Sec.~\ref{sec:viable}, we will not consider explicitly models whose low-energy mass spectrum would be the result of more than one fine-tuning between mass scale contributions much larger than the DM scale. The systematic determination of such cases would bring us too far, and this would not bring many new different cases. One interesting exception, i.e.~model we do not get with at most one large fine-tuning, is the "singlet-triplet $1+6$" scenario, involving at low scale one Dirac color triplet and one Majorana singlet, split by a tree-level mass contribution,
\begin{align}
m_{(\vec{3},\vec{1},2/3)} \neq m_{(\vec{1},\vec{1},0)} \equiv m_\text{DM}\,,
\end{align}
coming from the $(\vec{15},\vec{1},\vec{1})_\text{PS}$ in $\vec{45}$ or $(\vec{15},\vec{1},\vec{3})_\text{PS}$ in $\vec{210}$. Here DM is in the form of the color singlet co-annihilating with the color triplet, just as explained in App.~\ref{app:chem_dec}.

One could also think about scenarios where the DM relic density is produced non-thermally but this would bring us too far, too.
Just as an example note that the scenario of Sec.~\ref{sec:Octet-triplet-singlet_8+6+1} where all the 15 states would be degenerate at tree level, so an "octet-triplet-singlet 15 scenario" which does not appear in the list above because it cannot lead to the observed relic density thermally, could work with non-thermal DM production, see the brief discussion in Sec.~\ref{sec:45}.

\section{Detailed determination of the candidates}
\label{sec:viable}

Having discussed the many possible low-scale mass hierarchies among the various DM partners, we will now show in detail how we obtained these scenarios from the top-down perspective. To this end we will discuss the case of each possible real chiral DM representation from the smallest ones, from the $\vec{10}$ up to $\vec{210}'$. As said above we make the assumption that DM comes from a single $SO(10)$ representation. 
At the end of this section we will nevertheless also consider the complex $\vec{126}$ case, which requires both a $\vec{126}$ and a $\overline{\vec{126}}$, as it displays new interesting features.
All along this discussion, we will also see that it is always necessary to make the DM candidate light by means of a fine-tuned cancellation; otherwise it would be drawn to the scale of its heaviest GUT partner, which is at least as heavy as $\unit[10^5]{TeV}$.
As explained at the end of Sec.~\ref{sec:DMsplittings}, to systematically determine the possible mass spectra, we will proceed directly from the tree-level master mass formula, considering the various possible hierarchies of mass contributions in this formula, without specifying what are all the explicit breaking paths which can lead to such a hierarchy.
These can be obtained easily by adding scalar representations that do not contribute to the mass formula to the scalar representations assumed in the mass formula.

\subsection{Fermionic $\vec{10}$ DM candidate}
\label{sec:10}

\begin{figure*}[t]
\includegraphics[width=0.7\textwidth]{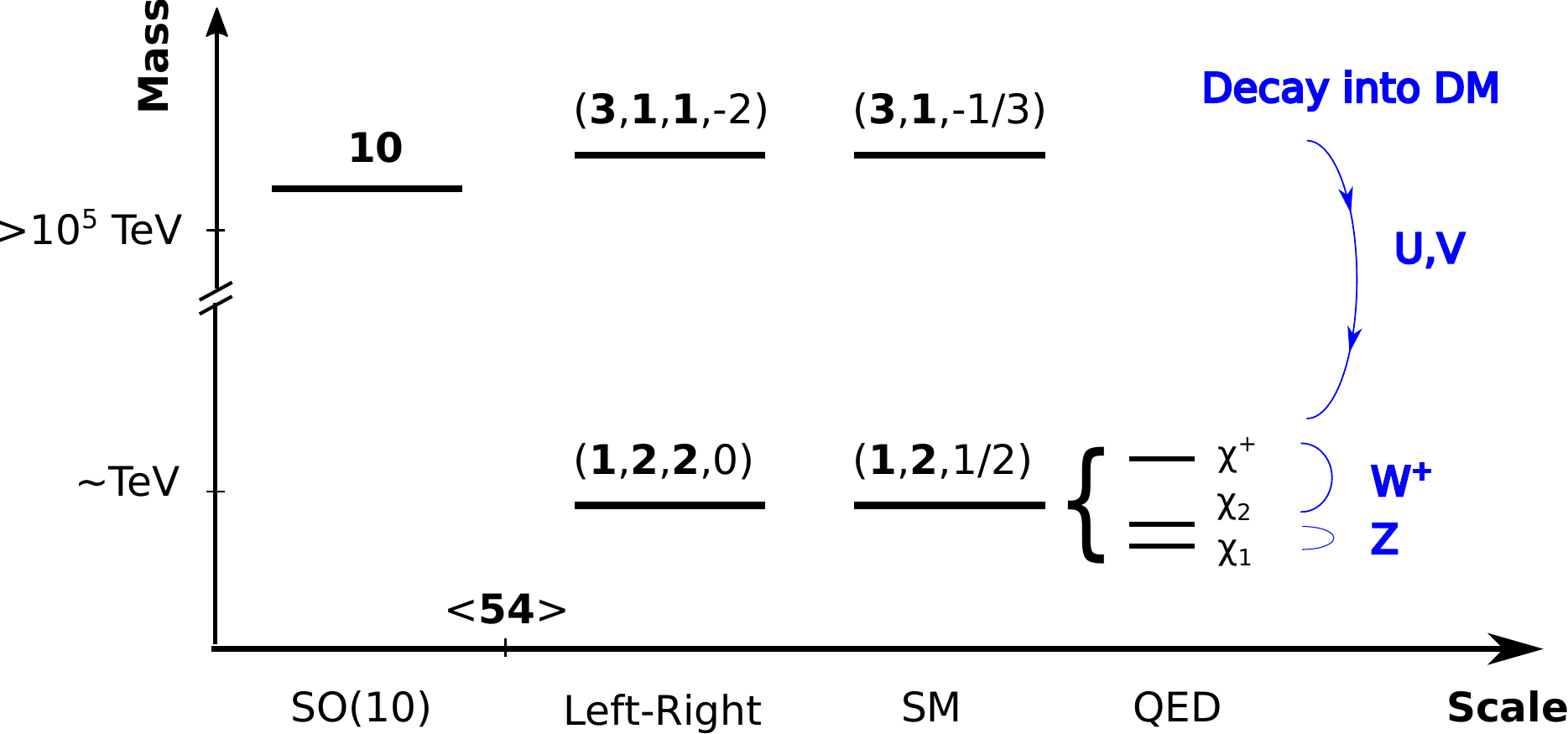}
\caption{Illustration of the fermionic $\vec{10}$ components mass from the $SO(10)$ scale down to low energies. There is only one tree-level mass splitting, by the $\langle \vec{54}\rangle$, and all other splittings arise radiatively. In blue we also indicate the massive gauge bosons relevant for the (tree-level) decays of the DM partners into the lightest DM state.}
\label{fig:spectrum_10}
\end{figure*}

We start our detailed discussion with the smallest non-trivial $SO(10)$ representation, a chiral $\vec{10}$. 
Under the Pati--Salam $G_{422}$, and its $G_{3221}$, $G_{3211}$, and SM subgroups the real $\vec{10}$ decomposes as follows:
\begin{align}
       \vec{10} &\stackrel{\text{PS}}{\to} \left(\vec{1},\vec{2},\vec{2}\right)\oplus \left(\vec{6},\vec{1},\vec{1}\right) \nonumber\\
	 &\stackrel{\text{LR}}{\to} \left(\vec{1},\vec{2},\vec{2},0\right)\oplus \left(\vec{3},\vec{1},\vec{1},-2\right) \oplus  \left(\overline{\vec{3}},\vec{1},\vec{1},2\right) \nonumber\\
	  &\stackrel{\text{3211}}{\to} \left(\vec{1},\vec{2},1,0\right)\oplus\left(\vec{1},\vec{2},-1,0\right) \\
	  &\qquad\oplus \left(\vec{3},\vec{1},0,-2\right)\oplus  \left(\overline{\vec{3}},\vec{1},0,2\right)  \nonumber\\
	&\stackrel{\text{SM}}{\to} \left(\vec{1},\vec{2},\tfrac{1}{2}\right)\oplus \left(\vec{1},\vec{2},-\tfrac{1}{2}\right)\oplus\left(\vec{3},\vec{1},-\tfrac{1}{3}\right) \oplus  \left(\overline{\vec{3}},\vec{1},\tfrac{1}{3}\right) ,\nonumber
\end{align}
(see also Fig.~\ref{fig:spectrum_10})
and eventually forms one massive Dirac color triplet $\left(\vec{3},\vec{1},-\tfrac{1}{3}\right)$ and one Dirac electroweak doublet $\left(\vec{1},\vec{2},\tfrac{1}{2}\right)$.
The product rule
\begin{align}
\vec{10}\otimes \vec{10} &= \vec{1}_S \oplus \vec{45}_A \oplus \vec{\vec{54}}_S
\end{align}
shows that the $\vec{10}$ is allowed an $SO(10)$ invariant mass term $m_\vec{1}$ and can be split at tree level only by a $\vec{54}$ VEV $v_\vec{54}$, seeing as a coupling to $\vec{45}$ is anti-symmetric and hence vanishes in our minimal setup.
We find the Dirac masses 
\begin{align}
	m_{\left(\vec{1},\vec{2},\tfrac{1}{2}\right)} &= m_\vec{1} +3\, y_{\vec{54}}\, v_\vec{54}\,, \label{10treelevelsplitting1}\\
	m_{\left(\vec{3},\vec{1},-\tfrac{1}{3}\right)} &= m_\vec{1} - 2\, y_\vec{54} \,v_\vec{54}\,,  \label{10treelevelsplitting2}
\end{align}
$y_{\vec{54}}$ being the conveniently normalized Yukawa coupling.\footnote{The Clebsch--Gordan coefficients match those of Ref.~\cite{Malinsky:2007qy}.}
The only possible DM candidate resides in the electroweak Dirac doublet $\left(\vec{1},\vec{2},\tfrac{1}{2}\right)$ with mass in the TeV range.  This leads to the \underline{LR bi-doublet 4 scenario} where the "4" refers to the number of fermionic degrees of freedom at low scale. Its low-energy phenomenology (see scenario~\ref{sec:LR_bi-doublet_4}), in particular constraints on mass splittings, implies an LR scale broken below $\unit[75]{TeV}$~\cite{Garcia-Cely:2015quu}.
This means also that paths proceeding through $G_{421}$ are excluded for this scenario because such a group must be broken above $\sim \unit[1000]{TeV}$, implying a $W_R$ with a mass of at least this scale, leading to a too small neutral mass splitting. To have a $G_{3211}$ symmetry at an intermediate scale between the LR breaking scale and the electroweak (EW) scale is a possibility.

This leaves the question of what happens to the Dirac fermion $\left(\vec{3},\vec{1},-\tfrac{1}{3}\right)$, the $SO(10)$ partner of the bi-doublet. Clearly, its presence at low scales would be very constrained from heavy nuclei searches etc., so we have to make sure it is heavier than the EW doublet and decays sufficiently fast. 
Since it does not belong to the same PS multiplet as DM, it is a "high-scale DM partner," as defined in Sec.~\ref{sec:DMsplittings}. This means that
its decay proceeds via the virtual GUT-scale gauge bosons $U$ and $V$ (see Tab.~\ref{tableLowerLimits} and Fig.~\ref{fig:spectrum_10}), 
which implies that its mass must be at least $m_{\left(\vec{3},\vec{1},-1/3\right)} > \unit[10^5]{TeV}$, in order that it decays before BBN.
As the GUT-gauge-boson induced radiative splitting of the $\vec{10}$ is insufficient for this purpose, one has to rely on a fine-tuned cancellation of the bare mass $m_\vec{1}$ and the  $\vec{54}$ VEV $v_\vec{54}$ in Eqs.~(\ref{10treelevelsplitting1}) and~(\ref{10treelevelsplitting2}).
It is thus mandatory that there exists a $\vec{54}$ representation which acquires a VEV.
Note, however, that $v_\vec{54}$ is not necessarily the only VEV contributing to $SO(10)$ breaking (seeing as we need some additional PS breaking VEVs anyway), so it could conceivably be lower than the naive GUT-scale $\unit[10^{13}]{TeV}$; since we need it above $\unit[10^5]{TeV}$, this would still imply a fine-tuning of $1$ in $10^5$ or more to obtain the required multiplet splitting.
This mass splitting problem turns out to be present in all of our simple GUT DM models. (A recent attempt at avoiding this problem requires two copies of the $\vec{10}$~\cite{Boucenna:2015sdg}.) This fine-tuning aside, the $\vec{10}$ makes for a perfectly viable $SO(10)$ DM candidate, in the form of the neutral component of a low-scale quasi-degenerate fermion bi-doublet~\cite{Garcia-Cely:2015quu}, requiring in summary  a $\vec{54}$ VEV above $\unit[10^5]{TeV}$ and a low-scale LR scale below $\unit[75]{TeV}$ [i.e.~a path not going through $SU(5)$ or $G_{421}$], with or without an extra $G_{3211}$ breaking step at a lower scale.

\subsection{Fermionic $\vec{45}$ DM candidates}
\label{sec:45}

The adjoint representation of $SO(10)$ contains several promising DM candidates, as can already be seen from the PS decomposition
\begin{align}
	\vec{45} &\stackrel{\text{PS}}{\to} \left(\vec{1},\vec{3},\vec{1}\right)\oplus \left(\vec{1},\vec{1},\vec{3}\right)\oplus  \left(\vec{15},\vec{1},\vec{1}\right)\oplus \left(\vec{6},\vec{2},\vec{2}\right),
\end{align}
where the first three components contain electrically neutral particles. The product rule
\begin{align}
\hspace{-1.9ex}\vec{45}\otimes \vec{45} &= \vec{1}_S \oplus \vec{45}_A \oplus \vec{54}_S\oplus \vec{210}_S\oplus \vec{770}_S\oplus \vec{945}_A
\end{align}
illustrates that the masses can be split at tree level by scalar VEVs $\langle\vec{54}\rangle$, $\langle\vec{210}\rangle$, and $\langle\vec{770}\rangle$. To avoid large representations, we will restrict ourselves to scalars of dimension $\leq 210$.

Our discussion is greatly simplified by the fact that we can always use the matter-parity even scalar representations $\vec{126}$ and $\vec{45}$ to go down the PS breaking path (Fig.~\ref{fig:so(10)_breaking_paths_PS}) without giving any mass contribution to the $\vec{45}$ DM representation.
It implies that to determine the possible mass spectra scenarios we can just discuss them looking directly at the mass formula,
considering the various possible hierarchies one could have between the various possible mass contributions, without needing to consider the breaking paths explicitly. This can be done even if for what concerns the mass scale of the various $SO(10)$ gauge bosons (and radiative mass splittings induced by one-loop diagrams involving these gauge bosons) what matters is the full breaking path. In practice to discuss the various possible low-energy spectra the mass formula can give, we will discuss them according to down to which subgroup we assume that tree-level mass contributions are generated, starting from $G_{422}$, considering subsequently $G_{3221}$ and $G_{3211}$.

\subsubsection{Tree-level masses down to $G_{422}$}

If the $SO(10)$ breaking path goes through the $G_{422}$ subgroup and is afterwards broken only by $\langle\vec{126}\rangle$ and/or $\langle\vec{45}\rangle$, the general tree-level mass formula is
\begin{align}
	m_{\left(\vec{1},\vec{3},\vec{1}\right)} &= m_\vec{1} -6 m_\vec{54} + m_{{\vec{210}}_{(\vec{1},\vec{1},\vec{1})}}\,,\\
        m_{\left(\vec{1},\vec{1},\vec{3}\right)}&= m_\vec{1} -6 m_\vec{54} - m_{{\vec{210}}_{(\vec{1},\vec{1},\vec{1})}}\,,\\
	m_{\left(\vec{15},\vec{1},\vec{1}\right)} &= m_\vec{1} +4 m_\vec{54} \,,\\
	m_{\left(\vec{6},\vec{2},\vec{2}\right)} &= m_\vec{1} - m_\vec{54} \,,
\label{eq:45PSmasses}
\end{align}
where everywhere in this work we will define
\begin{equation}
m_{X}\equiv y_X\, v_X \,,
\end{equation}
see App.~\ref{app:tables} for more details.
The mass contribution $m_{{\vec{210}}_{(\vec{1},\vec{1},\vec{1})}}$ from the $\vec{210}$ clearly breaks the left--right exchange $D$ parity~\cite{Chang:1984qr}, whereas $m_\vec{54}$ conserves it.
Whatever is the structure of the various terms in these equations, the $\left(\vec{6},\vec{2},\vec{2}\right)$ mass needs to be above $\unit[10^5]{TeV}$ because it is a "high-scale DM partner," as defined in Sec.~\ref{sec:DMsplittings}, so TeV-scale DM demands here too a fine-tuned cancellation. To discuss the various possibilities
we will first consider a single scalar VEV mass contribution and subsequently consider the general case.

\underline{Tree-level mass contribution only from $\vec{54}$}: If there is no $\vec{210}$ mass contribution but only a $\vec{54}$ one, the left- and right-handed triplets are degenerate at tree level due to $D$ parity. This common mass $m_{\left(\vec{1},\vec{3},\vec{1}\right)} = m_{\left(\vec{1},\vec{1},\vec{3}\right)} = \mathcal{O}(\unit{TeV})$ can be tuned, keeping the other $\vec{45}$ components above $\unit[10^5]{TeV}$.
The LR group breaking scale  must be below $\unit[10^6]{TeV}$ in order for the charged right-handed wino partner to have decayed by the BBN time, or even below $\sim \unit[100]{TeV}$ due to the relic density constraints (which given the fact that $G_{422}$ must be broken above $\sim \unit[10^3]{TeV}$ requires $G_{422}$ to be broken to $G_{3221}$ by an extra $\vec{45}$ in order to decouple the PS and LR scales), see scenario~\ref{sec:LR_triplet_6}. This is the "\underline{LR triplet 6 scenario}", where the "6" indicates that there are six low-energy states degenerate at tree level. 

Alternatively one could have the $\left(\vec{15},\vec{1},\vec{1}\right)$ to be the lightest $\vec{45}$ component (with all other $\vec{45}$ components above $\unit[10^5]{TeV}$), as this includes a total singlet after PS breaking:
\begin{align}
	\left(\vec{15},\vec{1},\vec{1}\right) \stackrel{\text{LR}}{\to} &\left(\vec{1},\vec{1},\vec{1},0\right)\oplus \left(\vec{8},\vec{1},\vec{1},0\right)\nonumber\\
	&\oplus \left(\vec{3},\vec{1},\vec{1},4\right)\oplus  \left(\overline{\vec{3}},\vec{1},\vec{1},-4\right).
\end{align}
This scenario cannot lead to the observed relic density in a thermal way, neither through annihilation nor through co-annihilation (the latter because the radiative splitting induced is too large for that), see the discussion of scenario~\ref{sec:Octet-triplet-singlet_8+6+1}. 

One could think about non-thermal production of the singlet with a low reheating temperature. This could be achieved assuming a reheating temperature sufficiently far below the PS scale so that the singlet is not put in equilibrium but is at most non-thermally produced~\cite{Hall:2009bx}. This is easier to achieve for a large PS scale, which in turn increases the mass splitting among the $\vec{15}$ components (see Fig.~\ref{fig:mass_ratio_toy}). This should work, but will of course lead to an unwelcome dependence on the initial conditions. Still, we end up with colored DM partners in the TeV range that can be searched for at the LHC, mimicking to some degree gluinos and squarks. We will call this scenario the "\underline{octet-triplet-singlet non-thermal 15 scenario}", which we mentioned as a possible "other scenario" in \ref{sec:Otherscenarios}. To be more explicit, if the reheating temperature is below the PS scale, the singlet $\chi_1$ only has dimension-six interactions of the form $\overline{\chi}_1\gamma^\mu \chi_3 \overline{q}\gamma_\mu \ell/v_\text{PS}^2$ with the color triplet $\chi_3$, ignoring the color octet partner for simplicity. As shown in Ref.~\cite{Elahi:2014fsa}, the UV freeze-in DM abundance from such a non-renormalizable operator should scale as
\begin{align}
\Omega_\text{DM} \propto m_\text{DM} m_\text{Pl} T_\text{RH}^3/v_\text{PS}^4\,,
\end{align}
assuming the hierarchy $m_\text{DM} \ll T_\text{RH} \ll v_\text{PS}$ and ignoring numerical prefactors.
Fixing $m_\text{DM} = \unit[1]{TeV}$ for illustration purposes, we can use the radiative mass splitting of $\chi_1$ and $\chi_3$ (App.~\ref{app:rge}) to put an upper bound of  $v_\text{PS} < \unit[7\times 10^5]{TeV}$ on the PS scale by demanding the triplet to decay before BBN. The triplet mass is then around $\unit[2]{TeV}$, depending on the exact value of $v_\text{PS}$. Note that the singlet is indeed out of chemical equilibrium in this case, see Fig.~\ref{fig:chem_eq}. 
Analogous to Ref.~\cite{Elahi:2014fsa} we can then estimate a required reheating temperature around $\unit[1]{TeV}$ in order to obtain the correct relic abundance, which is low due to the rather low PS scale. Properly taking the fermion masses into account would certainly change this number somewhat, but it seems likely that one can adjust the reheating temperature to obtain the desired DM abundance.
Let this suffice as an example for non-freeze-out DM from GUTs, it should be clear that there are far more scenarios that could be discussed. We will not consider any other examples, focusing instead on thermal scenarios.

\underline{Only $\vec{210}_{(\vec{1},\vec{1},\vec{1})}$}: The VEV of a ${{\vec{210}}_{(\vec{1},\vec{1},\vec{1})}}$ scalar field has an opposite mass contribution to both $L$ and $R$ triplets, manifestly breaking $D$ parity. In this case since the $SO(10)$ conserving mass term $m_\vec{1}$ must be at least of order  $\unit[10^5]{TeV}$ in order that the ${(\vec{6},\vec{2},\vec{2})}$ components are at least at this scale, the $(\vec{15},\vec{1},\vec{1})$ is also at this scale and the only possibility is to have the $(\vec{1},\vec{3},\vec{1})$ or the $(\vec{1},\vec{1},\vec{3})$ at low scale but not both. The former leads to the "\underline{$SU(2)_L$ triplet 3 scenario}" or "wino" scenario~\ref{sec:SU(2)_L_triplet_3}.
If instead the $SU(2)_R$ triplet is alone at low scale, this leads to a singlet DM scenario with a charged partner heavier by an amount of up to GeV, depending on the $W_R$ mass. This is the "\underline{$SU(2)_R$ triplet 3 scenario}" of Sec.~\ref{sec:SU(2)_R_triplet_3}.

\underline{General case: $\vec{54}$ and $\vec{210}_{(\vec{1},\vec{1},\vec{1})}$}: With respect to the case with only a $\vec{54}$ representation above, this case brings an additional
tree-level splitting between the $(\vec{1},\vec{3},\vec{1})$ and the $(\vec{1},\vec{1},\vec{3})$. If this splitting is far above the DM mass we get the same phenomenology as with only a $\vec{210}_{(\vec{1},\vec{1},\vec{1)}}$: a single \underline{$SU(2)_L$ triplet} or a single \underline{$SU(2)_R$ triplet} candidate.
Alternatively if the $\vec{210}_{(\vec{1},\vec{1},\vec{1})}$ mass contribution is of order of the DM mass, one ends up with both L and R triplets at low scale but with a splitting of order of their mass. With respect to the case above with only a $\vec{54}$ representation, this brings the possibility to have independent masses for both triplets, still with both neutral components of the triplets constituting the DM. This leads to much more freedom for the masses of these triplets and allows one to alleviate largely the strong indirect detection constraints which hold for a pure wino or for the case where both triplets are degenerate. This is the "\underline{LR triplet 3+3 scenario}"~\ref{sec:LR_triplet_3+3}. 
It can be mentioned here that with more than one large fine-tuning, we could also have at low scale
any of the triplets (but not both) together with the $(\vec{15},\vec{1},\vec{1})$ representation. This shows that with more fine-tuning we can sometimes get more degrees of freedom at low scale but not necessarily whatever states.

\subsubsection{Tree-level masses down to $G_{3221}$}

With respect to the previous cases above, this case necessarily involves a
$\vec{210}_{(\vec{15},\vec{1},\vec{1})}$ VEV, which conserves $D$~parity. The only net effect of this scalar field is to bring different mass contributions to the LR multiplets within the PS $(\vec{15},\vec{1},\vec{1})$ multiplet:
\begin{align}
\begin{split}
	m_{\left(\vec{1},\vec{3},\vec{1},0\right)} &= m_\vec{1} -6 m_\vec{54} + m_{{\vec{210}}_{(\vec{1},\vec{1},\vec{1})}}\,,\\
	m_{\left(\vec{1},\vec{1},\vec{3},0\right)} &= m_\vec{1} -6 m_\vec{54} - m_{{\vec{210}}_{(\vec{1},\vec{1},\vec{1})}}\,,\\
	m_{\left(\vec{1},\vec{1},\vec{1},0\right)} &= m_\vec{1} +4 m_\vec{54} - 2 m_{{\vec{210}}_{(\vec{15},\vec{1},\vec{1})}}\,,\\
	m_{\left(\vec{3},\vec{1},\vec{1},4\right)} &= m_\vec{1} +4 m_\vec{54} - m_{{\vec{210}}_{(\vec{15},\vec{1},\vec{1})}}\,,\\
	m_{\left(\vec{8},\vec{1},\vec{1},0\right)} &= m_\vec{1} +4 m_\vec{54} + m_{{\vec{210}}_{(\vec{15},\vec{1},\vec{1})}}\,,\\
	m_{\left(\vec{3},\vec{2},\vec{2},-2\right)} &= m_\vec{1} - m_\vec{54} \,.
\end{split}
\label{eq:45LRmasses}
\end{align}
This extra mass contribution does not change anything in the scenarios above which have $SU(2)_L$ and/or $SU(2)_R$ triplet(s) at low scale (except if we were allowing for more than one tuning in case one can get both triplets and the $(\vec{1},\vec{1},\vec{1},0)$ at low scale, with all other states at a high scale).

Thus, beside this special case, to add a $v_{{\vec{210}}_{(\vec{15},\vec{1},\vec{1})}}$ contribution is only relevant for the case where DM is made of the $\left(\vec{1},\vec{1},\vec{1},0\right)$ singlet within the $\left(\vec{15},\vec{1},\vec{1}\right)$. 
The addition of the $m_{{\vec{210}}_{(\vec{15},\vec{1},\vec{1})}}$ contribution can make the 
$\left(\vec{1},\vec{1},\vec{1},0\right)$ singlet a viable DM candidate, provided this contribution is below TeV.
In this case the singlet, triplet, and octet within the $\left(\vec{15},\vec{1},\vec{1}\right)$ are all present at low scale, with a mass splitting between the 
singlet and triplet which can have the value necessary for having the right amount of co-annihilation of the singlet into the triplet (see App.~\ref{app:chem_dec}). This is the ``\underline{octet-triplet-singlet 8+6+1 scenario}'' of Sec.~\ref{sec:Octet-triplet-singlet_8+6+1}. Note nevertheless that this requires
the mass splittings in the $\mathcal{O}(\unit[10]{GeV})$ range~\cite{Mitridate:2017izz} despite the fact that the pure radiative splitting is more of order TeV (see App.~\ref{app:rge}), which implies a cancellation of the tree-level and radiative mass splittings at the few percent level. This also requires a rather low PS scale in order to keep the singlet in chemical equilibrium during co-annihilation, roughly $m_X/g_4 \lesssim \unit[1500]{TeV} (m_\text{DM}/\unit[2]{TeV})^{3/4}$, fairly close to meson constraints, see App.~\ref{app:chem_dec}). 
The phenomenology is then similar to Refs.~\cite{deSimone:2014pda,Mitridate:2017izz}, although we have two co-annihilation partners, which change the numbers somewhat. 

With this low PS scale, the tiny mass splitting the lifetime of the $\left(\vec{8},\vec{1},\vec{1},0\right)$ and $\left(\vec{3},\vec{1},\vec{1},4\right)$ is just about enough to satisfy BBN constraints (see Fig.~\ref{fig:chem_eq}).\footnote{Assuming a mass splitting of $\unit[100]{GeV}$ ($\unit[10]{GeV}$), BBN gives a lower bound on the PS scale of $\unit[4\times 10^4]{TeV}$ ($\unit[2\times 10^3]{TeV}$), still compatible with the constraints of Tab.~\ref{tableLowerLimits}.} This makes the colored partners in particular stable on collider scales, which leads to new signatures discussed in Sec.~\ref{sec:Octet-triplet-singlet_8+6+1}.

\subsubsection{Tree-level masses down to $G_{3211}$}

So far we have ignored a possible VEV $v_{{\vec{210}}_{(\vec{15},\vec{1},\vec{3})}}$.
The presence of such a VEV implies that the lowest group down to which one gets tree-level mass contributions is $G_{3211}$.
Under this group the three LR multiplets which contain a DM candidate decompose as
\begin{align}
    (\vec{1},\vec{1},\vec{1},0)&\rightarrow (\vec{1},\vec{1},0,0) \,,\\
	(\vec{1},\vec{3},\vec{1},0)&\rightarrow (\vec{1},\vec{3},0,0) \,,\\
	(\vec{1},\vec{1},\vec{3},0)&\rightarrow (\vec{1},\vec{1},0,0)'\oplus (\vec{1},\vec{1},2,0)\,.
\end{align}
At this level four VEVs can contribute to the mass formula leading to a more involved mass pattern:
\begin{align}
\begin{split}
    m_{\left(\vec{1},\vec{1},0,0\right)'} &= m_\vec{1} +A+\frac{1}{2}\sqrt{B}\,,\\
    m_{\left(\vec{1},\vec{1},0,0\right)} &= m_\vec{1} +A-\frac{1}{2}\sqrt{B}\,,\\
	m_{\left(\vec{1},\vec{3},0,0\right)} &= m_\vec{1} -6 m_\vec{54} + m_{{\vec{210}}_{(\vec{1},\vec{1},\vec{1})}}\,,\\
	m_{\left(\vec{1},\vec{1},2,0\right)} &= m_\vec{1} -6 m_\vec{54} - m_{{\vec{210}}_{(\vec{1},\vec{1},\vec{1})}}\,,\\
	m_{\left(\vec{3},\vec{1},0,4\right)} &= m_\vec{1} +4 m_\vec{54} - m_{{\vec{210}}_{(\vec{15},\vec{1},\vec{1})}}\,,\\
	m_{\left(\vec{8},\vec{1},0,0\right)} &= m_\vec{1} +4 m_\vec{54} + m_{{\vec{210}}_{(\vec{15},\vec{1},\vec{1})}}\,,\\
	m_{\left(\vec{3},\vec{2},-1,-2\right)} &= m_\vec{1} - m_\vec{54} + m_{{\vec{210}}_{(\vec{15},\vec{1},\vec{3})}}\,,\\
	m_{\left(\vec{3},\vec{2},1,-2\right)} &= m_\vec{1} - m_\vec{54} - m_{{\vec{210}}_{(\vec{15},\vec{1},\vec{3})}}\,,
\end{split}
\end{align}
with
\begin{align}
	A &\equiv - m_\vec{54} -\frac{1}{2} m_{{\vec{210}}_{(\vec{1},\vec{1},\vec{1})}} - m_{{\vec{210}}_{(\vec{15},\vec{1},\vec{1})}}\,,\\
	B &\equiv (10 m_\vec{54} + m_{{\vec{210}}_{(\vec{1},\vec{1},\vec{1})}}-2 m_{{\vec{210}}_{(\vec{15},\vec{1},\vec{1})}})^2\nonumber\\
	&\quad +24 m_{{\vec{210}}_{(\vec{15},\vec{1},\vec{3})}}^2 \,.
\end{align}
Thus a $v_{{\vec{210}}_{(\vec{15},\vec{1},\vec{3})}}$ VEV modifies the masses of the LR singlet, of the neutral component of the $SU(2)_R$ triplet, and of two color triplets out of the PS sextet. These four contributions are different, as the CG coefficient is different for each one.
Most importantly, the $v_{{\vec{210}}_{(\vec{15},\vec{1},\vec{3})}}$ VEV also \emph{mixes} the singlets, which leads to the square root expression in their masses from a matrix diagonalization.
The discussion of this case depends on whether 
the mass contribution $m_{\vec{210}_{(\vec{15},\vec{1},\vec{3})}}$ is smaller than any of the other tree-level mass contributions. 

\underline{Leading $m_{\vec{210}_{(\vec{15},\vec{1},\vec{3})}}$}: By leading $m_{\vec{210}_{(\vec{15},\vec{1},\vec{3})}}$ contribution,
we here mean leading with respect to the
contribution of the other tree-level mass contribution in the $\sqrt{B}$ term (including for instance if $SO(10)$ is broken only by $\vec{45}$ representations on top of this one). 
A leading $m_{\vec{210}_{(\vec{15},\vec{1},\vec{3})}}\sim \hbox{TeV}$ contribution is excluded because, in this case, either all contributions from all scalars are small, which leaves all $\vec{45}$ particles at low scale [including high-scale DM partners such as the $(\vec{6}, \vec{2}, \vec{2})_\text{PS}$], or there is fine-tuned cancellation of the various other contributions in the $\sqrt{B}$ term, which may leave one of the DM candidates at a low scale only at the price of several tunings.
To have instead $m_{\vec{210}_{(\vec{15},\vec{1},\vec{3})}}\gg\unit{TeV}$
is compatible with a low-scale DM singlet, $(\vec{1},\vec{1},0,0)$ or $(\vec{1},\vec{1},0,0)'$, alone at low scale, but in this case this singlet has nothing to annihilate into and nothing to co-annihilate with.
A large leading $m_{\vec{210}_{(\vec{15},\vec{1},\vec{3})}}\gg\unit{TeV}$ is also compatible with having the $SU(2)_L$ triplet at low scale. In this case we go back to the situations with low-scale triplets above but with the neutral $SU(2)_R$ triplet component necessarily at a much higher scale.
This may leave the charged $SU(2)_R$ triplet components at a lower scale, but this is excluded because in this case they have nothing to decay into. Thus one concludes that a leading $m_{\vec{210}_{(\vec{15},\vec{1},\vec{3})}}\gg\unit{TeV}$ contribution is compatible only with having  the $SU(2)_L$ triplet at low scale, i.e.~the "\underline{$SU(2)_L$ triplet 3 scenario}"~\ref{sec:SU(2)_L_triplet_3}, which also requires a large $m_{\vec{210}_{(\vec{1},\vec{1},\vec{1})}}$ contribution in order to split the charged R triplet components from the L triplet.

\underline{Subleading $m_{\vec{210}_{(\vec{15},\vec{1},\vec{3})}}$}: If $m_{\vec{210}_{(\vec{15},\vec{1},\vec{3})}}$ is smaller than the other tree-level mass contributions in the $\sqrt{B}$ term,
the $(\vec{1},\vec{1},0,0)$ or $(\vec{1},\vec{1},0,0)'$ singlets receive a "seesaw" suppressed contribution from this $m_{\vec{210}_{(\vec{15},\vec{1},\vec{3})}}$:
\begin{align}
	m_{\left(\vec{1},\vec{1},0,0\right)} &\simeq m_\vec{1} -6 m_\vec{54} - m_{\vec{210}_{(\vec{1},\vec{1},\vec{1})}}\label{seesawplus}\\
	&\quad-\frac{6 (m_{\vec{210}_{(\vec{15},\vec{1},\vec{3})}})^2}{10 m_\vec{54}-2 m_{\vec{210}_{(\vec{15},\vec{1},\vec{1})}} +m_{\vec{210}_{(\vec{1},\vec{1},\vec{1})}}} \,,\nonumber\\
	m_{\left(\vec{1},\vec{1},0,0\right)'} &\simeq m_\vec{1} +4 m_\vec{54} -2 m_{\vec{210}_{(\vec{15},\vec{1},\vec{1})}}\label{seesawminus}\\
	&\quad+\frac{6 (m_{\vec{210}_{(\vec{15},\vec{1},\vec{3}})})^2}{10 m_\vec{54}-2 m_{\vec{210}_{(\vec{15},\vec{1},\vec{1})}} +m_{\vec{210}_{(\vec{1},\vec{1},\vec{1})}}}\,.\nonumber
\end{align}
To have this contribution be subleading is somewhat to be expected as it breaks a smaller group than the other contributions.
This allows an interesting situation where this seesaw contribution is of order the DM mass or lower, even though all breaking scales are much larger. For example, for the denominator of order the GUT scale one gets a contribution of order, say $\unit[1]{GeV}$ or $\unit[1]{TeV}$, if  $m_{\vec{210}_{(\vec{15},\vec{1},\vec{3})}}\sim \text{few }\unit[10^4]{TeV}$ or $\sim \unit[10^6]{TeV}$, respectively.
In this case this seesaw contribution allows one to shift the mass of each of both singlets accordingly.
This does not affect much the scenarios above which, without this contribution, lead to the
"\underline{$SU(2)_L$ triplet 3 scenario}"~\ref{sec:SU(2)_L_triplet_3}, but does affect
the frameworks above which, without this contribution, lead one of these two singlets at a low scale.

In practice this means that the scenarios above where the neutral component of the $SU(2)_R$ triplet was DM now receives an extra seesaw contribution for this neutral component, leading to a DM candidate whose mass is given by Eq.~\eqref{seesawminus}.
Thus these scenarios, which are the "{LR triplet 6 scenario}"~\ref{sec:LR_triplet_6}, as well as the "{$SU(2)_R$ triplet 3 scenario}"~\ref{sec:SU(2)_R_triplet_3} above, as well as the "{LR triplet 3+3 scenario}"~\ref{sec:LR_triplet_3+3}  become 
a "\underline{LR triplet 5+1 scenario}"~\ref{sec:LR_triplet_5+1} (where
the $SU(2)_L$ triplet which is degenerate at tree level with the charged $SU(2)_R$ triplet components), a "\underline{$SU(2)_R$ triplet 2+1 scenario}"~\ref{sec:SU(2)_R_triplet_2+1}, and a "\underline{LR triplet $3+2+1$ scenario}"~\ref{sec:LR_triplet_3+2+1}, respectively.
Note that, given the fact that these scenarios require the $W_R$ mass to be below $\sim \unit[100]{TeV}$ (in order that the R triplet component does not overclose the Universe, see~\ref{sec:SU(2)_R_triplet_3}), this gives an accordingly upper bound on $m_{\vec{210}_{(\vec{15},\vec{1},\vec{3})}}$ because $\langle \vec{210}_{(\vec{15},\vec{1},\vec{3})}\rangle$ contributes to $m_{W_R}$.
To have a seesaw contribution not larger than $\sim \unit[10]{GeV}$ (as required by the relic density constraint, Fig.~\ref{fig:right-handed_wino}) requires that the denominator of the seesaw contribution is of order $\unit[10^7]{TeV}$ (for a LR scale of order $\unit[100]{TeV}$), 
which is compatible with the fact that some of the mass contributions in the denominator must be larger than $\unit[10^5]{TeV}$  to make the ``high-scale partners'' heavy enough (with $m_\vec{1}+4 m_{\vec{54}}-2 m_{\vec{210}_{(\vec{15},\vec{1},\vec{1})}}$ in Eq.~(\ref{seesawminus}) tuned to be of order TeV).

Let us give one quantitative example on how to obtain the "{$SU(2)_R$ triplet 2+1 scenario}"~\ref{sec:SU(2)_R_triplet_2+1} in this setup. To minimize fine-tuning we use the $\vec{45}$ VEV $\langle\vec{45}_{(\vec{15},\vec{1},\vec{1})}\rangle >\unit[10^{12}]{TeV}$ to break $SO(10)$ to $G_{3221}^{\slashed{D}}$ (see Fig.~\ref{fig:so(10)_breaking_paths_PS}) without contributing to the DM masses. We further pick $m_{\vec{210}_{(\vec{1},\vec{1},\vec{1})}}=\unit[10^{6}]{TeV}$ and $m_{\vec{210}_{(\vec{15},\vec{1},\vec{1})}}=\unit[4\times 10^{5}]{TeV}$.

The $SO(10)$-invariant mass term $m_\vec{1}$ has to be fine-tuned to push the $(\vec{1},\vec{1},{\vec{3}},0)$ mass scale below TeV: $m_\vec{1} = \unit[(10^{6} -0.25)]{TeV}$. This gives us the $SU(2)_R$ triplet with mass $\unit[0.25]{TeV}$ and all other $\vec{45}$ partners with mass above $\unit[10^{5}]{TeV}$. Finally we turn on $m_{\vec{210}_{(\vec{15},\vec{1},\vec{3})}}=\unit[15]{TeV}$ in order to split the $SU(2)_R$ triplet components by $\unit[6]{GeV}$ through the seesaw contribution. As far as gauge boson masses are concerned, the colored $U$, $V$, and $X$ have GUT-scale masses, whereas $W_R$ is potentially at the $\unit[10]{TeV}$ scale. 
A full breaking to the SM requires furthermore a $\langle \vec{126}\rangle$, which should be below $\unit[10]{TeV}$ in order not to contribute too much to $m_{W_R}$.
In this case we realize the "{$SU(2)_R$ triplet 2+1 scenario}"~\ref{sec:SU(2)_R_triplet_2+1}. 
Note that the DM mass splittings depend on the mass terms $m_X = y_X v_X$, whereas the gauge boson masses do not depend on the Yukawa couplings but rather on the gauge couplings. Above we have essentially taken all Yukawas and gauge couplings to be of order one for simplicity, but with these additional parameters the relevant scales are subject to much more freedom.

Coming back to our systematic discussion of subleading $m_{\vec{210}_{(\vec{15},\vec{1},\vec{3})}}$, the only other possibility is to consider an order TeV seesaw contribution in Eq.~(\ref{seesawminus}), i.e.~considering as DM candidate the singlet in the $(\vec{15},\vec{1},\vec{1})$. In this case the case where all the 15 components were degenerate at tree level, Eq.~(\ref{eq:45PSmasses}) (which was not working as explained above), now becomes an "\underline{octet-triplet-singlet 14+1 scenario}" \ref{sec:Octet-triplet-singlet_14+1} where the triplet and octet are still degenerate at tree level. The "octet-triplet-singlet 8+6+1 scenario", Eq.~\eqref{eq:45LRmasses}, remains an "\underline{octet-triplet-singlet 8+6+1 scenario}", but with more sources of tree-level mass splitting.
The "octet-triplet-singlet 14+1 scenario" is viable through co-annihilation, since it allows one to split the singlet from the triplet by just the right amount at the price of a tuning at the few percent level, see the discussion for the "octet-triplet-singlet 8+6+1 scenario" above (requiring the seesaw induced splitting to be of order $\sim$~TeV).

Finally note that the last possibility is to have a seesaw contribution in Eqs.~(\ref{seesawplus}) and~(\ref{seesawminus}) much larger than $\sim\unit{TeV}$. This necessarily splits any SM singlet DM candidates in the $\vec{45}$ from any other multiplet (by such a large seesaw mass contribution), 
excluding these singlets as DM candidates (at least again if one invokes only one large fine-tuning in the mass formula). Thus this is a viable option only along the pure "\underline{$SU(2)_L$ triplet 3 scenario}"~\ref{sec:SU(2)_L_triplet_3}.

\subsection{Fermionic $\vec{54}$ DM candidates}
\label{sec:54}

The $\vec{54}$ only receives tree-level mass splittings from a $\vec{54}$ VEV (Tab.~\ref{tab:422and3221}), making the discussion very simple:
\begin{align}
\begin{split}
	m_{\left(\vec{1},\vec{1},\vec{1}\right)} &= m_\vec{1} +2 m_\vec{54} \,,\\
	m_{\left(\vec{1},\vec{3},\vec{3}\right)} &= m_\vec{1} +6 m_\vec{54} \,,\\
	m_{\left(\vec{6},\vec{2},\vec{2}\right)} &= m_\vec{1} +  m_\vec{54} \,,\\
	m_{\left(\vec{20}',\vec{1},\vec{1}\right)} &= m_\vec{1} -4 m_\vec{54} \,.
\end{split}
\end{align}
It contains one PS singlet, which only has GUT-suppressed annihilations and thus overcloses the Universe if it was ever in thermal equilibrium. Freeze-in should work, but will not be discussed here.
The other candidate is the bi-triplet $\left(\vec{1},\vec{3},\vec{3}\right)$~\cite{Garcia-Cely:2015quu}, leading to the \underline{bi-triplet 9 scenario}~\ref{sec:LR_bi-triplet_9}.
This requires a rather low LR scale, below $\unit[100]{TeV}$.

\subsection{Fermionic $\vec{120}$ DM candidates}
\label{sec:120}

The $\vec{120}$ fermion representation with decomposition
\begin{align}
	\vec{120} &\stackrel{\text{PS}}{\to} \left(\vec{1},\vec{2},\vec{2}\right)\oplus \left(\vec{6},\vec{1},\vec{3}\right)\oplus  \left(\vec{6},\vec{3},\vec{1}\right)\nonumber \\
	&\quad \quad  \oplus \left(\vec{10},\vec{1},\vec{1}\right) \oplus \left(\vec{15},\vec{2},\vec{2}\right),
\end{align}
contains two colorless $SU(2)_L\times SU(2)_R$ bi-doublets whose neutral components are DM candidates.
Masses for the various components come from
\begin{align}
\begin{split}
\hspace{-1ex}\vec{120}\otimes \vec{120} &= \vec{1}_S \oplus \vec{45}_A \oplus \vec{54}_S\oplus \vec{210}_S\\
&\quad\oplus \vec{210}_A\oplus \vec{770}_S\oplus \dots
\end{split}
\end{align}
For the fermionic $\vec{120}$ representation there is no mass contribution coming from a $\langle\vec{45}\rangle$ or $\langle\vec{126}\rangle$ scalar VEV, which greatly simplifies the discussion, just as for the $\vec{45}$ fermion DM representation.
Here too we will discuss the $\vec{120}$ case as a function of the last subgroup down to which tree-level mass contributions are generated.

\subsubsection{Tree-level masses down to $G_{422}$}

If the $SO(10)$ breaking chain goes through the $G_{422}$ subgroup, the various components may receive, on top of a universal $SO(10)$ invariant mass $m_\vec{1}$, two mass contributions, from $v_\vec{54}$ and $v_{\vec{210}_{(\vec{1},\vec{1},\vec{1})}}$, the latter breaking $D$ parity explicitly:
\begin{align}
\begin{split}
 m_{(\vec{1}, \vec{2}, \vec{2})} & =  m_\vec{1}-9 m_\vec{54} \,,\\
 m_{(\vec{6}, \vec{1}, \vec{3})} & =  m_\vec{1}-4 m_\vec{54}-m_{\vec{210}_{(\vec{1},\vec{1},\vec{1})}} \,,\\
 m_{(\vec{6}, \vec{3}, \vec{1})} & =  m_\vec{1}-4 m_\vec{54}+m_{\vec{210}_{(\vec{1},\vec{1},\vec{1})}} \,,\\
 m_{(\vec{10}, \vec{1}, \vec{1})} & =  m_\vec{1}+6 m_\vec{54} \,,\\
 m_{(\vec{15}, \vec{2}, \vec{2})} & =  m_\vec{1}+m_\vec{54} \,.
\end{split}
\label{eq:120underG422}
\end{align}
The $(\vec{10},\vec{1},\vec{1})$ as well as the $(\vec{6},\vec{1},\vec{3})$ and  $(\vec{6},\vec{3 },\vec{1})$ are high-scale DM partners, so that
they must lie above $\unit[10^5]{TeV}$.
Since both bi-doublets as well as the $(\vec{10},\vec{1},\vec{1})$ receive mass contributions only from $m_\vec{1}$ and the $\vec{54}$, a TeV scale DM candidate necessarily requires a $\vec{54}$ contribution above $\unit[10^5]{TeV}$ fine-tuned with $m_\vec{1}$.

\underline{Tree-level mass contribution only from $\vec{54}$}: in this case one can get either of both bi-doublets at low scale.
If the bi-doublet singled out at low energy is the $(\vec{1},\vec{2},\vec{2})$ one gets the same low-energy "\underline{bi-doublet 4 scenario}"~\ref{sec:LR_bi-doublet_4} as for the $\vec{10}$ representation above. If instead it is the bi-doublet in $(\vec{15},\vec{2},\vec{2})$ which appears at low scale,
it comes with the entire  $(\vec{15},\vec{2},\vec{2})$ multiplet at low scale, that is to say with a color-octet bi-doublet and two color-triplet bi-doublets, for a total of 60 tree-level degenerate particles at low scale. In this case one gets radiative splittings between the octet, triplet, and singlet which are similar to the ones reported above for the $(\vec{15},\vec{1},\vec{1})$ in the $\vec{45}$ representation (see also App.~\ref{app:rge}). This means that the radiative splittings are large enough for these color partners to decay before they could co-annihilate with the singlet DM particles. Thus these color partners do not play any role for the relic density, and the relic density is obtained in the same way as for the $(\vec{1},\vec{2},\vec{2})$ candidate, i.e.~the fact that
these color singlets form a bi-doublet allows one to get the relic density again along the lines of the "{bi-doublet 4 scenario}"~\ref{sec:LR_bi-doublet_4}.
This distinguishes this $(\vec{15},\vec{2},\vec{2})$ in $\vec{120}$ case from the $(\vec{15},\vec{1},\vec{1})$ of PS in the $\vec{45}$ case above. We call this scenario the "\underline{octet-triplet-singlet bi-doublet 60 scenario}"~\ref{sec:Octet-triplet-singlet_bi-doublet_60}. As already said in Sec.~\ref{sec:Octet-triplet-singlet_bi-doublet_60}, any bi-doublet scenarios must have the LR scale below $\unit[75]{TeV}$ in order to split the resulting neutral Dirac state into two Majorana ones which satisfy the direct detection constraints~\cite{Garcia-Cely:2015quu}.

\underline{General case: $\vec{54}$ and $\vec{210}_{(\vec{1},\vec{1},\vec{1})}$}: With respect to the case with only a $\vec{54}$ representation, this case brings an extra mass contribution to the sextets of PS. But since these do not contain any DM candidates and must still have a large mass, to add such a contribution does not bring any new feature.

\subsubsection{Tree-level masses down to $G_{3221}$}

The mass splittings for this case are given in Tab.~\ref{tab:422and3221}, repeated here for convenience:
\begin{align}
\begin{split}
 m_{(\vec{1}, \vec{2}, \vec{2}, 0)} & =  m_\vec{1}-9 m_\vec{54} \,,\\
 m_{(\vec{3}, \vec{1}, \vec{3}, -2)} & =  m_\vec{1}-4 m_\vec{54}-m_{\vec{210}_{(\vec{1},\vec{1},\vec{1})}} \,,\\
 m_{(\vec{3}, \vec{3}, \vec{1}, -2)} & =  m_\vec{1}-4 m_\vec{54}+m_{\vec{210}_{(\vec{1},\vec{1},\vec{1})}} \,,\\
 m_{(\vec{1}, \vec{1}, \vec{1}, -6)} & =  m_\vec{1}+6 m_\vec{54}-3 m_{\vec{210}_{(\vec{15},\vec{1},\vec{1})}} \,,\\
 m_{(\vec{3}, \vec{1}, \vec{1}, -2)} & =  m_\vec{1}+6 m_\vec{54}-m_{\vec{210}_{(\vec{15},\vec{1},\vec{1})}} \,,\\
 m_{(\vec{6}, \vec{1}, \vec{1}, -2)} & =  m_\vec{1}+6 m_\vec{54}+m_{\vec{210}_{(\vec{15},\vec{1},\vec{1})}} \,,\\
 m_{(\vec{1}, \vec{2}, \vec{2}, 0)'} & =  m_\vec{1}+m_\vec{54}-2 m_{\vec{210}_{(\vec{15},\vec{1},\vec{1})}} \,,\\
 m_{(\vec{3}, \vec{2}, \vec{2}, 4)} & =  m_\vec{1}+m_\vec{54}-m_{\vec{210}_{(\vec{15},\vec{1},\vec{1})}} \,,\\
 m_{(\vec{8}, \vec{2}, \vec{2}, 0)} & =  m_\vec{1}+m_\vec{54}+m_{\vec{210}_{(\vec{15},\vec{1},\vec{1})}} \,.
 \end{split}
 \label{eq:120underG3221}
\end{align}
This case necessarily involves a $m_{\vec{210}_{(\vec{15},\vec{1},\vec{1})}}$ which gives a mass contribution to all $\vec{120}$ states except to the $(\vec{1},\vec{2},\vec{2})$ bi-doublet and to the color triplets from the PS sextets. 

\underline{Tree-level mass contribution only from $m_{\vec{210}_{(\vec{15},\vec{1},\vec{1})}}$}: This could leave the $(\vec{1},\vec{2},\vec{2},0)'$ from $(\vec{15},\vec{2},\vec{2})_\text{PS}$ as the only low-scale multiplet, in which case we get the viable "\underline{bi-doublet 4 scenario}"~\ref{sec:LR_bi-doublet_4}.

\underline{Tree-level mass contribution only from $m_{\vec{210}_{(\vec{1},\vec{1},\vec{1})}}$ and} \underline{$m_{\vec{210}_{(\vec{15},\vec{1},\vec{1})}}$}: This case is interesting for two different reasons. First of all, even if this case involves two VEVs, it remains ``minimal'' in the sense that it involves two different mass contributions which can come from a single scalar $\vec{210}$ field. Second, in this case, all states receive mass contributions from the VEVs of scalar fields except one, the PS singlet bi-doublet $(\vec{1},\vec{2},\vec{2},0)$, which receives a contribution only from 
the $SO(10)$ invariant $m_\vec{1}$. Since the latter has no relation to the GUT scale, the bi-doublet can easily sit at the TeV scale, while all other states naturally receive mass contributions of order of the GUT scale from $\langle \vec{210}\rangle$. 
Thus, here there is seemingly no need for a cancellation between two large contributions to have a DM candidate at low scale, i.e.~no need for tree-level fine-tuning. However, still one will need a fine-tuning at the level of the loop-mass contribution because this particular fermion mass is not protected by a symmetry, i.e.~it is not protected from renormalization effects proportional to the higher scales, similar to the well-known hierarchy problem for scalars.

\begin{figure}[t]
\includegraphics[width=0.4\textwidth]{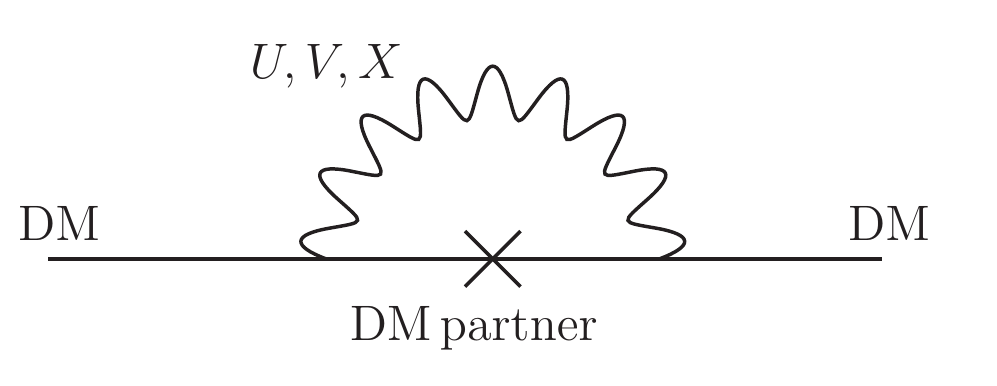}
\caption{A massless or light DM fermion will pick up the large mass of its partners (denoted as a cross) since it is connected to them via the GUT gauge bosons $U$, $V$, and $X$ as well as scalars (not shown).
}
\label{fig:mass_contribution_from_GUT}
\end{figure}

To understand this better let us symbolically write the Lagrangian in terms of a Weyl field $\xi_{\vec{120}}$ and a real scalar field $\phi_{\vec{210}}$,
\begin{align}
\begin{split}
\L &= \i \xi_{\vec{120}}^\dagger 	\overline{\sigma}^\mu D_\mu \xi_{\vec{120}} \\
&\quad - \tfrac12\left( m_\vec{1} \xi_{\vec{120}}\xi_{\vec{120}} +  y \xi_{\vec{120}}\xi_{\vec{120}}\phi_{\vec{210}} + \hc\right) ,
\end{split}
\end{align}
suppressing all indices and Clebsch--Gordon coefficients. For $m_\vec{1}=0=y$, the Lagrangian has an extra global $U(1)$ symmetry $\xi_{\vec{120}} \to e^{\i \alpha}\xi_{\vec{120}}$ that commutes with $SO(10)$ and ensures that the $\xi_{\vec{120}}$ remains massless. For $m_\vec{1}\neq 0 =y$, all loop corrections to the mass are then necessarily proportional to $m_\vec{1}$, the only $U(1)$-breaking coupling. For $m_\vec{1}= 0 \neq y$, the $U(1)$ is broken even if we had a complex $\phi_{\vec{210}}$ with transformation $\phi_{\vec{210}}\to e^{-2\i \alpha}\phi_{\vec{210}}$, because the $\vec{\vec{210}}$ obtains a VEV by assumption. Some of the fermions then obtain masses of order $y \langle \phi_{\vec{210}}  \rangle$, while others might remain massless at tree level due to vanishing Clebsch--Gordon coefficients.  Clearly there is no remaining global symmetry that could protect the masslessness of these fermions, so they should obtain loop-induced masses proportional to $y \langle \phi_{\vec{210}}  \rangle$ (see Fig.~\ref{fig:mass_contribution_from_GUT}), reintroducing the necessity for fine-tuning for a light DM candidate with heavy partners.
(Since we started off with a real representation $\vec{120}$, all fermions are Majorana or bring their own Dirac partners.)

The above argument relies on the fact that any global symmetry that could protect $m=0$ should commute with the gauge symmetry, i.e.~all fermions in the $\xi_{\vec{120}}$ feel the same chiral symmetry, or otherwise it would already be broken by gauge interactions. However, in some cases accidental $U(1)$ symmetries emerge that can keep some fermions massless at least to some loop level. An example is already provided by the $\xi_{\vec{120}}$ under PS, as given by Eq.~\eqref{eq:120underG422}. Setting the $SO(10)$ invariant mass $m_\vec{1}$ to zero as well as $m_\vec{54}=0$, we see that the only non-vanishing masses are $m_{(\vec{6},\vec{1},\vec{3})}=-m_{(\vec{6},\vec{3},\vec{1})}$. The bi-doublet $(\vec{1},\vec{2},\vec{2})$ is thus massless at tree level, but has couplings to the massive fermions and bosonic representations $(\vec{6},\vec{2},\vec{2})$ (both scalar and vector). Hence one can draw one-loop diagrams such as Fig.~\ref{fig:mass_contribution_from_GUT} that would generate a bi-doublet mass $\propto m_{(\vec{6},\vec{1},\vec{3})}$, which, however, happen to cancel between the $(\vec{6},\vec{1},\vec{3})$ and $(\vec{6},\vec{3},\vec{1})$ contributions. Indeed, one can put the two massive Majorana fermions into one Dirac-like combination $\chi_\pm \equiv \xi_{(\vec{6},\vec{1},\vec{3})}\pm \i \xi_{(\vec{6},\vec{3},\vec{1})}$ that allows us to identify a $U(1)$ symmetry $\xi_{(\vec{1},\vec{2},\vec{2})} \to e^{\i \alpha}\xi_{(\vec{1},\vec{2},\vec{2})} $, $\chi_\pm \to e^{\mp \i \alpha}\chi_\pm $ that happens to be conserved in all one-loop diagrams that would give a mass to the bi-doublet. This $U(1)$ is of course broken by other terms in the Lagrangian, so at higher loop level one will indeed find a bi-doublet mass $\propto m_{(\vec{6},\vec{1},\vec{3})}$, albeit further suppressed by loop factors.
Note that such a curious cancellation no longer holds at the LR level; as soon as the $(\vec{15},\vec{2},\vec{2})$ picks up a mass $m_{(\vec{15},\vec{2},\vec{2})}$ from $m_{\vec{210}_{(\vec{15},\vec{1},\vec{1})}}$ [see Eq.~\eqref{eq:120underG3221}], one has one-loop diagrams that should raise the bi-doublet mass from zero to $m_{(\vec{15},\vec{2},\vec{2})}$. The phenomenology of this "less-tuned" case (but unfortunately still loop level tuned) is the one of the "\underline{bi-doublet 4 scenario}"~\ref{sec:LR_bi-doublet_4}.
Note that, consequently, this case requires a LR breaking scale below $\sim\unit[75]{TeV}$ to be viable.

Finally, note that adding to the $m_{\vec{210}_{(\vec{15},\vec{1},\vec{1})}}$ contribution a $m_{\vec{210}_{(\vec{1},\vec{1},\vec{1})}}$ contribution
leaves intact the possibility  to have the
bi-doublet in the $\vec{15}$ of PS as the only state at low scale, leading again to the "\underline{bi-doublet 4 scenario}"~\ref{sec:LR_bi-doublet_4}.

\underline{Tree-level mass contribution from $m_{\vec{54}}$ and $m_{\vec{210}_{(\vec{15},\vec{1},\vec{1})}}$}: if $m_{\vec{210}_{(\vec{15},\vec{1},\vec{1})}}$ is much 
larger than $\sim\unit{TeV}$, this case can lead to a low-scale DM candidate in the form of any of both bi-doublets, with nothing else at low scale. Both of them give again the "\underline{bi-doublet 4 scenario}." If instead $m_{\vec{210}_{(\vec{15},\vec{1},\vec{1})}}$ is a $\sim\unit{TeV}$ perturbation with respect to the case above where one has only a $\vec{54}$ contribution, this does not change anything to the $(\vec{1},\vec{2},\vec{2})$ ``\underline{bi-doublet 4 scenario}'' one can have in this case, but does change the $(\vec{15},\vec{2},\vec{2})$ scenario: the color octet, triplet, and singlet in the $(\vec{15},\vec{2},\vec{2})$ are now split at tree level, which allows one more possibilities. In particular, it allows to get
the right tiny amount of splitting between the color singlet and the color triplet and octet to get the observed relic density through co-annihilation of the singlet with the triplet and octet (see the discussion for the "octet-triplet-singlet scenario" above, but now each of these multiplets is a bi-doublet of the LR symmetry, not a singlet anymore). This leads to the \underline{octet-triplet-singlet bi-doublet 32+24+4 scenario} of Sec.~\ref{sec:Octet-triplet-singlet_bi-doublet_32+24+4}.

\underline{General case}: With all three VEVs $m_{\vec{54}}$, $m_{\vec{210}_{(\vec{1},\vec{1},\vec{1})}}$, and $m_{\vec{210}_{(\vec{15},\vec{1},\vec{1})}}$ present there are no important differences with respect to the case with only 
$m_{\vec{54}}$ and $m_{\vec{210}_{(\vec{15},\vec{1},\vec{1})}}$ because, again, an additional $m_{\vec{210}_{(\vec{1},\vec{1},\vec{1})}}$ brings contributions only to the color triplets out of the PS sextet which do not contain a DM candidate.

\subsubsection{Tree-level masses down to $G_{3211}$}

This case necessarily involves a $m_{\vec{210}_{(\vec{15},\vec{1},\vec{3})}}$ VEV, leading to the pattern of masses given in Tab.~\ref{tab:421and3211_45_120}. 
This contribution splits the components of the various $SU(2)_R$ multiplets, but importantly not the two $SU(2)_L$ doublets in the DM bi-doublet candidates, because these doublets are conjugate (see above). This contribution also mixes both bi-doublets.
Its main practical effect is to split the color-singlet bi-doublet $(\vec{1},\vec{2},\vec{2},0)$ in the $(\vec{15},\vec{2},\vec{2})$ from the other $(\vec{15},\vec{2},\vec{2})$ components.
This case essentially depends on how this $m_{\vec{210}_{(\vec{15},\vec{1},\vec{3})}}$ contribution compares with respect to the other contributions. More precisely one can distinguish two cases:

\underline{Subleading $m_{\vec{210}_{(\vec{15},\vec{1},\vec{3})}}$}: If this extra contribution is subleading with respect to the other contribution in the 
square root term appearing in the mass formula (Tab.~\ref{tab:421and3211_45_120}) of the $(\vec{1},\vec{2},1,0)$ $G_{3211}$ multiplet (inside the $(\vec{1},\vec{2},\vec{2})$ of PS) and of the  $(\vec{1},\vec{2},1,0)'$ $G_{3211}$ multiplet (inside the $(\vec{15},\vec{2},\vec{2})$), then the $m_{\vec{210}_{(\vec{15},\vec{1},\vec{3})}}$ contribution is seesaw suppressed and the mass formula for these two states reduces to:
\begin{align}
	m_{\left(\vec{1},\vec{2},1,0\right)} &\simeq m_\vec{1}-9 \,m_\vec{54} \nonumber\\
	&\quad-\frac{3}{8}\frac{ (m_{\vec{210}_{(\vec{15},\vec{1},\vec{3})}})^2}{m_\vec{1}-4 \,m_\vec{54}- m_{\vec{210}_{(\vec{15},\vec{1},\vec{1})}} }\,,\label{seesawplus2}\\
	m_{\left(\vec{1},\vec{2},1,0\right)'} &\simeq m_\vec{1} + m_\vec{54} -2 \,m_{\vec{210}_{(\vec{15},\vec{1},\vec{1})}}\nonumber\\
	&\quad-\frac{3}{8}\frac{ (m_{\vec{210}_{(\vec{15},\vec{1},\vec{3})}})^2}{m_\vec{1}-4 \,m_\vec{54}- m_{\vec{210}_{(\vec{15},\vec{1},\vec{1})}} }\,.\label{seesawminus2}
\end{align}
To have this extra contribution to be subleading is somewhat to be expected since this contribution breaks
$G_{3221}$, unlike the other ones that break a larger subgroup.
Interestingly, in this case, the $v_{\vec{210}_{(\vec{15},\vec{1},\vec{3})}}$ contribution can be of order TeV or less, even though $m_{\vec{210}_{(\vec{15},\vec{1},\vec{3})}}$ is larger than TeV.

In this case, any of the two bi-doublets can be present at low scale. This does not change much the scenarios above which involve the $(\vec{1},\vec{2},\vec{2})$ bi-doublet at low scale because in all these scenarios this bi-doublet is already alone at this scale.
However, for the $(\vec{15},\vec{2},\vec{2})$ case, a $v_{\vec{210}_{(\vec{15},\vec{1},\vec{3})}}$ seesaw contribution that shifts the mass of the bi-doublet only by $\sim\unit{TeV}$ or less shifts the masses of the color triplets in 
this $(\vec{15},\vec{2},\vec{2})$ by a much larger contribution since for the triplet the dependence on $v_{\vec{210}_{(\vec{15},\vec{1},\vec{3})}}$ is linear, not quadratic (see Tab.~\ref{tab:421and3211_45_120}):
\begin{align}
 m_{(\vec{3}, \vec{2}, -1, 4)} & =  m_\vec{1}+m_\vec{54}-m_{\vec{210}_{(\vec{15},\vec{1},\vec{1})}}-m_{\vec{210}_{(\vec{15},\vec{1},\vec{3})}} \,, \nonumber\\
 m_{(\vec{3}, \vec{2}, 1, 4)} & =  m_\vec{1}+m_\vec{54}-m_{\vec{210}_{(\vec{15},\vec{1},\vec{1})}}+m_{\vec{210}_{(\vec{15},\vec{1},\vec{3})}} \,, \nonumber\\
 m_{(\vec{8}, \vec{2}, 1, 0)} & =  m_\vec{1}+m_\vec{54}+m_{\vec{210}_{(\vec{15},\vec{1},\vec{1})}} \,.
\end{align}
Consequently, this does affect the scenarios above where the $(\vec{15},\vec{2},\vec{2})$ multiplet is at low scale, by removing the color triplets from the low-energy world, but not the color octet (since the latter does not receive any contribution from a $v_{\vec{210}_{(\vec{15},\vec{1},\vec{3})}}$).
Thus the "octet-triplet-singlet bi-doublet 60 scenario" of Sec.~\ref{sec:Octet-triplet-singlet_bi-doublet_60} above where all 60 states are degenerate at tree level above becomes an "\underline{octet-singlet  bi-doublet 32+4 scenario}"~\ref{sec:Octet-singlet_bi-doublet_32+4} (with 36 light states instead of 60), which, due to the $v_{\vec{210}_{(\vec{15},\vec{1},\vec{3})}}$ contribution, has different tree-level masses for the bi-doublet and for the color-octet bi-doublet. This scenario can lead to the observed relic density through bi-doublet annihilations similar to Sec.~\ref{sec:LR_bi-doublet_4}, but not through co-annihilations between the colored partners since the color singlet does not couple directly to the color octet through $X$ boson exchange.
Given the fact that the bi-doublet (annihilation) scenario requires a LR breaking scale below $\sim \unit[75]{TeV}$, the seesaw mass difference 
is, for example, of order $\unit[0.1]{TeV}$ or less if we take the denominator to be above $\unit[10^5]{TeV}$ (as we could expect, since at least some of the contributions in the denominator must be of this order to send the high-scale DM partners to this scale or above).
As for the  octet-triplet-singlet bi-doublet $32+24+4$ scenario above it also becomes an "\underline{octet-singlet bi-doublet 32+4  scenario}"~\ref{sec:Octet-singlet_bi-doublet_32+4}, with tree-level masses also from $v_{\vec{210}_{(\vec{15},\vec{1},\vec{1})}}$.

\underline{Leading $m_{\vec{210}_{(\vec{15},\vec{1},\vec{3})}}$}. 
If there are no other VEVs than $\langle (\vec{15},\vec{1},\vec{3})\rangle$ (breaking $SO(10)$ directly to $G_{3211}$ or going through additional subgroups via an extra $\vec{45}$ representation), we get an interesting situation where both the $(\vec{1},\vec{2},1,0)$ and the $(\vec{1},\vec{2},1,0)'$ are degenerate at low scale.
However, this is excluded by the fact that in this case $m_{\vec{210}_{(\vec{15},\vec{1},\vec{3})}}$, on the one hand, must be larger than $\unit[10^5]{TeV}$ to send the "high-scale DM partners" to at least this scale, and on the other hand, must be smaller than $\sim\unit[75]{TeV}$ so as not to induce a $W_R$ mass above this scale (as necessary for bi-doublets). Therefore, one needs another mass contribution of order $\unit[10^5]{TeV}$. Thus, for the $m_{\vec{210}_{(\vec{15},\vec{1},\vec{3})}}$ contribution not to be seesaw suppressed, i.e.~to be leading in the 
square root term appearing in the mass formula of the $(\vec{1},\vec{2},1,0)$ and of the $(\vec{1},\vec{2},1,0)'$ $G_{3211}$ multiplets (Tab.~\ref{tab:421and3211_45_120}), and also to have a candidate at low scale, one would need several tunings.

\subsection{Fermionic $\vec{210}$ DM candidates}
\label{sec:210}

There are many possibilities for DM if one turns to a representation as large as $\vec{210}$. This representation contains four DM candidates  (see Tab.~\ref{tab:DM_B})---one PS singlet $(\vec{1},\vec{1},\vec{1})$ and three candidates coming from the three different PS $\vec{15}$ representations: $(\vec{15},\vec{1},\vec{1})$, $(\vec{15},\vec{1},\vec{3})$, and $(\vec{15},\vec{3},\vec{1})$. Thus one $\vec{15}$ is a $SU(2)_L$ triplet, another one is a $SU(2)_R$ triplet, and the third one is a singlet of both of these $SU(2)$ groups. The components of the $\vec{210}$ can receive tree-level contributions from not less than six different scalar representations up to $\vec{210}$, 
\begin{align}
\hspace{-1.9ex}\vec{210}\otimes \vec{210} =& \vec{1}_S \oplus \vec{45}_S \oplus \vec{45}_A \oplus \vec{54}_S
\oplus \vec{210}_S \nonumber\\&\oplus \vec{210}_A\oplus \vec{770}_S+\dots ,
\end{align}
that is to say from a $\vec{54}$ VEV, from the two $\vec{45}$ VEVs, and from the three $\vec{210}$ VEVs (see Tabs.~\ref{tab:422and3221} and~\ref{tab:421and3211_210}).
The PS singlet DM candidate $(\vec{1},\vec{1},\vec{1})$ has interactions only with GUT-scale gauge bosons and thus cannot proceed through  freeze-out. It could be produced through freeze-in, but we will not look at this possibility.
The $(\vec{15},\vec{1},\vec{1})$ DM candidate was already present in the $\vec{45}$ DM representation discussed above. However, the  DM multiplets $\left(\vec{15},\vec{3},\vec{1}\right)$ and $\left(\vec{15},\vec{1},\vec{3}\right)$ are new.
Note that, unlike for the previous DM representations, basically all scalar representations that allow us to go down the GUT breaking path (Fig.~\ref{fig:so(10)_breaking_paths_PS}) contribute to the DM mass formula, including the scalar representation $\vec{45}$. This largely complicates the discussion, because in this case it is not sufficient to look just at the possible mass hierarchies that can emerge from the mass formula. In many cases, there is no breaking path leading to a given mass hierarchy obtained from the 
mass formula.

As for the previous representations, we will discuss the various cases according to the last group down to which one assumes that there are 
mass contributions, i.e.~down to the Pati--Salam group $G_{422}$,  the LR group $G_{3221}$, or  the $G_{421}$ group. Note that, contrary to all cases above, for which to pass or not through $G_{421}$ did not change anything to the
tree-level mass formula (because to pass through this group requires a ${\vec{45}_{(\vec{1},\vec{1},\vec{3})}}$ VEV that did not contribute to the masses), here this makes a difference, since the existence of a  ${\vec{45}_{(\vec{1},\vec{1},\vec{3})}}$ VEV would contribute. As for the very involved cases passing through $G_{3211}$ (Tab.~\ref{tab:421and3211_210}), we will not study them in detail, but just make a few general remarks.

\subsubsection{Tree-level masses down to $G_{422}$}

If the PS group is broken to the SM only by a scalar $\vec{126}$  representation, then the fermionic $\vec{210}$ components receive tree-level masses only from 
a ${\vec{54}_{(\vec{1},\vec{1},\vec{1})}}$ or a ${\vec{210}_{(\vec{1},\vec{1},\vec{1})}}$, 
which gives the mass pattern (see also table~\ref{tab:422and3221} in App.~\ref{app:tables})
\begin{align}
\begin{split}
	m_{\left(\vec{1},\vec{1},\vec{1}\right)} &= m_\vec{1} +12 m_\vec{54} \,,\\
	m_{\left(\vec{6},\vec{2},\vec{2}\right)} &= m_\vec{1} +7 m_\vec{54} \,,\\
	m_{\left(\vec{10},\vec{2},\vec{2}\right)} &= m_\vec{1} -3 m_\vec{54} \,,\\
	m_{\left(\vec{15},\vec{1},\vec{1}\right)} &= m_\vec{1} -8 m_\vec{54} \,,\\
	m_{\left(\vec{15},\vec{1},\vec{3}\right)} &= m_\vec{1} +2 m_\vec{54} + m_{{\vec{210}}_{(\vec{1},\vec{1},\vec{1})}}\,,\\
	m_{\left(\vec{15},\vec{3},\vec{1}\right)} &= m_\vec{1} +2 m_\vec{54} - m_{{\vec{210}}_{(\vec{1},\vec{1},\vec{1})}}\,.
\end{split}
\end{align}

\underline{Tree-level mass contribution only from ${\vec{210}_{(\vec{1},\vec{1},\vec{1})}}$}: A $D$-parity breaking ${\vec{210}_{(\vec{1},\vec{1},\vec{1})}}$  VEV alone allows at low scale a $\left(\vec{15},\vec{3},\vec{1}\right)$ or a $\left(\vec{15},\vec{1},\vec{3}\right)$, but not both.
If the $\left(\vec{15},\vec{3},\vec{1}\right)$ is alone at low scale, then the wino-like Majorana particle and its two colored partners split in mass radiatively (see App.~\ref{app:rge}). Given the fact that the mass splittings are determined by the PS scale, for a singlet mass of $\unit[2.7]{TeV}$, the triplet and octet masses are at least $\unit[4.6]{TeV}$ and $\unit[6.8]{TeV}$, respectively, due to the $\sim \unit[10^3]{TeV}$ lower bound on the PS scale. 
A larger PS scale increases the mass splitting further but also increases the lifetime of the color partners.
Imposing the BBN constraint of a lifetime below $\unit[0.1]{s}$ on the triplet partner, we find an upper bound on the  PS scale around $\unit[10^6]{TeV}$, which also implies upper bounds of the triplet and octet of $\unit[6.2]{TeV}$ and $\unit[11.9]{TeV}$, respectively. 
Due to the large mass splitting, co-annihilation is not efficient and the colored partners decay immediately after their freeze-out into the wino, which is still in equilibrium at that point. The DM abundance and (in)direct detection phenomenology is thus the same as for a wino and requires $m_\text{DM}\simeq\unit[2.7]{TeV}$. 
To distinguish this scenario from a standard wino, one has to observe the colored partners at a collider. Consequently one  has an \underline{octet-triplet-singlet $SU(2)_L$ triplet 45 scenario}~\ref{sec:Octet-triplet-singlet_SU(2)_L_triplet_45}. 
 
Having instead the $\left(\vec{15},\vec{1},\vec{3}\right)$ alone at a low scale does not work, simply because, here too, with only the ${\vec{210}_{(\vec{1},\vec{1},\vec{1})}}$ contributing to the masses, the PS group can only be broken directly to the SM via a $\vec{126}$, which gives with Eq.~\eqref{eq:gauge_boson_masses} $m_{W_R} \simeq m_X >\unit[10^3]{TeV}$. Since the radiative splitting of the PS $\vec{15}$ is too large to allow for colored co-annihilations, we have to obtain the correct relic abundance from the $SU(2)_R$ triplet interactions.
From Fig.~\ref{fig:right-handed_wino} we see immediately that this is not possible for an LR scale $m_{W_R} > \unit[10^3]{TeV}$, due to constraints on long-lived \emph{charged} DM partners.
  
\underline{Tree-level mass contribution only from $\vec{54}$}: The $\vec{54}$ VEV alone isolates at low scale either the $\left(\vec{1},\vec{1},\vec{1}\right)$ DM candidate or the $\left(\vec{15},\vec{1},\vec{1}\right)$ or both the $\left(\vec{15},\vec{3},\vec{1}\right)$ and the $\left(\vec{15},\vec{1},\vec{3}\right)$. Disregarding, as said above,  the singlet candidate, this leads to the $\left(\vec{15},\vec{1},\vec{1}\right)$ DM candidate as in the $\vec{45}$ case above (i.e.~a non-viable candidate assuming thermal freeze-out) or to a new case where
no less than 90 particles are tree-level degenerate at low scale, from both the $\left(\vec{15},\vec{3},\vec{1}\right)$ and the $\left(\vec{15},\vec{1},\vec{3}\right)$. This $D$-parity conserving scenario is, however, excluded as it involves the $\left(\vec{15},\vec{1},\vec{3}\right)$ at low scale, 
leading to the same problem as with this representation alone at low scale:
with only $\langle\vec{126}\rangle$ at our disposal to break PS $\to$ SM, the $W_R$ ends up to be too heavy to sufficiently dilute the  $\left(\vec{15},\vec{1},\vec{3}\right)$ abundance.

\underline{General case: $\vec{54}$ and ${\vec{210}_{(\vec{1},\vec{1},\vec{1})}}$}: Adding a smaller $\vec{54}$ VEV on top of a large ${\vec{210}_{(\vec{1},\vec{1},\vec{1})}}$ VEV does not change anything compared to the case with only 
large ${\vec{210}_{(\vec{1},\vec{1},\vec{1})}}$ VEV. 
Adding a smaller ${\vec{210}_{(\vec{1},\vec{1},\vec{1})}}$ VEV on top of a large $\vec{54}$ VEV instead allows the possibility
to have both $\left(\vec{15},\vec{3},\vec{1}\right)$ and $\left(\vec{15},\vec{1},\vec{3}\right)$ at the TeV scale, but with a tree-level mass difference from the ${\vec{210}_{(\vec{1},\vec{1},\vec{1})}}$ VEV, again excluded in the same way. 
Finally, the case where both ${\vec{210}_{(\vec{1},\vec{1},\vec{1})}}$ and $\vec{54}$ VEVs are large allows for any of the four DM candidates in the $\vec{210}$ to be alone at low scale,
with no difference with respect to the case in which any of these candidates is alone at low scale from only a $\vec{54}$ or only a 
${\vec{210}_{(\vec{1},\vec{1},\vec{1})}}$, thus leading again to the \underline{octet-triplet-singlet $SU(2)_L$ triplet 45 scenario}~\ref{sec:Octet-triplet-singlet_SU(2)_L_triplet_45} as the only possibility.

\subsubsection{Tree-level masses down to $G_{3221}$}

This case introduces two more tree-level mass contributions, from ${\vec{210}_{(\vec{15},\vec{1},\vec{1})}}$ and from
${\vec{45}_{(\vec{15},\vec{1},\vec{1})}}$, allowing one to have an intermediate LR symmetry step. The resulting mass pattern is given in table~\ref{tab:422and3221} in App.~\ref{app:tables}.
The four DM candidates are now in $\left(\vec{1},\vec{1},\vec{1}, 0\right)$, $\left(\vec{1},\vec{1},\vec{1}, 0\right)'$, $\left(\vec{1},\vec{1},\vec{3}, 0\right)$,
and $\left(\vec{1},\vec{3},\vec{1}, 0\right)$ representations of $G_{3221}$.
Here too, it is interesting to start with the case where there is only one of these VEVs.

\underline{Leading tree-level mass contribution from ${\vec{45}_{(\vec{15},\vec{1},\vec{1})}}$}: 
It is interesting to note that a $D$-parity breaking ${\vec{45}_{(\vec{15},\vec{1},\vec{1})}}$ VEV alone splits any of these four representations from any other $G_{3221}$ one in the $\vec{210}$ multiplets.  
In this case, this VEV cannot be small because the whole $\vec{210}$ multiplet would be present at low scale.
It must be larger than $\unit[10^3]{TeV}$ to have the PS breaking scale of at least this order, and, furthermore, larger than $\unit[10^5]{TeV}$ to send the high-scale DM partners to at least this scale.
Thus one needs this VEV to be large, and  one is left with any of these four multiplets alone at low scale.
In this case, the two singlet DM candidates again cannot be thermal, whereas the other two simply lead to the \underline{$SU(2)_L$ triplet 3 scenario}~\ref{sec:SU(2)_L_triplet_3} or \underline{$SU(2)_R$ triplet 3 scenario}~\ref{sec:SU(2)_R_triplet_3} already encountered for the $\vec{45}$ representation, see above. The latter is allowed here because the ${\vec{45}_{(\vec{15},\vec{1},\vec{1})}}$ breaks PS into LR, which can then be broken into the SM by a $\vec{126}$ at a lower scale. Actually, in the latter case, since we need the $W_R$ lighter than $\sim \unit[100]{TeV}$, and since we assume here sub-leading ${\vec{54}}$ and sub-leading ${\vec{210}_{(\vec{15}, \vec{1},\vec{1})}}$, the ${\vec{45}_{(\vec{15},\vec{1},\vec{1})}}$ must be of order the GUT scale in order to break $SO(10)$ (directly into the LR group). 

\underline{Leading tree-level mass contribution from ${\vec{210}_{(\vec{15},\vec{1},\vec{1})}}$}: Again, this ($D$-parity conserving) VEV must be larger than $\unit[10^3]{TeV}$ scale to have the PS breaking scale of at least this order, and above $\unit[10^5]{TeV}$ to send the high-scale DM partners to at least this scale. This  leads either to both DM triplet candidates, 
$\left(\vec{1},\vec{1},\vec{3}, 0\right)$
and $\left(\vec{1},\vec{3},\vec{1}, 0\right)$ to be degenerate at low scale with the PS singlet,
$\left(\vec{1},\vec{1},\vec{1}, 0\right)$, or to the $\left(\vec{1},\vec{1},\vec{1}, 0\right)'$ to be degenerate at low scale with the $\left(\vec{3},\vec{2},\vec{2}, -2\right)'$.
Both cases are excluded because the singlets do not have efficient (co-)annihilation channels.
More generally,
with a dominant ${\vec{210}_{(\vec{15},\vec{1},\vec{1})}}$ VEV contribution, adding any other contribution will send the singlet $\left(\vec{1},\vec{1},\vec{1}, 0\right)$ to a high scale and lead to the \underline{$SU(2)_L$ triplet 3 scenario} \ref{sec:SU(2)_L_triplet_3}, the \underline{$SU(2)_R$ triplet 3 scenario} \ref{sec:SU(2)_R_triplet_3}, the \underline{LR triplet 6 scenario} \ref{sec:LR_triplet_6} or the \underline{LR triplet 3+3 scenario} \ref{sec:LR_triplet_3+3} (see the $\vec{45}$ DM case above). In the last three cases, since we need the $W_R$ to be lighter than $\sim \unit[100]{TeV}$, and since we assume here subleading ${\vec{54}}$ and subleading ${\vec{210}_{(\vec{15}, \vec{1},\vec{1})}}$, the ${\vec{210}_{(\vec{15},\vec{1},\vec{1})}}$ must be of order the GUT scale so as to break $SO(10)$ (directly into the LR group). 

\underline{Subleading ${\vec{210}_{(\vec{15},\vec{1},\vec{1})}}$ and subleading ${\vec{45}_{(\vec{15},\vec{1},\vec{1})}}$}:
If both these VEVs are smaller than the $\vec{54}$ and/or ${\vec{210}_{(\vec{1},\vec{1},\vec{1})}}$ contributions (as could be expected, as the latter two break a larger group than the former two), their contribution to the mass of the DM candidates in the $\vec{210}$ is obviously subleading. We will not discuss the resulting possibilities in full details, but merely sketch them.

One possibility that emerges easily is to have only a wino at low scale, leading to the \underline{$SU(2)_L$ triplet 3 scenario} \ref{sec:SU(2)_L_triplet_3}. The other possibilities of low-energy spectra to get the observed relic density are again to invoke co-annihilations between the DM color singlet and its color triplet partners within the same PS multiplet and/or to have a $SU(2)_R$ triplet at low scale. 
Once again, a low-scale $SU(2)_R$ triplet requires the $W_R$ mass below $\sim\unit[100]{TeV}$.
And once again co-annihilation of the singlet with its color triplet partner requires the PS breaking scale to be at most of order a few $\unit[10^3]{TeV}$, so that the ${\vec{210}_{(\vec{15},\vec{1},\vec{1})}}$ and  ${\vec{45}_{(\vec{15},\vec{1},\vec{1})}}$ VEVs must be at most at this scale.  
Furthermore it requires
the mass splitting between the color triplet and singlet to be below TeV so that the mass contributions of these two scalar fields must be at most of this scale (implying Yukawa couplings below $\sim 10^{-3}$). This means that $SO(10)$ must be broken into PS by either the $\vec{54}$ or the  ${\vec{210}_{(\vec{1},\vec{1},\vec{1})}}$, so that at least one of these VEVs must be of order of the GUT scale. Putting all that together, it turns out that it leaves enough freedom to get the following viable low-energy spectra. For the $\left(\vec{15},\vec{1},\vec{1}\right)_\text{PS}$ 
DM candidate there is only one possibility,  the \underline{octet-triplet-singlet 8+6+1 scenario}~\ref{sec:Octet-triplet-singlet_8+6+1} (but not $14+1$ spectrum because it cannot give a viable seesaw contribution).
For the $\left(\vec{15},\vec{3},\vec{1}\right)$ and/or $\left(\vec{15},\vec{1},\vec{3}\right)$, many scenarios are possible:
beside the \underline{$SU(2)_L$ triplet 3 scenario} \ref{sec:SU(2)_L_triplet_3} scenario, one has the \underline{$SU(2)_R$ triplet 3 scenario} \ref{sec:SU(2)_R_triplet_3}, the \underline{LR triplet 6 scenario} \ref{sec:LR_triplet_6}, the \underline{LR triplet 3+3 scenario}.
More complicated scenarios are also possible, such as the ``\underline{octet-triplet-singlet $SU(2)_L$ triplet 24+18+3 scenario}'', where the entire $\left(\vec{15},\vec{3},\vec{1}\right)$ is present at low scale, and similarly the ``\underline{octet-triplet-singlet $SU(2)_R$ triplet 24+18+3 scenario}''.
Combinations of the last two scenarios are also possible, leading to the ``\underline{octet-triplet-singlet $SU(2)_{L+R}$ triplet $n$ scenarios}'' with $n$ equal to $48+36+6$ or $3+3+18+18+24+24$ or $3+3+42+42$ (with in the latter case degenerate color triplet and octet). 
We will just mention these complicated possibilities here and will not put them in our list of Sec.~\ref{sec:lowscale}.

\subsubsection{Tree-level masses down to $G_{421}$}

As already mentioned above, whether the breaking path has a $G_{421}$ step matters for the $\vec{210}$ DM representation, because the  ${\vec{45}_{(\vec{1},\vec{1},\vec{3})}}$ VEV, which is necessary for this intermediate step, contributes to the masses at tree level.
Here the four DM candidates are in  $\left(\vec{1},\vec{1}, 0\right)$, $\left(\vec{15},\vec{1}, 0\right)'$, $\left(\vec{15},\vec{1}, 0\right)$,
and $\left(\vec{15},\vec{3}, 0\right)$ representations of $G_{421}$, with masses given in Tab.~\ref{tab:421and3211_210}. 
The first candidate comes from the $(\vec{1},\vec{1},\vec{1})$ PS singlet and is again not interesting in a thermal setup.
The next two 
are a mixture of the $(\vec{15},\vec{1},\vec{1})$ and $(\vec{15},\vec{1},\vec{3})$ of PS. The mixing is induced
by the ${\vec{45}_{(\vec{1},\vec{1},\vec{3})}}$ VEV which breaks PS [or directly $SO(10)$] to $G_{421}$. These two candidates are not viable because one has only radiative splittings between the color singlet and triplet, which are too large for co-annihilation to be effective.
The last candidate comes out of the $(\vec{15},\vec{3},\vec{1})$ of PS. For this candidate there is no mass contribution from the ${\vec{45}_{(\vec{1},\vec{1},\vec{3})}}$ so that we get back to the \underline{octet-triplet-singlet $SU(2)_L$ triplet 45 scenario}~\ref{sec:Octet-triplet-singlet_SU(2)_L_triplet_45} we got with just a ${\vec{210}_{(\vec{1},\vec{1},\vec{1})}}$ VEV (and possibly also a $\vec{54}$ VEV).

\subsubsection{Tree-level masses down to $G_{3211}$}

We will not study in details this complex case, which can bring two new scalar field mass contributions, from ${\vec{210}_{(\vec{15},\vec{1},\vec{3})}}$ and ${\vec{45}_{(\vec{1},\vec{1},\vec{3})}}$, see Tab.~\ref{tab:421and3211_210}.
We will just make two remarks. We first note that, obviously, when these new contributions are subleading with respect to other ones, these VEVs can split the various components of an LR multiplet (and hence of a PS multiplet too) by contributions of order TeV, or less. This can split, in particular, any low-scale $\vec{15}$ of $SU(4)_c$  into $8+3+1$, allowing one to have the right amount of splitting for getting the relic density through co-annihilations.
Second, if $SO(10)$ is directly broken into $G_{3211}$, so that there are contributions only from these ${\vec{210}_{(\vec{15},\vec{1},\vec{3})}}$ and ${\vec{45}_{(\vec{1},\vec{1},\vec{3})}}$ VEVs,  we end up with scenarios in which the \emph{singlet} DM candidates have no co-annihilation partner and are thus over-abundant, while the \emph{wino} DM candidate comes with a myriad of unwelcome long-lived low-scale partners.

\subsection{Fermionic $\vec{210}$' DM candidates}
\label{sec:210prime}

The $\vec{210}'$ only couples to a $\vec{54}$ VEV, so the discussion is very simple:
\begin{align}
\begin{split}
 m_{(\vec{1}, \vec{2}, \vec{2})} & =  m_\vec{1}+12 m_\vec{54} \,, \\
 m_{(\vec{1}, \vec{4}, \vec{4})} & =  m_\vec{1}+27 m_\vec{54} \,, \\
 m_{(\vec{6}, \vec{1}, \vec{1})} & =  m_\vec{1}+2 m_\vec{54}  \,,\\
 m_{(\vec{6}, \vec{3}, \vec{3})} & =  m_\vec{1}+12 m_\vec{54}  \,,\\
 m_{(\vec{20'}, \vec{2}, \vec{2})} & =  m_\vec{1}-3 m_\vec{54}  \,,\\
 m_{(\vec{50}, \vec{1}, \vec{1})} & =  m_\vec{1}-18 m_\vec{54}  \,.
\end{split}
\end{align} 
There are two potential DM candidates, one  bi-doublet $\left(\vec{1},\vec{2},\vec{2}\right)$ and one bi-quadruplet $\left(\vec{1},\vec{4},\vec{4}\right)$. The bi-doublet has a similar phenomenology to the $\vec{10}$ but this possibility is here excluded because the bi-doublet is accidentally degenerate with a $\left(\vec{6},\vec{3},\vec{3}\right)$, so that the latter cannot be a ``high scale partner,'' as required.

The other DM candidate is the bi-quadruplet $\left(\vec{1},\vec{4},\vec{4}\right)$, leading to the "\underline{LR bi-quadruplet 16 scenario}"~\ref{sec:LR_bi-quadruplet_16}.

\subsection{Fermionic $\vec{126}\oplus\overline{\vec{126}}$ DM candidates}
\label{sec:126}

These candidates bring the largest number of new particles yet because we have to include two copies of the complex $\vec{126}$ in order to make all new particles massive. Equivalently, we can discuss the vector-like representation $\vec{126}\oplus\overline{\vec{126}}$. As can be seen from  Tables~\ref{tab:126_422}, \ref{tab:126_421}, \ref{tab:126_3221}, and~\ref{tab:126_3211} the expressions for the masses are unwieldy. From the many possibilities, we will only pick out one interesting scenario that brings new qualitative features compared to the previous candidates. Let us assume a low-scale $G_{3211}$ and pick VEVs so that $\Psi \sim (\vec{1},\vec{1},2,-6)$ is the lightest (Dirac) fermion inside the $\vec{126}\oplus\overline{\vec{126}}$ (see Tab.~\ref{tab:126_3211}). This requires scalars in the representations $\vec{210}$ or $\vec{45}$. The fermion originates from $(\vec{10},\vec{1},\vec{3})\to (\vec{10},\vec{1},2)\to (\vec{1},\vec{1},2,-6)$ via PS or $(\vec{1},\vec{1},\vec{3},-6)\to (\vec{1},\vec{1},2,-6)$ under LR.

According to Eq.~\eqref{eq:hypercharge} this fermion has no hypercharge and is hence an SM singlet, but still has couplings to the low-scale $Z'$ of $G_{3211}$. These interactions can then be used to achieve the correct DM abundance~\cite{Arcadi:2017atc}. A light $Z'$ from $G_{3211}$ has been proposed long ago as an interesting and well-motivated benchmark~\cite{Deshpande:1979df,Robinett:1981yz}. The $Z'$ couplings are orthogonal to the hypercharge combination $Y=T_3^R + (B-L)/2$ but still depend on the $SU(2)_R$ gauge coupling $g_R$. Taking the LR-motivated case $g_R = g_L =e/\sin \theta_W$ gives the explicit expression for the $Z'$ couplings~\cite{Deshpande:1979df,Carena:2004xs,Langacker:2008yv}
\begin{align}
\L \supset g_{Z'} Z^\prime_\mu \left[ (1-\tan^2\theta_W) j_{3,R}^\mu - \tfrac12 \tan^2\theta_W j_{B-L}^\mu \right]
\end{align}
with the usual $B-L$ current $j_{B-L}$, diagonal $SU(2)_R$ current $j_{3,R}^\mu=\sum_{f} \overline{f}\gamma^\mu T_3^R f$, and the coupling strength is fixed to $g_{Z'} = e/(\tan\theta_W \sqrt{\cos (2\theta_W)})\simeq 0.8$. The couplings of the SM fermions are not particularly illuminating, but our DM candidate couples simply with $g_{Z'}\overline{\Psi}\slashed{Z}^\prime \Psi$, i.e.~the $\tan^2\theta_W$ terms cancel in the coupling.
Here we have neglected any $Z$--$Z'$ mixing for simplicity, which is in any case required to be small to satisfy direct detection constraints \cite{Langacker:2008yv}.

Using for example the formulae from Ref.~\cite{Alves:2015mua} it is easy to calculate the relevant annihilation cross sections for freeze-out. Summing over all SM fermions (but not right-handed neutrinos, which we assume to be heavier than the DM) we find the non-relativistic annihilation cross section $\sigma v (\overline{\Psi}\Psi \to \overline{f}f)$ via $s$-channel $Z'$ to be
\begin{align}
\sigma v  \simeq {g_{Z'}^4 m_\Psi^2\over 8 \pi} {(21-48 \tan^2\theta_W+40 \tan^4\theta_W)\over \left(4 m_\Psi^2-m_{Z'}^2\right)^2 +m_{Z'}^2 \Gamma_{Z'}^2 } \,,
\end{align}
neglecting the SM fermion masses.\footnote{If the right-handed neutrinos are light as well, we have an additional contribution $(3g_{Z'}^4 m_\Psi^2/ 8 \pi)/\left[\left(4 m_\Psi^2-m_{Z'}^2\right)^2 +m_{Z'}^2 \Gamma_{Z'}^2 \right]$.} 
For DM masses around the $Z'$ resonance $m_\Psi \sim m_{Z'}/2$ we can easily obtain the correct relic abundance even for the multi-TeV masses necessary to evade experimental constraints~\cite{Alves:2015mua}, as shown already in Fig.~\ref{fig:126plot} in Sec.~\ref{sec:ZprimeDM}.

Note that $g_R\neq g_L$ as well as RGE running can change the $Z'$ couplings and modify the DM phenomenology. The qualitative behavior of Fig.~\ref{fig:126plot} should remain the same, and one could at best hope to lower the $Z'$ mass bounds to open up parameter space away from the resonance. This happens also if additional light, unstable particles are present,  so that the $Z'$ can decay into them, which simultaneously weakens the dilepton bounds and increases the DM annihilation cross section.

An alternative path to this DM scenario is to break $SO(10)$ via $SU(5)\times U(1)_\chi\to G_\text{SM}\times U(1)_\chi$, leaving the $U(1)_\chi$ at low scales~\cite{Masiero:1980dd,Ma:2018uss,Ma:2018zuj}. Since $\Psi\sim (\vec{1},-10)$ under $SU(5)\times U(1)_\chi$, it is clear that it is again an SM singlet with $Z'_\chi$ interactions. This scenario is actually part of the previous one, changing simply $\sin^2\theta_W \to 3/8$~\cite{Langacker:2008yv} in the couplings. In the end one finds almost the same phenomenology as in Fig.~\ref{fig:126plot}, in particular still strong LHC constraints of order $\unit[4.2]{TeV}$ on the $Z'_\chi$~\citep{Sirunyan:2018exx}.

This concludes our discussion of fermionic dark matter candidates.

\subsection{Scalar DM candidates}

Scalar $SO(10)$ multiplets can also be stabilized by matter parity and thus form WIMPs. In addition to the gauge interactions, one, however, always has Higgs-portal interactions with all other $SO(10)$ scalars, which make a thorough discussion involved. Still, one finds some well known candidates, such as the inert-doublet-like (or sneutrino-like) $(\vec{1},\vec{2},1/2)\subset G_\text{SM}$~\cite{Ma:2006km,Barbieri:2006dq,LopezHonorez:2006gr,Hambye:2009pw}. These doublets can be found in the scalar $\vec{16}$~\cite{Kadastik:2009dj,Kadastik:2009cu} or $\vec{144}$, see Tab.~\ref{tab:ListOfEWMultiplets}.
Similar to the bi-doublet fermion case, the neutral complex scalar here needs to split into two real scalars with a mass difference above $\unit[200]{keV}$ in order to survive direct detection bounds. Focusing on gauge interactions, a discussion of this candidate under a low-scale LR group can be found in Ref.~\cite{Garcia-Cely:2015quu}.

Other interesting scalar DM candidates within the $\vec{16}$~\cite{Kadastik:2009dj,Kadastik:2009cu} or $\vec{144}$ are the total singlets $(\vec{1},\vec{1},0)\subset G_\text{SM}$ that come from the LR representation $(\vec{1},\vec{1},\vec{2},3)$. Much as the case discussed in Sec.~\ref{sec:ZprimeDM}, these complex scalars do not have any SM gauge interactions, but still couple to the $Z'$ of $G_{3211}$. Once again we can easily obtain the correct DM abundance around the $Z'$ resonance without violating constraints from direct detection or LHC dilepton searches, qualitatively similar to Fig.~\ref{fig:126plot}. The additional Higgs-portal couplings lead to a much wider parameter space, beyond the scope of this paper.

Finally, the $\vec{144}$ also contains a complex electroweak triplet $(\vec{1},\vec{3},0)\subset G_\text{SM}$, part of the LR multiplet $(\vec{1},\vec{3},\vec{2},3)$ (Tab.~\ref{tab:DM_B}). Using the SM gauge interactions, this can lead to the correct relic abundance for multi-TeV masses \cite{Cirelli:2005uq,Hambye:2009pw}. Similar to the previous paragraph, this candidate has no hypercharge but still $Z'$ interactions with a low-scale $G_{3211}$ group, which can be used to push the mass further up around the $Z'$ resonance. This should lead to interesting indirect detection signatures due to Sommerfeld-enhanced photon fluxes.

\section{Accidental DM stability}
\label{sec:indirect_way}

So far we have looked at scenarios where the DM stability is directly explained by the gauge symmetries of the model, leading to absolute DM stability. 
There are also scenarios where DM stability is only indirectly induced by the gauge symmetries, i.e.~where DM stability results accidentally from the gauge symmetries and the particle content of the model, such as baryon number conservation for the proton. In our setup this could occur when matter parity $P_M$ is spontaneously broken by scalar VEVs $\langle\vec{16}\rangle_{\slashed{P}_M}$ or $\langle\vec{144}\rangle_{\slashed{P}_M}$. These VEVs, although not mandatory, can be useful in the GUT symmetry breaking, so it is worthwhile to consider their effect on DM multiplets.  With matter parity broken, all of our DM multiplets can in principle decay into the SM fermions residing in their $\vec{16}_\text{SM}$, but whether this actually happens depends on the particle content of the model. 
If matter parity is broken spontaneously, most of our fermionic DM candidates $\vec{R}_\text{DM}$ are allowed a destabilizing Yukawa interaction, $\langle\vec{16}\rangle_{\slashed{P}_M}\vec{16}_\text{SM}\vec{R}_\text{DM}$ or $\langle\vec{144}\rangle_{\slashed{P}_M}\vec{16}_\text{SM}\vec{R}_\text{DM}$ on account of the branching rules
\begin{align}
\begin{split}
\vec{16}\otimes \overline{\vec{16}} &= \vec{1}\oplus\vec{45}\oplus\vec{210} \,,\\
\vec{16}\otimes \vec{16} &= \vec{10}\oplus\vec{120}\oplus\overline{\vec{126}} \,,\\
\vec{16}\otimes \overline{\vec{144}} &= \vec{45}\oplus\vec{54}\oplus\vec{210}\oplus\vec{945}\oplus\overline{\vec{1050}} \,,\\
\vec{16}\otimes \vec{144} &= \vec{10}\oplus\vec{120}\oplus\vec{126}\oplus\vec{320}\oplus\vec{1728} \,,
\end{split}
\end{align}
which then typically leads to fast DM decay, unless the corresponding Yukawa couplings are minuscule.

Exceptions are the DM fermion multiplets $\vec{54}$ and $\vec{210}'$, which have no Yukawa couplings with the $\vec{16}_\text{SM}$ and a matter-parity breaking $\langle\vec{16}\rangle_{\slashed{P}_M}$.
Thus they can be stable \emph{accidentally}, provided there are no other states, low scale or high scale, which could destabilize them. The presence of a $\vec{144}$  scalar representation VEV $\langle\vec{144}\rangle_{\slashed{P}_M}$ would for example destabilize the $\vec{54}$ candidate through a Yukawa coupling $\langle\vec{144}\rangle_{\slashed{P}_M}\vec{16}_\text{SM}\vec{54}_\text{DM}$. The $\vec{210}'_\text{DM}$, on the other hand, would still remain stable even in the presence of both $\langle\vec{16}\rangle_{\slashed{P}_M}$ and $\langle\vec{144}\rangle_{\slashed{P}_M}$, at least at the renormalizable level. At the non-renormalizable level one can write down operators such as $\overline{\vec{144}}_{\slashed{P}_M}\vec{16}_{\slashed{P}_M}\vec{16}_{\slashed{P}_M} \vec{16}_\text{SM}\vec{210}'_\text{DM}$ that would lead to $\vec{210}'_\text{DM}$ decay.
Such destabilizing higher-dimensional operators could be induced by other (heavy) scalars or fermions.
It is well known that a dimension-five operator, generically leads to a far too fast DM decay (as for instance, here, an operator $\vec{16}_{\slashed{P}_M} \vec{10}_\text{SM} \vec{16}_\text{SM} \vec{54}_\text{DM} $). A dimension-six operator also leads to fast decay, unless suppressed by a scale of the order of the GUT scale or more. Thus, as is well appreciated, the accidental DM stability highly depends also on the high-energy content of the model.
If stable, the resulting DM phenomenology of the $\vec{54}$ or $\vec{210}'$ does of course not change; only the $SO(10)$ breaking path becomes more flexible.

\section{Discussion and summary}
\label{sec:summary}

Grand unified theories are well-motivated extensions of the SM. They shed light on its fundamental interaction structure and on the
quantum numbers of the SM particles under these fundamental interactions. They also provide hints toward the understanding of the flavor structure of the SM. Furthermore, by incorporating the abelian hypercharge group into a larger non-abelian group, they can solve the Landau pole problem.
Grand unified theories do not automatically lead to dark matter candidates.  However, GUTs based on $SO(10)$  contain the discrete matter-parity symmetry $Z_2^{3(B-L)}$ as a subgroup, thus offering a particularly natural and simple explanation for the existence of DM: if the $SO(10)$ symmetry breaking path preserves matter parity, the lightest parity-even fermion or parity-odd scalar is automatically stable. In supersymmetric theories, this allows us to understand the origin of $R$~parity, which predicts that the lightest supersymmetric particle could be a good DM candidate. In non-supersymmetric theories, the same mechanism can be invoked to stabilize a DM particle as the lightest component of a given $SO(10)$ multiplet. 
This multiplet has gauge interactions and is hence a prime example for a WIMP. Here, we have studied in detail
how $SO(10)$ multiplets can be broken apart in such a way that the lightest component is a viable DM candidate, leading to a systematic classification of DM scenarios.

As is well known, several issues, such as gauge unification or the fermion mass structure, become extremely complex as soon as one departs from minimal grand unified schemes. This stems from the fact that they largely depend on the masses and interactions of the components of basically any extra multiplet that could lie between the electroweak and the GUT scale, depending on the subgroup symmetry breaking scales, Yukawa couplings, mixing of particles, etc. Instead, we have shown that the various DM scenarios resulting from matter-parity stabilization are less dependent on all of these ingredients. As we discuss at length, the possible viable DM scenarios depend mostly (but crucially) on the path along which $SO(10)$ is broken, and in particular on the values of the various subgroup symmetry breaking scales. These breaking scales affect the masses of the various $SO(10)$ gauge bosons which in turn determine the DM relic density via thermal (co-)annihilation.
They also affect the masses of the heavy partners in the DM multiplet, which must be sufficiently heavy so as to decay fast enough,
as well as the tree-level and radiative mass splittings between the  various components of the DM multiplet. 
In other words, each viable DM candidate requires a certain set of scalar VEVs and $SO(10)$-breaking paths, which thus impose conditions on the full model.

Apart from the dependence on the subgroup symmetry breaking scales, the DM candidates have only limited dependence on UV physics, such as on the masses and interactions of other heavy states. 
This is related to the fact that the DM phenomenology we consider (relic density, direct and indirect detection, etc.) is infrared dominated, modulo the presence of  heavier states, which, as mediators,  could play a role for co-annihilation processes. Here, we have assumed that, apart from possible $SO(10)$ gauge bosons (which can lead to successful co-annihilation if they are lighter than  $\sim  \text{few } \unit[10^{3}]{TeV}$), there are no such states, or that their interactions are small compared to unity (e.g.~small Yukawa couplings of fermion DM to scalar states). For fermionic DM scenarios, the relevant interactions are essentially  gauge interactions, whose couplings 
are basically known. This is unlike the scalar DM scenarios, which also involve unknown quartic scalar couplings. Thanks to all these facts, and adopting a list of simple criteria (see Sec.~\ref{sec:constraints}), such as to assume a single DM multiplet, we could establish the above systematic determination of DM scenarios.

A scenario which shows up in many cases from this systematic study is a wino-like DM candidate. This is not surprising, as this is the only scenario that does not require the existence of any subgroup broken at an intermediate scale. The presence of intermediate subgroups, in particular the Pati--Salam and/or LR group at  relatively low scales, allows, however, a large variety of other possible DM candidates, leading in particular to many different scenarios with low-scale partners. These partners, be they charged or colored, could be discovered much more easily than the DM particle. In particular, due to the fact that DM candidates have a relic density set by gauge interactions, the masses of the DM particles and partners are predicted to lie around the $\sim$~TeV scale, leading to potentially many discovery opportunities at the HL-LHC or at a future $\unit[100]{TeV}$ collider.
Simple examples  include triplets under $SU(2)_R$, which bring charged DM partners, or an adjoint representation of Pati--Salam's $SU(4)$, which brings squark and gluinolike colored partners, or LR a bi-doublet/triplet system.

We should emphasize that, even if the number of possible scenarios is relatively large, the ways they can account for the observed relic density are very limited:  basically, there are the wino way, the $SU(2)_R$ triplet, the co-annihilation with color triplets and the bi-doublet or bi-triplet ways. Combinations of these basic relic density mechanisms are also possible in some cases, allowing for example the wino multiplet to have a mass different from the $\unit[2.7]{TeV}$ it must have in the pure wino scenario. In this sense, as these relic density mechanisms require specific partners and mass scales, the $SO(10)$ DM setup is quite predictive (at least for scenarios driven by gauge interactions, i.e.~fermion scenarios and possibly also some scalar DM scenarios). 
Many of these scenarios are already constrained experimentally by collider and (in)direct detection data.

Another clear trend that this analysis reveals is that the $SO(10)$ framework does not favor SM singlet DM! Indeed, in all (fermionic) cases, with one exception, DM belongs to a SM multiplet, so that it has low-scale partners, and a related potentially rich phenomenology.
The exception shows up from the only complex representation under study, where DM belongs to $\vec{126}\oplus\overline{\vec{126}}$ representations and whose relic density is determined from annihilation driven by a low-scale $Z'$ gauge boson. 
While this scenario can emerge from a Pati--Salam symmetry breaking path, it is the only viable fermion scenario showing up through a Georgi--Glashow path, given our assumptions (together with a simple wino scenario with all beyond-the-SM gauge bosons at the GUT scale). 

Note that, if all possible $SO(10)$-breaking scalar VEVs are considered, the masses of the DM partners can in most cases be chosen to be independent of the DM mass,  but predictive relations between masses of the low scale DM multiplet components arise in cases with a limited number of scalar VEVs or, if there are no VEVs that split these components, in the case of pure radiative mass splitting.
Another conclusion is that all the possible scenarios require a fine-tuning qualitatively similar to the doublet--triplet splitting problem, although a quantitatively weaker one. Even if absent at tree level, tuning is necessary at loop level because the components of a multiplet do not have individual protective chiral symmetries.
Ultimately, we stress that a more refined analysis would obviously become mandatory if a hint for $SO(10)$ were to be observed, be it via proton decay or via a WIMP through (in)direct detection signals in upcoming experiments such as Hyper-Kamiokande, LZ or CTA.
Beyond the scan of possible scenarios performed in this work, such an analysis should in particular combine DM phenomenology with the request of successful unification and the SM fermion mass constraints.

\section*{Acknowledgements}

TH thanks G.~D'Ambrosio and G.~Mangano at Universita Federico~II in Naples for hospitality during an early stage of this project,
and the Galileo Galilei Institute for Theoretical Physics and INFN for support and hospitality during the completion of this work. 
JH would like to thank James McKay for useful discussions on radiative mass splittings and Renato Fonseca for his invaluable help with \texttt{Susyno}.
MT would like to thank Michele Redi for useful discussions, the Galileo Galilei Institute for Theoretical Physics for hospitality and the Laboratoire de Physique Th\'eorique at Orsay for support and hospitality. 
This work is supported by the F.R.S.-FNRS, the IISN convention 4.4503.15, the "Probing dark matter with neutrinos" ULB-ARC grant, and the Excellence of Science (EoS) convention 30820817. 
JH is a postdoctoral researcher of the F.R.S.-FNRS and furthermore supported, in part, by the National Science Foundation under Grant No.~PHY-1620638 and by a Feodor Lynen Research Fellowship of the Alexander von Humboldt Foundation. 

\appendix

\section{Renormalization group and mass splitting}
\label{app:rge}

For the convenience of the reader, we collect formulae used in the renormalization group running. For each gauge group factor $G$, the associated coupling constant~$g$ evolves according to the differential equation
\begin{align}
 \frac{\dd g(\mu)}{\dd \log (\mu)} = \frac{[g(\mu)]^3}{16 \pi^2} b(\mu)\,,\label{eq:beta_function}
\end{align}
$\mu$ being the renormalization scale. We restrict ourselves to one-loop calculations, in which case each gauge coupling runs independently, $b (\mu)=b$ being just a number that depends on the quantum numbers of the particles with masses below $\mu$, which are fixed once we specify the group $G$, the representation $F$ of the chiral fermions and the representation $S$ of the real scalars~\cite{Machacek:1983tz,Jones:1974pg}:
\begin{align}
 b = -\frac{11}{3} \, S_2 (\text{adj})  + \frac{4}{6} S_2 (F) + \frac{1}{6} S_2 (S)\,.
\end{align}
\emph{Dirac} fermions and \emph{complex} scalars simply count twice as much as their chiral/real counterparts. 
For $G=SU(N)$, the Dynkin index of the adjoint (fundamental) is $S_2 (\text{adj}) = N$ ($S_2 (N) = 1/2$). More generally, 
\begin{align}
	 S_2 (R) \delta^{a b} &= \tr \left[ T^a (R) T^b (R)\right],\\
	 S_2 (R) &= \frac{\dim (R)}{\dim (\text{adj})} C_2 (R)\,,
\end{align}
with the representation matrices $T^a (R)$ and quadratic Casimir operator $C_2 (R)$. A list of these group coefficients is given in Tab.~\ref{tab:casimirs} and can also be obtained efficiently from \texttt{Susyno}~\cite{Fonseca:2011sy}.

Eq.~\eqref{eq:beta_function} can easily be solved analytically, introducing the variable $\alpha\equiv g^2/(4\pi)$:
\begin{align}
 \frac{1}{\alpha (\mu_1)} = \frac{1}{\alpha (\mu_2)} + \frac{b}{2\pi} \log \left(\frac{\mu_2}{\mu_1}\right).
\end{align}

\begin{table}[t]
	\begin{tabular}{ll}
		Group & $(R, C_2 (R), S_2 (R))$ \\
		\hline
		$U(1)$ & $(Q,Q^2,Q^2)$\\	
		$SU(2)$ & $(\vec{2},\tfrac{3}{4},\tfrac{1}{2}), (\vec{3},2,2), (\vec{4},\tfrac{15}{4},5), (\vec{5},6,10), (\vec{6},\tfrac{35}{4},\tfrac{35}{2})$\\		
		$SU(3)$ & $(\vec{3},\tfrac{4}{3},\tfrac{1}{2}), (\vec{6},\tfrac{10}{3},\tfrac{5}{2}), (\vec{8},3,3), (\vec{10},6,\tfrac{15}{2}), (\vec{15},\tfrac{16}{3},10)$\\
		$SU(4)$ & $(\vec{4}, \tfrac{15}{8}, \tfrac{1}{2}), (\vec{6}, \tfrac{5}{2}, 1), (\vec{10}, \tfrac{9}{2},3), (\vec{15}, 4, 4), (\vec{20}, \tfrac{39}{8}, \tfrac{13}{2})$\\
		$SU(5)$ & $(\vec{5}, \tfrac{12}{5}, \tfrac{1}{2}), (\vec{10}, \tfrac{18}{5}, \tfrac{3}{2}), (\vec{15}, \tfrac{28}{5}, \tfrac{7}{2}), (\vec{24}, 5, 5), (\vec{35}, \tfrac{48}{5}, 14)$\\
		\hline
	\end{tabular}
	\caption{Quadratic Casimir and Dynkin coefficients for some groups. Note that the adjoint of $U(1)$ has $Q=0$.
	\label{tab:casimirs}
	}
\end{table}

At one-loop order, the $\overline{\text{MS} }$ mass $M$ of a Majorana or Dirac fermion in representation $R$ runs according to
\begin{align}
	\mu \frac{\dd M(\mu)}{\dd \mu} = -\frac{3}{2\pi} C_2 (R) \alpha (\mu) M (\mu)\,.
\end{align}
Using the one-loop result for $\alpha (\mu)$, we can solve this to
\begin{align}
 M(\mu) = M(\mu_0) \left[\frac{\alpha (\mu_0)}{\alpha (\mu)} \right]^{\frac{3 C_2 (R)}{b}} .
 \label{eq:MSbar_running_mass}
\end{align}
The scale-independent pole mass $m$ can then be calculated as
\begin{align}
m = M(m) \left(1 +  \frac{C_2 (R) \alpha (m)}{\pi}\right),
\label{eq:pole_vs_MSbar}
\end{align}
evaluated at $\mu = m \simeq M(m)$ to suppress large logs. The connection between the pole mass $m$ and the $\overline{\text{MS} }$ mass $M$ at some high scale $\mu$ can then be obtained by combining Eq.~\eqref{eq:MSbar_running_mass} and Eq.~\eqref{eq:pole_vs_MSbar},
\begin{align}
\frac{m}{M(\mu)} = \left(1 +   \frac{C_2 (R) \alpha (m)}{\pi}\right) \left[\frac{\alpha (\mu)}{\alpha (m)} \right]^{\frac{3 C_2 (R)}{b}},
\label{eq:pole_vs_GUT}
\end{align}
which can also be written as a Taylor series in $\alpha (m) \log \left(\mu/m\right)$ to illustrate that we are properly resumming the leading log terms~\cite{Manohar:2018aog}. If the fermion is charged under several gauge groups,  the right-hand side of Eq.~\eqref{eq:pole_vs_GUT} becomes a product over all group factors.

The extension of this approach to three loops with the most general gauge group and particle content can be found in Ref.~\cite{Martin:2006ub}.

\begin{figure}[t]
\includegraphics[width=0.48\textwidth]{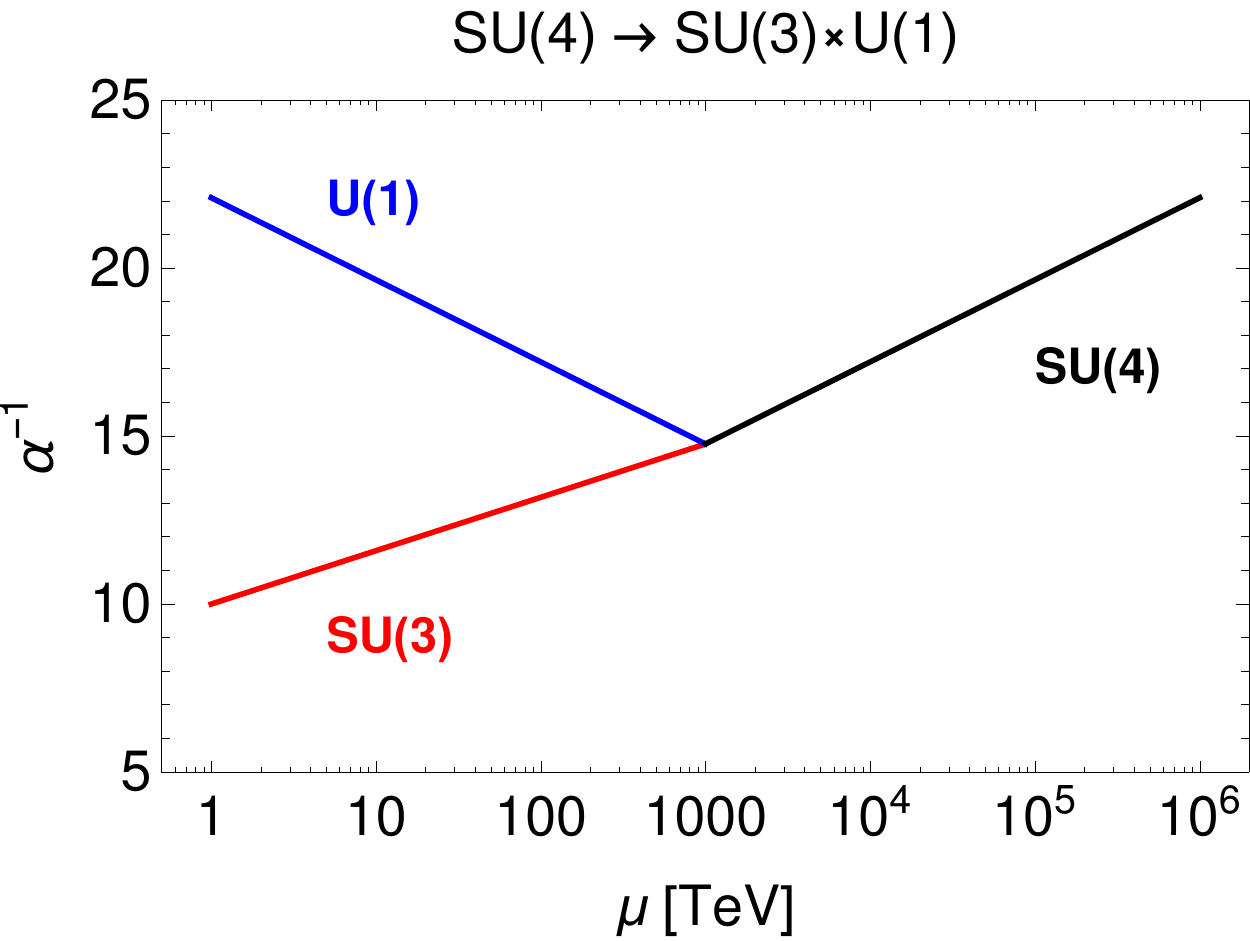}\\
\hspace{1ex}
\includegraphics[width=0.48\textwidth]{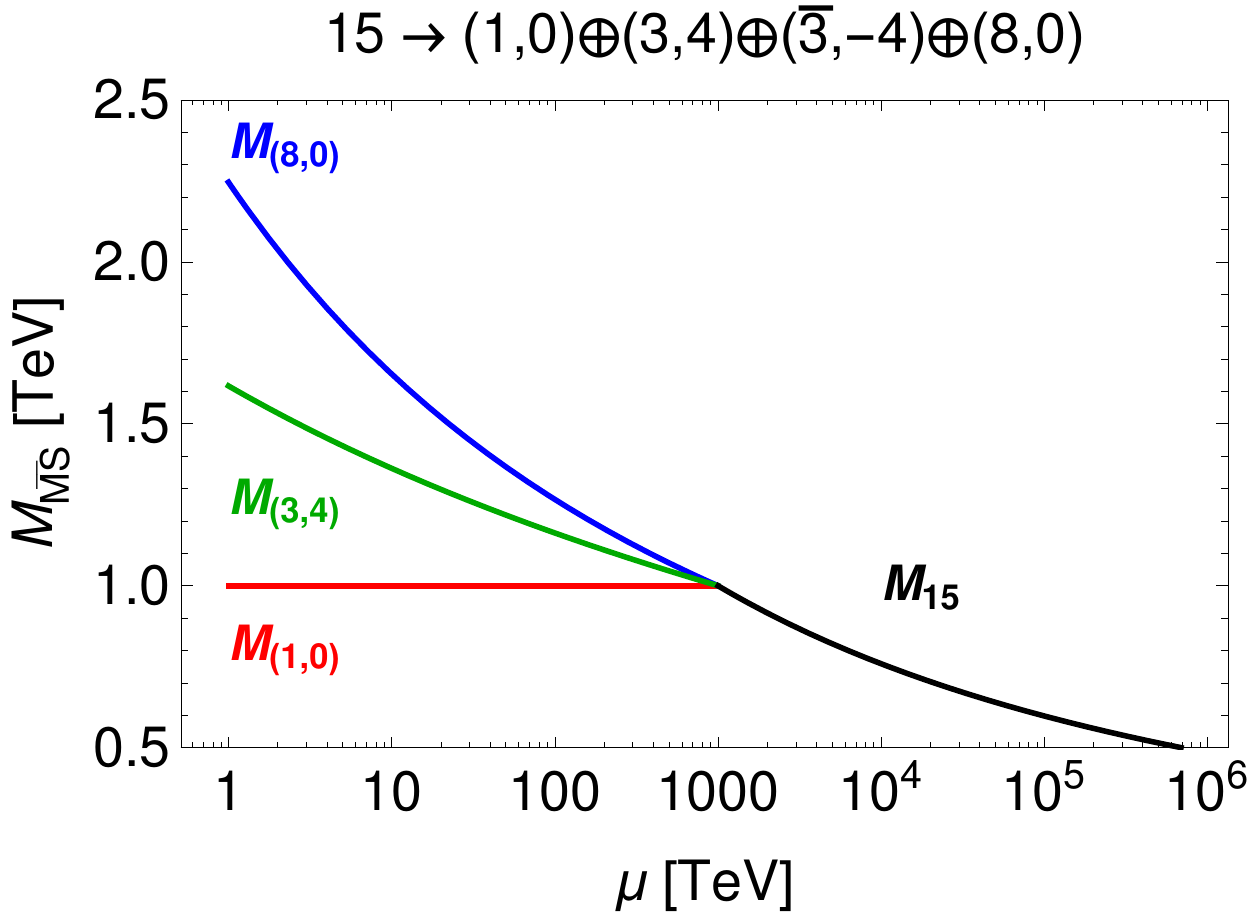}
\caption{Upper panel: running of gauge couplings assuming an $SU(4)$ breaking scale of $\unit[1000]{TeV}$.
Lower panel: running of $\overline{\text{MS} }$ masses of an $SU(4)$ representation $\vec{15}$.
}
\label{fig:rge_toy}
\end{figure}

\begin{figure}[tb]
\includegraphics[width=0.48\textwidth]{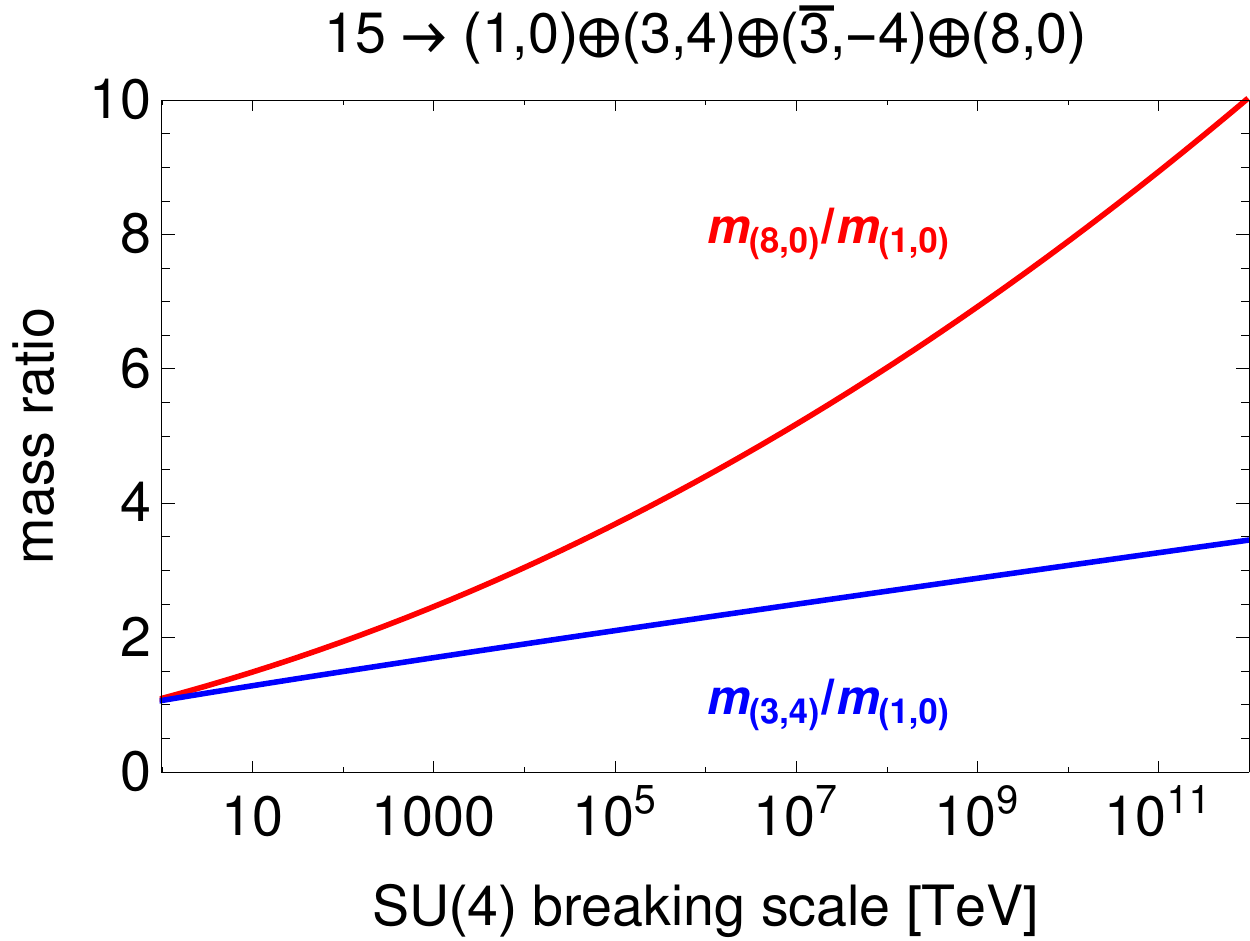}
\caption{Pole-mass ratios $m_{(\vec{8},0)}/m_{(\vec{1},0)}$ and $m_{(\vec{3},4)}/m_{(\vec{1},0)}$ of the particles in a $\vec{15}$ as a function of the $SU(4)$ breaking scale.
}
\label{fig:mass_ratio_toy}
\end{figure}

As an example, let us consider the breaking of $SU(4)$ to $SU(3)\times U(1)$ at a scale of $\unit[1000]{TeV}$, having the Pati--Salam group in mind. Below the breaking scale, the $SU(4)$ gauge coupling splits into two couplings that evolve differently (Fig.~\ref{fig:rge_toy} (upper panel)). For definiteness we assume six copies of massless Dirac fermions in the representation $\vec{4}$, which corresponds to the SM fermion content in the Pati--Salam model. As DM we add one chiral $\vec{15}$, which has an $SU(4)$ invariant $\overline{\text{MS} }$ mass $M_\vec{15}$ in the  $SU(4)$ phase, and we assume no additional couplings to $SU(4)$ breaking scalars. Below the $SU(4)$ breaking scale, the $\vec{15}$ breaks into one Dirac fermion $(\vec{3},4)$ and two Majorana particles, $(\vec{1},0)$ and $(\vec{8},0)$, all with differently running masses, as shown in Fig.~\ref{fig:rge_toy} (lower panel). The boundary condition for the running is simply
\begin{align}
\begin{split}
M_\vec{15} (\mu_{SU(4)})&=M_{(\vec{3},4)} (\mu_{SU(4)})\\
&=M_{(\vec{1},0)} (\mu_{SU(4)})\\
&=M_{(\vec{8},0)} (\mu_{SU(4)})
\end{split}
\end{align}
for the $\overline{\text{MS} }$ masses at the $SU(4)$ breaking scale $\mu_{SU(4)}$. The singlet $(\vec{1},0)$ mass $M_{(\vec{1},0)}$ is of course constant for $\mu <\mu_{SU(4)}$, while the octet $(\vec{8},0)$ experiences strong running due to the large Casimir operator and strong $SU(3)$ running. The ratio of pole masses $m_{(\vec{8},0)}/m_{(\vec{1},0)}$ thus grows if the $SU(4)$ breaking scale $\mu_{SU(4)}$ is pushed to higher scales, as shown in Fig.~\ref{fig:mass_ratio_toy}. The ratios can be explicitly given by
\begin{align}
\frac{m_{(\vec{8},0)}}{m_{(\vec{1},0)}} &= \frac{(\pi+ 3 \alpha_S ) (6 \pi + 13 \alpha_S \log \left(\frac{\mu_{SU(4)}}{\unit{TeV}}\right) )^{27/13}}{36 \times 6^{1/13}\pi ^{40/13}}\,,\nonumber\\
\frac{m_{(\vec{3},4)}}{m_{(\vec{1},0)}} &= \frac{(3 \pi + 4 \alpha_S) (6 \pi + 13 \alpha_S \log \left(\frac{\mu_{SU(4)}}{\unit{TeV}}\right))^{81/130} }{9\times 2^{12/13} 3^{81/130} \pi ^{25/13} } \nonumber\\
&\quad \times \frac{6 \pi + 4 \alpha_S+33 \alpha_S \log \left(\frac{\mu_{SU(4)}}{\unit{TeV}}\right)}{(2 \pi +11 \alpha_S \log \left(\frac{\mu_{SU(4)}}{\unit{TeV}}\right))^{7/10}}\,, 
\label{eq:mass_splitting_15}
\end{align}
where $\alpha_S$ is the strong coupling constant at the TeV scale. 

Notice that these mass ratios remain approximately valid even if the $\vec{15}$ is charged under additional gauge groups such as $SU(2)_L\times SU(2)_R$ because Eq.~\eqref{eq:pole_vs_GUT} is a simple product of group factors. The $\vec{15}$ components then of course form $SU(2)_L\times SU(2)_R$ multiplets that are themselves radiatively split. For example, a $(\vec{15},\vec{3})$ under $SU(4)\times SU(2)_L\to SU(3)\times SU(2)_L\times U(1)$ brings the massive fermions  $(\vec{1},\vec{3},0)$, $(\vec{3},\vec{3},4)$,  and $(\vec{8},\vec{3},0)$ with mass ratios as in Fig.~\ref{fig:mass_ratio_toy}; however, each $SU(2)_L$ triplet is further split by the non-zero $Z$ and $W$ masses in complete analogy to the wino.

\section{Co-annihilation and chemical equilibrium}
\label{app:chem_dec}

\begin{figure}[t]
\includegraphics[width=0.4\textwidth]{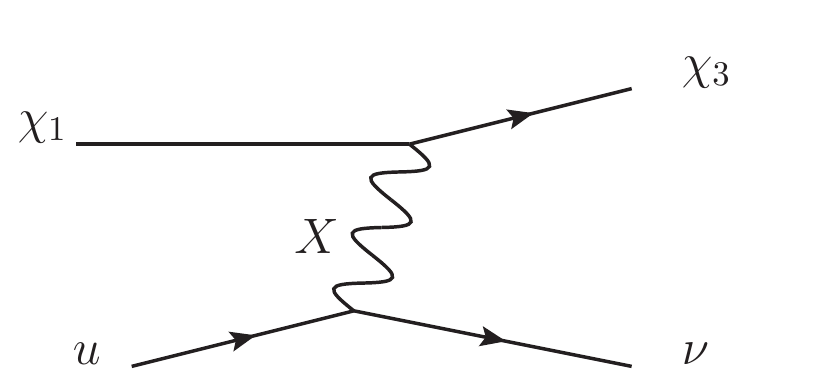}
\caption{One possible conversion rate process relevant for co-annihilation involving the heavy Pati--Salam vector leptoquark $X$. $\chi_1$ is the SM-singlet DM candidate and $\chi_3$ its color-triplet partner that will annihilate into gluons.}
\label{fig:coannihilation_via_X}
\end{figure}

In this appendix, we provide some details regarding chemical equilibrium between DM and its partners in co-annihilation scenarios. Suppose a set of $n$ $Z_2$-odd particles $\chi_i$ with $i= 1, \ldots , n$.  The lightest such particle, say $\chi_1$, is \emph{a priori} the DM candidate but the abundance of the $\chi_i$ particles is in full generality determined by a set of Boltzmann equations that schematically read  \cite{Griest:1990kh,Edsjo:1997bg}
\begin{align}
\label{eq:setBol}
{\dd n_i\over \dd t} &= - 3 H n_i - \sum_{j,f,f^\prime} \left\{ \langle \sigma_{ij} v\rangle (n_i n_j - n_i^{\rm eq}n_j^{\rm eq})\right. \\
&\quad - (\langle \sigma^\prime_{ij} v\rangle n_i n_f^{\rm eq} - \langle \sigma^\prime_{ji} v \rangle n_j n_{f^\prime}^{\rm eq})  \left. - \Gamma_{ij} (n_i - n_i^{\rm eq})\right\} .\nonumber
\end{align}
Here $f$ refers loosely to SM degrees of freedom. The second term on the right-hand side involves all reactions of the type $\sigma_{ij} = \sigma (\chi_i \chi_j - f f^\prime)$, thus including annihilation of heavier odd particles. Such processes may play a role if these particles are nearly as abundant as the lightest odd particle, which translates into a condition on the mass degeneracy, roughly $(m_{\chi_j} - m_{\chi_1})/m_{\chi_1} \lesssim 10 \%$~\cite{Griest:1990kh}. If this is so, they are particularly relevant if the heavier partner states have stronger interactions and so may be the driving channel that drains the DM abundance. A good example is colored DM partners, as in scenarios \ref{sec:Octet-triplet-singlet_8+6+1} or \ref{sec:Otherscenarios} of the present work (see also e.g.~Refs.~\cite{deSimone:2014pda,Giacchino:2015hvk,ElHedri:2018atj,Davoli:2018mau}). The key condition is that the $\chi_i$ particles are in chemical equilibrium amongst {themselves}. This is dictated by transition cross sections of the type $\sigma^\prime_{ij} = \sigma(\chi_i f \rightarrow \chi_j f^\prime)$, which appear in the third term of Eq.~\eqref{eq:setBol}. If such transitions are fast enough compared to the Hubble rate,  then $n_i/n_i^{\rm eq} = n_j/n_j^{\rm eq}$. This leads to a drastic simplification, as the set of Boltzmann equations \eqref{eq:setBol} reduces to a single equation for the total density of $Z_2$-odd particles, $n = \sum_i n_i$, whose abundance is driven by an effective annihilation cross section \cite{Griest:1990kh,Edsjo:1997bg}.  The condition of chemical equilibrium is in general taken for granted. For instance, it is implicit in numerical codes such as {\sc MicrOMEGAs}~\cite{Belanger:2006is}. In principle, however, one must check whether chemical equilibrium is satisfied. In scenarios similar to \ref{sec:Octet-triplet-singlet_8+6+1} co-annihilation plays a crucial role in determining the abundance of DM. For instance, in scenario \ref{sec:Otherscenarios}, DM is a singlet Majorana fermion, $\chi_1$, which comes from a $\vec{15}$ of $SU(4)_{\rm PS}$ (relevant Lagrangian given in Eq.~\eqref{eq:Xinteractions}). Its abundance may be driven by co-annihilation of its color-triplet Dirac partner  $\chi_3$ into gluons. The transitions between $\chi_1$ and $\chi_3$ involve the heavy PS lepto-quark gauge boson $X$ (see Fig.~\ref{fig:coannihilation_via_X}), and the requirement of chemical equilibrium sets an upper limit on $m_X$. This condition stems from the requirement that $
\Gamma_{\chi_1 \chi_3^{(c)}} \gtrsim H(T_{\rm fo})
 $
where $\Gamma_{\chi_1 \chi_3^{(c)}} = \langle\Gamma\rangle_{\chi_1 f \rightarrow \chi_3^{(c)} f^\prime}$ is the thermally averaged transition rate of a $\chi_1$ into the colored $\chi_3$ (or, since $\chi_1$ is Majorana, its charge conjugate $\chi_3^{(c)}$) and $H$ is the Hubble rate at the time/temperature of freeze-out $T_{\rm fo}$. 
If this condition is not satisfied, co-annihilation is ineffective. In the present case, this would imply that the $\chi_1$ particle would be over-abundant if it  was in thermal equilibrium, as it has only very feeble interactions with SM degrees of freedom on its own.

For concreteness we consider the transition process $\chi_1 +  u \rightarrow \chi_3 + \nu$ of Fig.~\ref{fig:coannihilation_via_X}, where $u$ is a up-like quark and $\nu$ its associated neutrino. 
In the limit in which the $\chi_1$ and $\chi_3$ particles are highly non-relativistic, the rate in the thermal bath is  given by\footnote{The $\nu$ Fermi-blocking factor is for preciousness, as it plays no fundamental role in the sequel. We  keep it because we can obtain a closed analytical expression for the case of massless SM fermions  and also because quantum statistic effects are numerically significant for relativistic degrees of freedom.}
\begin{align}
\langle\Gamma\rangle_{\chi_1 u \rightarrow \chi_3 \nu} &= {N_c\over 2 m_1} \int {g_u \dd^3 p_u \over (2 \pi)^3 2 E_u}{g_\nu \dd^3 p_\nu \over (2 \pi)^3 2 E_\nu}{g_3\dd^3 p_{\chi_3} \over (2 \pi)^3 2 m_{3}}\nonumber\\
&\quad \times \vert {\cal M}\vert^2 (2 \pi)^4 \delta^4(p_{\chi_1} + p_u - p_{\chi_3} - p_\nu) \nonumber\\
&\quad \times  f_u(E_u)(1 - f_\nu(E_\nu)) \,,\label{eq:rateChemical}
\end{align}
where $g_u, g_3$, and $g_\nu$ are the spin degeneracy factors. The  squared matrix element is averaged over both initial and final spins (see \cite{Kolb:1990vq}) and for $\chi_1 +  u \rightarrow \chi_3 + \nu$ is given by
\begin{align}
\vert {\cal M}\vert^2 \simeq {2  \over 3} {g_4^4\over m_{X}^4} m_1 m_3 E_\nu E_u ( 1+ \cos\theta) \,.
\end{align} 
This is taking into account that the interaction with  the PS lepto-quark boson of mass $m_X \gg m_{1,3}$ is purely vectorial, with coupling $g_4 \sqrt{2/3}$ at the $\chi_1\chi_3$ vertex, and $g_4 \sqrt{1/2}$ for the $u \nu$ one (see Eq.~\eqref{eq:Xinteractions}).\footnote{We neglect the heavy right-handed neutrinos here for simplicity, so we have couplings only to the left-handed state, $g_\nu = 1$. This together with the average over both the initial and the final state particles properly takes into account the factor of $1/2$ in the cross section for $\chi_1 + u \rightarrow \chi_3 + \nu$ compared to $\chi_1 + d \rightarrow \chi_3 + e$, which is obtained using $g_d = g_e = 2$.} 
Neglecting the masses of all SM fermions we may rewrite (\ref{eq:rateChemical}) as
\begin{align}
\label{eq:rateInt}
\langle\Gamma\rangle_{\chi_1 u \rightarrow \chi_3 \nu}  =  {N_c \over 12 \pi^2} {g_4^4\over m_X^4} \Delta m^5 \, {F}(\Delta m/T) \,,
\end{align}
with mass splitting $\Delta m \equiv m_3 - m_1$ and
\begin{align}
{F}(x) \equiv \int_1^\infty \dd z {z^2 (z- 1)^2\over (1 + e^{z x})(1 + e^{x (1-z)})} \,.
\end{align}
The prefactor in Eq.~\eqref{eq:rateInt} is akin to the decay rate of $\chi_3$ in Eq.~\eqref{eq:decay_of_15}, which is $\propto g_4^4 \Delta m^5/m_X^4$. The function $F$ captures the phase space that is available at finite temperature for the process $\chi_1 + u \rightarrow \chi_3 + \nu$, provided $E_u -E_\nu \geq \Delta m$.\footnote{This is analogous to $p + e \rightarrow n + \nu$, which is relevant to set the initial conditions for primordial nucleosynthesis, see Kolb \& Turner  (there is a missing factor of $1/2m_p$ in Eq.~(4.13) of \cite{Kolb:1990vq}).} 
The integration  is over $z = E_u/\Delta m$. At high temperatures, $T \gg \Delta m$, the integral can be solved in closed form and equals $F(\Delta m/T) \simeq 7 \pi^4/30 \, T^5/\Delta m^5$, which leads to the familiar $\propto T^5$ behavior of contact interactions at high temperature:
\begin{align}
\langle\Gamma\rangle_{\chi_1 u \rightarrow \chi_3 \nu} \simeq N_c\, {7 \pi^2 \over 360} {g_4^4\over m_X^4} T^5 \quad \mbox{for}\; T \gg \Delta m \,.
\label{eq:small_splitting}
\end{align}
Neglecting quantum statistics would change the rate by a factor of $720/(7\pi^4)\simeq 1.06$.
For freeze-out, we may need to consider also the regime $T \sim \Delta m$, in which case the phase-space integral is given by the following expression:
\begin{align}
\label{eq:rate_high_T}
F\left(\frac{\Delta m}{T}\right) &\simeq {3\over 2} {T^5\over \Delta m^5} e^{-{\Delta m}/{T}} \\
&\quad \times \left(15 \zeta(5) + 7 \zeta(4) {\Delta m\over T}  + \zeta(3){\Delta m^2\over T^2} \right) , \nonumber
\end{align} 
with  $\zeta(3) \simeq 1.20$,  $\zeta(4) = \pi^4/90$, and $\zeta(5)\simeq 1.04$. Since $45 \zeta(5)/2 \simeq 23.33$ and $7 \pi^4/30 \simeq 22.73$, this expression actually gives an excellent fit even in the high temperature regime. 

\begin{figure}[t]
\includegraphics[width=0.45\textwidth]{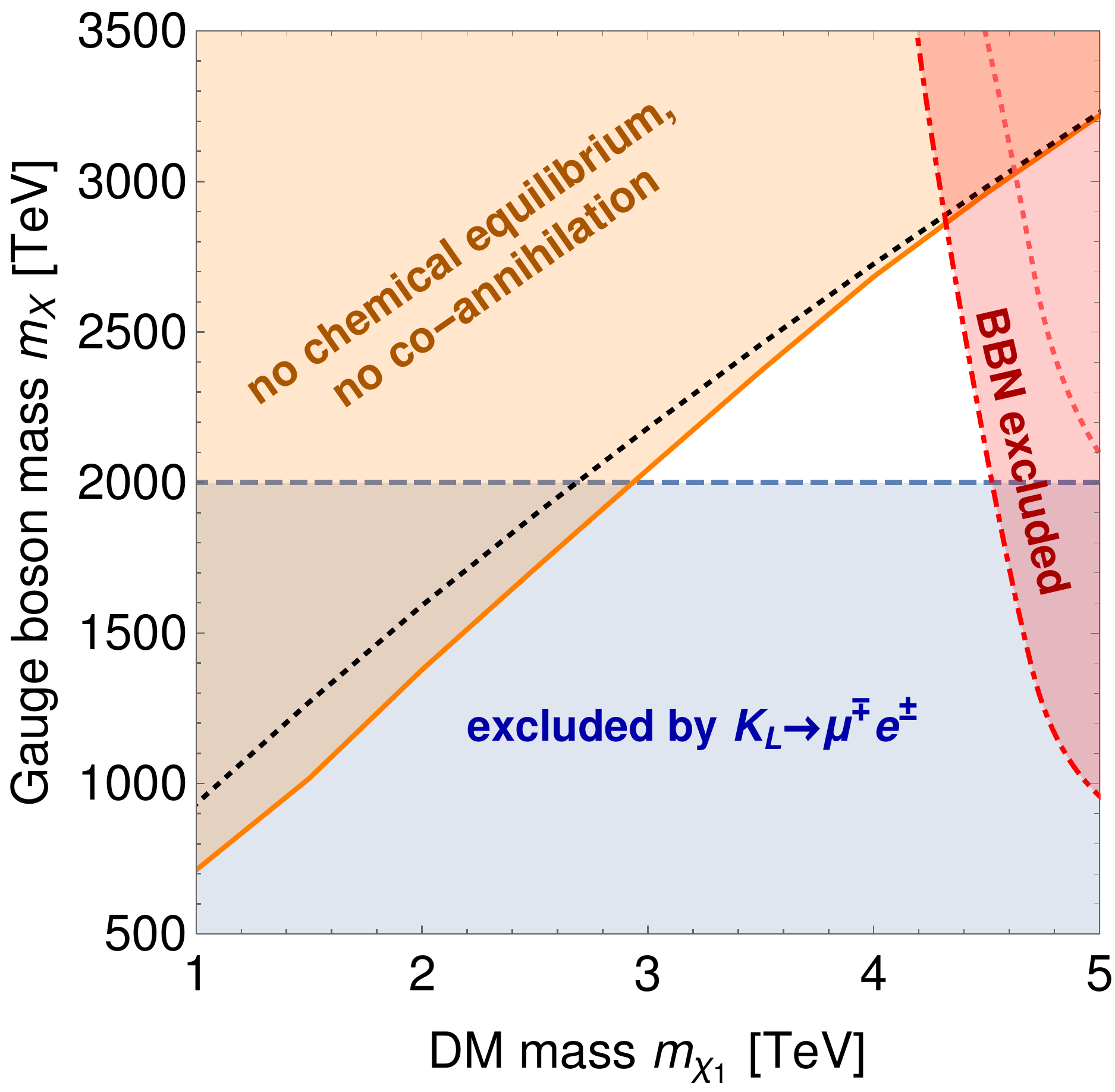}
\caption{Constraints from chemical equilibrium in the $m_{\chi_1}$--$m_X$ plane. Co-annihilation is ineffective in the orange region based on condition \eqref{eq:chem_fo}; the black dotted line is  the same but uses the simple high-temperature limit of the rate, Eq.~\eqref{eq:rate_high_T}. 
The dashed (blue) horizontal line is the experimental lower bound on $m_X$. For parameters in the region bounded by the dot-dashed (red) line the $\chi_3$ lifetime is longer than $\unit[0.1]{s}$ and may spoil BBN; the dotted red line is an estimate of bound-state effects. The viable parameter space corresponds to the white region in the middle, where the correct $\chi_1$ DM abundance can be obtained by picking the right mass splitting~\cite{deSimone:2014pda}.}
\label{fig:chem_eq}
\end{figure}

To get a bound on $m_X$, we take into account all the processes that contribute to the transitions $\chi_1 \rightarrow \chi_3$. For a given family, these are {\em a priori}
\begin{align}
\begin{split}
\chi_1  + u \rightarrow \chi_3 + \nu \,, \\ 
\chi_1 + \bar \nu \rightarrow \chi_3 + \bar u \,,
\end{split}
\end{align}
and, with a rate 2 times larger due to available right-handed SM fermions, 
\begin{align}
\begin{split}
\chi_1  + d \rightarrow \chi_3 + e \,, \\ 
\chi_1 + \bar e \rightarrow \chi_3 + \bar d \,.
\end{split}
\end{align}
For the third generation some of these processes are absent on account of the top quark being too heavy. Neglecting the top-quark channels, the transition rate may be approximated to be $g_{\rm eff} = \left\{(N_f = 2) \times 2 + (N_f = 3) \times 4\right\} = 16$ times the rate of Eq.~\eqref{eq:rateInt}. If $T_{\rm fo} \gtrsim m_{\rm top}$, the factor is instead $g_{\rm eff} = 18$. Chemical equilibrium then amounts to requesting
\begin{align}
\label{eq:chem_fo}
g_{\rm eff} \langle\Gamma\rangle_{\chi_1 u \rightarrow \chi_3 \nu}\, \gtrsim \, H(T_{\rm fo}) \,,
\end{align}
where $H(T_{\rm fo})$ is the expansion rate at the time of freeze-out, taken here to be the freeze-out of the annihilation process $\chi_3 \bar \chi_3\rightarrow g g$. Concretely we take $T_{\rm fo} \simeq m_{\chi_3}/25$ \cite{deSimone:2014pda}. 
Assuming the high temperature limit of the rate of Eq.~(\ref{eq:rate_high_T}) gives a bound
\begin{align}
\label{eq:bound_mx_approx}
m_X/g_4 \lesssim \unit[900]{TeV} \left( \frac{m_1}{\unit{TeV}} \right)^{3/4}
\end{align}
on the mass of the heavy Pati--Salam  particle $X$. Taking $g_4 \simeq 1$, we see that the experimental lower bound on $m_X \gtrsim \unit[2000]{TeV}$ of~\cite{Valencia:1994cj,Kuznetsov:1994tt,Ambrose:1998us,Smirnov:2007hv} requires considering DM masses $m_1 \gtrsim \unit[3]{TeV}$. Note that the lower bound on $m_X$ can be pushed down by an order of magnitude when fermion mixing is taken into account~\cite{Kuznetsov:1994tt,Smirnov:2008zzb,Smirnov:2018ske}, which of course opens up our parameter space. A more precise bound is obtained by taking into account the dependence of the rate (\ref{eq:rateInt}) on the mass splitting $\Delta m= m_{\chi_3} - m_{\chi_1} > 0$. This calculation is depicted in Fig.~\ref{fig:chem_eq} in the plane $m_{\chi_1}$--$m_X$. The dotted line corresponds to the bound  (\ref{eq:bound_mx_approx}) in the high temperature $T_{\rm fo} \gg \Delta m$ approximation. 

To have specific values for $\Delta m$, we refer to Fig.~2 of Ref.~\cite{deSimone:2014pda}, lower-left panel. In this reference, the abundance of a singlet DM candidate through the annihilation of a color-triplet fermion--anti-fermion pair into gluons is determined taking into account various effects, in particular the Sommerfeld effect. Specifically, we make use of the light green curve to extract values of $\Delta m$. For instance, for $m_{\chi_1} = \unit[2]{TeV}$ we have taken $\Delta m = \unit[85]{GeV}$. In this case, $\Delta m/T_{\rm fo} \approx 1$ and finite $\Delta m$ effects are important. For higher DM masses, the required mass splitting is smaller and the high temperature rate provides a good approximation. The result is shown as the solid (orange) line in Fig.~\ref{fig:chem_eq}. In the shaded (orange) region, the transitions $\chi_1 \rightarrow \chi_3$ are out-of-equilibrium and co-annihilation is ineffective.
An immediate outcome is that the DM abundance is too large  to match the cosmological observations. The intermediate regime, in which chemical equilibrium is barely realized, is interesting by itself, see \cite{Garny:2017rxs}.
Clearly, finite $\Delta m$ effects tend to decrease the rate for $\chi_1 \rightarrow \chi_3$ and thus lower the bound on $m_X$. 

We should emphasize that the possible formation of $\chi_3$--$\bar \chi_3$ bound states at the time of annihilation may affect the annihilation rate and so the precise value of the mass splitting. Such effects were not taken into account in Ref.~\cite{deSimone:2014pda} but have been studied for the case of co-annihilation of squarks in Ref.~\cite{Mitridate:2017izz}. The impact of bound states on the mass splitting $\Delta m$ is mild, ${\cal O}(15\%)$, except for large DM masses for which $\Delta m$ decreases more slowly than expected, see Fig.~12 and explanations in \cite{Mitridate:2017izz}. As bound state formation tends to increase $\Delta m$, in the absence of explicit calculations in the case of fermionic colored partners, we have merely checked that modifying the $\Delta m$ by relative factors that follow the pattern in Fig.~12, left panel, of  \cite{Mitridate:2017izz} does not change the bounds on $m_X$ shown in Fig.~\ref{fig:chem_eq} much (i.e.~by at most a few percent). 

In Fig.~\ref{fig:chem_eq} we also report the constraints on the lifetime of the $\chi_3$ particle from BBN that requires $\tau \lesssim \unit[0.1]{s}$~\cite{Kawasaki:2004qu}.  The shaded region bounded by the red dot-dashed line is obtained using the mass splitting given in Ref.~\cite{deSimone:2014pda}, as discussed in the previous paragraph.  As the lifetime of $\chi_3$ is very sensitive to the mass splitting $\Delta m$ (see Eq.~\eqref{eq:decay_of_15}), the other red dotted curve shows the limit on $\tau$ for the higher values of the mass splitting than in Fig.~2 of Ref.~\cite{deSimone:2014pda}, anticipating possible bound-state effects. Specifically, we have taken $\Delta m = \unit[30]{GeV}$ at $m_{\chi_1} = \unit[4]{TeV}$ and  $\Delta m = \unit[10]{GeV}$ at $m_{\chi_1} = \unit[5]{TeV}$. Finally, the dashed (blue) curve corresponds to the lower bound $m_X \gtrsim \unit[2000]{TeV}$ of~\cite{Valencia:1994cj,Ambrose:1998us,Smirnov:2007hv}, which could, however, be reduced. Altogether, the viable candidates are confined between $m_{\chi_1} \simeq \unit[3]{TeV}$ and about $\unit[5]{TeV}$, with an upper bound on the PS scale of $m_X \lesssim \unit[3000]{TeV}$. Notice finally that the lifetime of the $\chi_{3}$ particle for $m_{\chi_1} \simeq \unit[2]{TeV}$ (for which we take $\Delta m = \unit[85]{GeV}$) is about $5 \times \unit[10^{-7}]{s}$, corresponding to $c \tau \sim \unit[150]{m}$. For such values, $\chi_3$ particles produced at the LHC would appear as stable $R$~hadrons in the detectors. The current limits  are  $\sim \unit[1.8]{TeV}$ for a stable gluino and $\sim \unit[900]{GeV}$ for a stable stop  \cite{Aaboud:2016uth}. Limits on a long-lived $\chi_3$ should be somewhere in between, below the minimum of $\unit[3]{TeV}$ for $m_{\chi_3} \sim m_{\chi_1}$  reported in Fig.~\ref{fig:chem_eq}.  

The discussion of the present appendix brings support to the fact that the colored triplet-DM singlet (i.e.~$6+1$ of Sec.~\ref{sec:Otherscenarios}) scenario is viable, provided the mass splitting and the PS scale are tuned appropriately. 
Notice that there might be additional enhancement factors (top-quark or right-handed neutrino channels, $k$-factor enhanced next-to-leading order corrections, bound-state effects) and that the PS-scale limits can be weakened by playing with fermion mixing, so the parameter space can be opened up a bit more. 

Co-annihilation of a singlet DM directly with an \emph{octet} partner will not work by itself due to the lack of direct DM--octet--gauge-boson vertices, 
but it could do so together with an intermediate colored triplet. 
This is precisely the scenario \ref{sec:Octet-triplet-singlet_8+6+1}. As it involves two mass splittings, which affect the formation of bound states and/or the Sommerfeld effects, we leave the determination of the relic abundance  in this scenario for other work. Finally, the formalism may be applied to other scenarios, for instance the case of co-annihilation of an $SU(2)_R$  triplet (e.g.~the triplet 3 scenario of Sec.~\ref{sec:SU(2)_R_triplet_3}).

\section{Tables}
\label{app:tables}

A chiral fermion $F$ in an irreducible representation $R$ under $SO(10)$ is split into a sum of representations under the subgroup $G$ when $SO(10)$ is broken $SO(10)\to G$. If $F$ does not couple to the scalars $S$ that break $SO(10)$, it will remain degenerate at tree level, but if it has a Yukawa coupling $y \overline{F}^c F S$, it will receive mass splittings $y \langle S \rangle$ proportional to the relevant Clebsch--Gordan coefficients (calculable for example with  \texttt{Susyno}~\cite{Fonseca:2011sy}). All our candidates have an $SO(10)$ symmetric mass term $m_\vec{1}$, not connected to $SO(10)$ breaking. They can furthermore couple to scalars $\vec{45}$, $\vec{54}$, and $\vec{210}$, all of which can obtain $SO(10)$ breaking VEVs, splitting the masses within the fermion multiplet. We will denote the product of Yukawa coupling times VEV as $m_S \propto y \langle S \rangle$, normalized for each multiplet in order to obtain simple expressions with few fractions and square roots. If more than one Yukawa coupling exists, we distinguish them with indices $A$ and $B$, e.g.~$m_{\vec{54},A}$ and $m_{\vec{54},B}$. If a scalar has VEVs in several components, we distinguish them according to their $G_{422}$ quantum numbers, e.g.~$m_{\vec{210}_{(\vec{1},\vec{1},\vec{1})}}$ and $m_{\vec{210}_{(\vec{15},\vec{1},\vec{1})}}$.

All our fermions are in a real representation of $SO(10)$ and thus massive. Depending on the subgroup, the massive states are Dirac or Majorana fermions. In the following, $m_R$ denotes the mass of a Majorana (Dirac) fermion if the representation $R$ is real (complex) under the relevant subgroup. For a Dirac particle, there is a choice whether one labels the particle and mass by $R$ or $\overline{R}$ (only relevant for $SU(3)$ and $SU(4)$); we always pick the one without a bar.
Finally, the mass terms $m_R$ we give here can be negative or even complex in general, in which case the physical masses correspond to the absolute values (or singular values in case of mass matrices).

There are four $SO(10)$ subgroups of relevance for us: $G_{422}$, $G_{3221}$, $G_{421}$, and $G_{3211}$. Rather than discuss them separately, we will discuss them two at a time, starting with $G_{422}$ and $G_{3221}$.

\subsection{$G_{422}$ and $G_{3221}$}

We start with the $SO(10)$ subgroups $G_{422}$ and $G_{3221}$; to keep the expressions readable, we pick a $U(1)$ normalization that avoids fractions, namely $U(1)_{3 (B-L)}$.
Tab.~\ref{tab:422and3221} lists the chiral representations $\vec{10}$, $\vec{45}$, $\vec{54}$, $\vec{120}$, $\vec{210}$, and $\vec{210}'$ together with their tree-level mass splittings due to VEVs down to the subgroups $G_{422}$ and $G_{3221}$. The considerably lengthier representation $\vec{126}\oplus\overline{\vec{126}}$ is listed in Tab.~\ref{tab:126_422} for $G_{422}$ and in Tab.~\ref{tab:126_3221} for $G_{3221}$.

\onecolumngrid

\begingroup
\begin{table}
\begin{tabular}{l|l|l}
$SO(10)$ & $SU(4)\times SU(2)_L\times SU(2)_R$ & $SU(3)_C\times SU(2)_L\times SU(2)_R\times U(1)_{3 (B-L)}$\\
\hline
\rotatebox[origin=c]{90}{$\vec{10}$} & $\begin{array}{lll}
 m_{(\vec{1}, \vec{2}, \vec{2})} & = & m_\vec{1}+3 m_\vec{54} \\
 m_{(\vec{6}, \vec{1}, \vec{1})} & = & m_\vec{1}-2 m_\vec{54} \\
\end{array}$ & $\begin{array}{lll}
 m_{(\vec{1}, \vec{2}, \vec{2}, 0)} & = & m_\vec{1}+3 m_\vec{54} \\
 m_{(\vec{3}, \vec{1}, \vec{1}, -2)} & = & m_\vec{1}-2 m_\vec{54} \\
\end{array}$ \\
\hline
\rotatebox[origin=c]{90}{$\vec{45}$} & $\begin{array}{lll}
 m_{(\vec{1}, \vec{1}, \vec{3})} & = & m_\vec{1}-6 m_\vec{54}-m_{\vec{210}_{(\vec{1},\vec{1},\vec{1})}} \\
 m_{(\vec{1}, \vec{3}, \vec{1})} & = & m_\vec{1}-6 m_\vec{54}+m_{\vec{210}_{(\vec{1},\vec{1},\vec{1})}} \\
 m_{(\vec{6}, \vec{2}, \vec{2})} & = & m_\vec{1}-m_\vec{54} \\
 m_{(\vec{15}, \vec{1}, \vec{1})} & = & m_\vec{1}+4 m_\vec{54} \\
\end{array}$ & $\begin{array}{lll}
 m_{(\vec{1}, \vec{1}, \vec{3}, 0)} & = & m_\vec{1}-6 m_\vec{54}-m_{\vec{210}_{(\vec{1},\vec{1},\vec{1})}} \\
 m_{(\vec{1}, \vec{3}, \vec{1}, 0)} & = & m_\vec{1}-6 m_\vec{54}+m_{\vec{210}_{(\vec{1},\vec{1},\vec{1})}} \\
 m_{(\vec{3}, \vec{2}, \vec{2}, -2)} & = & m_\vec{1}-m_\vec{54} \\
 m_{(\vec{1}, \vec{1}, \vec{1}, 0)} & = & m_\vec{1}+4 m_\vec{54}-2 m_{\vec{210}_{(\vec{15},\vec{1},\vec{1})}} \\
 m_{(\vec{3}, \vec{1}, \vec{1}, 4)} & = & m_\vec{1}+4 m_\vec{54}-m_{\vec{210}_{(\vec{15},\vec{1},\vec{1})}} \\
 m_{(\vec{8}, \vec{1}, \vec{1}, 0)} & = & m_\vec{1}+4 m_\vec{54}+m_{\vec{210}_{(\vec{15},\vec{1},\vec{1})}} \\
\end{array}$ \\
\hline
\rotatebox[origin=c]{90}{$\vec{54}$} & $\begin{array}{lll}
 m_{(\vec{1}, \vec{1}, \vec{1})} & = & m_\vec{1}+2 m_\vec{54} \\
 m_{(\vec{1}, \vec{3}, \vec{3})} & = & m_\vec{1}+6 m_\vec{54} \\
 m_{(\vec{6}, \vec{2}, \vec{2})} & = & m_\vec{1}+m_\vec{54} \\
 m_{(\vec{20'}, \vec{1}, \vec{1})} & = & m_\vec{1}-4 m_\vec{54} \\
\end{array}$ & $\begin{array}{lll}
 m_{(\vec{1}, \vec{1}, \vec{1}, 0)} & = & m_\vec{1}+2 m_\vec{54} \\
 m_{(\vec{1}, \vec{3}, \vec{3}, 0)} & = & m_\vec{1}+6 m_\vec{54} \\
 m_{(\vec{3}, \vec{2}, \vec{2}, -2)} & = & m_\vec{1}+m_\vec{54} \\
 m_{(\vec{6}, \vec{1}, \vec{1}, 4)} & = & m_\vec{1}-4 m_\vec{54} \\
 m_{(\vec{8}, \vec{1}, \vec{1}, 0)} & = & m_\vec{1}-4 m_\vec{54} \\
\end{array}$ \\
\hline
\rotatebox[origin=c]{90}{$\vec{120}$} & $\begin{array}{lll}
 m_{(\vec{1}, \vec{2}, \vec{2})} & = & m_\vec{1}-9 m_\vec{54} \\
 m_{(\vec{6}, \vec{1}, \vec{3})} & = & m_\vec{1}-4 m_\vec{54}-m_{\vec{210}_{(\vec{1},\vec{1},\vec{1})}} \\
 m_{(\vec{6}, \vec{3}, \vec{1})} & = & m_\vec{1}-4 m_\vec{54}+m_{\vec{210}_{(\vec{1},\vec{1},\vec{1})}} \\
 m_{(\vec{10}, \vec{1}, \vec{1})} & = & m_\vec{1}+6 m_\vec{54} \\
 m_{(\vec{15}, \vec{2}, \vec{2})} & = & m_\vec{1}+m_\vec{54} \\
\end{array}$ & $\begin{array}{lll}
 m_{(\vec{1}, \vec{2}, \vec{2}, 0)} & = & m_\vec{1}-9 m_\vec{54} \\
 m_{(\vec{3}, \vec{1}, \vec{3}, -2)} & = & m_\vec{1}-4 m_\vec{54}-m_{\vec{210}_{(\vec{1},\vec{1},\vec{1})}} \\
 m_{(\vec{3}, \vec{3}, \vec{1}, -2)} & = & m_\vec{1}-4 m_\vec{54}+m_{\vec{210}_{(\vec{1},\vec{1},\vec{1})}} \\
 m_{(\vec{1}, \vec{1}, \vec{1}, -6)} & = & m_\vec{1}+6 m_\vec{54}-3 m_{\vec{210}_{(\vec{15},\vec{1},\vec{1})}} \\
 m_{(\vec{3}, \vec{1}, \vec{1}, -2)} & = & m_\vec{1}+6 m_\vec{54}-m_{\vec{210}_{(\vec{15},\vec{1},\vec{1})}} \\
 m_{(\vec{6}, \vec{1}, \vec{1}, -2)} & = & m_\vec{1}+6 m_\vec{54}+m_{\vec{210}_{(\vec{15},\vec{1},\vec{1})}} \\
 m_{(\vec{1}, \vec{2}, \vec{2}, 0)'} & = & m_\vec{1}+m_\vec{54}-2 m_{\vec{210}_{(\vec{15},\vec{1},\vec{1})}} \\
 m_{(\vec{3}, \vec{2}, \vec{2}, 4)} & = & m_\vec{1}+m_\vec{54}-m_{\vec{210}_{(\vec{15},\vec{1},\vec{1})}} \\
 m_{(\vec{8}, \vec{2}, \vec{2}, 0)} & = & m_\vec{1}+m_\vec{54}+m_{vec{210}_{(\vec{15},\vec{1},\vec{1})}} \\
\end{array}$ \\
\hline
\rotatebox[origin=c]{90}{$\vec{210}$} & $\begin{array}{lll}
 m_{(\vec{1}, \vec{1}, \vec{1})} & = & m_\vec{1}+12 m_\vec{54} \\
 m_{(\vec{6}, \vec{2}, \vec{2})} & = & m_\vec{1}+7 m_\vec{54} \\
 m_{(\vec{10}, \vec{2}, \vec{2})} & = & m_\vec{1}-3 m_\vec{54} \\
 m_{(\vec{15}, \vec{1}, \vec{1})} & = & m_\vec{1}-8 m_\vec{54} \\
 m_{(\vec{15}, \vec{1}, \vec{3})} & = & m_\vec{1}+2 m_\vec{54}+m_{\vec{210}_{(\vec{1},\vec{1},\vec{1})}} \\
 m_{(\vec{15}, \vec{3}, \vec{1})} & = & m_\vec{1}+2 m_\vec{54}-m_{\vec{210}_{(\vec{1},\vec{1},\vec{1})}} \\
\end{array}$ & $\begin{array}{lll}
 m_{(\vec{1}, \vec{1}, \vec{1}, 0)} & = & m_\vec{1}+2 m_\vec{54}+2 m_{\vec{210}_{(\vec{15},\vec{1},\vec{1})}} \\
 & & +\sqrt{3 m_{\vec{45}_{(\vec{15},\vec{1},\vec{1})}}^2+4 \left(-5 m_\vec{54}+m_{\vec{210}_{(\vec{15},\vec{1},\vec{1})}}\right)^2} \\
 m_{(\vec{3}, \vec{2}, \vec{2}, -2)'} & = & 5 m_\vec{54}-m_{\vec{210}_{(\vec{15},\vec{1},\vec{1})}} \\
 & & +\sqrt{-2 m_{\vec{45}_{(\vec{15},\vec{1},\vec{1})}}^2+\left(m_\vec{1}+2 m_\vec{54}+m_{\vec{210}_{(\vec{15},\vec{1},\vec{1})}}\right)^2} \\
 m_{(\vec{1}, \vec{2}, \vec{2}, -6)} & = & m_\vec{1}-3 m_\vec{54}+6 m_{\vec{210}_{(\vec{15},\vec{1},\vec{1})}} \\
 m_{(\vec{3}, \vec{2}, \vec{2}, -2)} & = & -5 m_\vec{54}+m_{\vec{210}_{(\vec{15},\vec{1},\vec{1})}} \\
 & & +\sqrt{-2 m_{\vec{45}_{(\vec{15},\vec{1},\vec{1})}}^2+\left(m_\vec{1}+2 m_\vec{54}+m_{\vec{210}_{(\vec{15},\vec{1},\vec{1})}}\right)^2} \\
 m_{(\vec{6}, \vec{2}, \vec{2}, -2)} & = & m_\vec{1}-3 m_\vec{54}-2 m_{\vec{210}_{(\vec{15},\vec{1},\vec{1})}} \\
 m_{(\vec{1}, \vec{1}, \vec{1}, 0)'} & = & m_\vec{1}+2 m_\vec{54}+2 m_{\vec{210}_{(\vec{15},\vec{1},\vec{1})}} \\
 & & -\sqrt{3 m_{\vec{45}_{(\vec{15},\vec{1},\vec{1})}}^2+4 \left(-5 m_\vec{54}+m_{\vec{210}_{(\vec{15},\vec{1},\vec{1})}}\right)^2} \\
 m_{(\vec{3}, \vec{1}, \vec{1}, 4)} & = & m_\vec{1}-8 m_\vec{54}+2 m_{\vec{210}_{(\vec{15},\vec{1},\vec{1})}} \\
 m_{(\vec{8}, \vec{1}, \vec{1}, 0)} & = & m_\vec{1}-8 m_\vec{54}-2 m_{\vec{210}_{(\vec{15},\vec{1},\vec{1})}} \\
 m_{(\vec{1}, \vec{1}, \vec{3}, 0)} & = & m_\vec{1}+2 m_\vec{54}+2 m_{\vec{45}_{(\vec{15},\vec{1},\vec{1})}}+m_{\vec{210}_{(\vec{1},\vec{1},\vec{1})}}+4 m_{\vec{210}_{(\vec{15},\vec{1},\vec{1})}} \\
 m_{(\vec{3}, \vec{1}, \vec{3}, 4)} & = & m_\vec{1}+2 m_\vec{54}+m_{\vec{45}_{(\vec{15},\vec{1},\vec{1})}}+m_{\vec{210}_{(\vec{1},\vec{1},\vec{1})}}+2 m_{\vec{210}_{(\vec{15},\vec{1},\vec{1})}} \\
 m_{(\vec{8}, \vec{1}, \vec{3}, 0)} & = & m_\vec{1}+2 m_\vec{54}-m_{\vec{45}_{(\vec{15},\vec{1},\vec{1})}}+m_{\vec{210}_{(\vec{1},\vec{1},\vec{1})}}-2 m_{\vec{210}_{(\vec{15},\vec{1},\vec{1})}} \\
 m_{(\vec{1}, \vec{3}, \vec{1}, 0)} & = & m_\vec{1}+2 m_\vec{54}-2 m_{\vec{45}_{(\vec{15},\vec{1},\vec{1})}}-m_{\vec{210}_{(\vec{1},\vec{1},\vec{1})}}+4 m_{\vec{210}_{(\vec{15},\vec{1},\vec{1})}} \\
 m_{(\vec{3}, \vec{3}, \vec{1}, 4)} & = & m_\vec{1}+2 m_\vec{54}-m_{\vec{45}_{(\vec{15},\vec{1},\vec{1})}}-m_{\vec{210}_{(\vec{1},\vec{1},\vec{1})}}+2 m_{\vec{210}_{(\vec{15},\vec{1},\vec{1})}} \\
 m_{(\vec{8}, \vec{3}, \vec{1}, 0)} & = & m_\vec{1}+2 m_\vec{54}+m_{\vec{45}_{(\vec{15},\vec{1},\vec{1})}}-m_{\vec{210}_{(\vec{1},\vec{1},\vec{1})}}-2 m_{\vec{210}_{(\vec{15},\vec{1},\vec{1})}} \\
\end{array}$ \\
\hline
\rotatebox[origin=c]{90}{$\vec{210}'$} & $\begin{array}{lll}
 m_{(\vec{1}, \vec{2}, \vec{2})} & = & m_\vec{1}+12 m_\vec{54} \\
 m_{(\vec{1}, \vec{4}, \vec{4})} & = & m_\vec{1}+27 m_\vec{54} \\
 m_{(\vec{6}, \vec{1}, \vec{1})} & = & m_\vec{1}+2 m_\vec{54} \\
 m_{(\vec{6}, \vec{3}, \vec{3})} & = & m_\vec{1}+12 m_\vec{54} \\
 m_{(\vec{20'}, \vec{2}, \vec{2})} & = & m_\vec{1}-3 m_\vec{54} \\
 m_{(\vec{50}, \vec{1}, \vec{1})} & = & m_\vec{1}-18 m_\vec{54} \\
\end{array}$ & $\begin{array}{lll}
 m_{(\vec{1}, \vec{2}, \vec{2}, 0)} & = & m_\vec{1}+12 m_\vec{54} \\
 m_{(\vec{1}, \vec{4}, \vec{4}, 0)} & = & m_\vec{1}+27 m_\vec{54} \\
 m_{(\vec{3}, \vec{1}, \vec{1}, -2)} & = & m_\vec{1}+2 m_\vec{54} \\
 m_{(\vec{3}, \vec{3}, \vec{3}, -2)} & = & m_\vec{1}+12 m_\vec{54} \\
 m_{(\vec{6}, \vec{2}, \vec{2}, 4)} & = & m_\vec{1}-3 m_\vec{54} \\
 m_{(\vec{8}, \vec{2}, \vec{2}, 0)} & = & m_\vec{1}-3 m_\vec{54} \\
 m_{(\vec{10}, \vec{1}, \vec{1}, -6)} & = & m_\vec{1}-18 m_\vec{54} \\
 m_{(\vec{15}, \vec{1}, \vec{1}, -2)} & = & m_\vec{1}-18 m_\vec{54} \\
\end{array}$ \\
\hline
\end{tabular}
\caption{\label{tab:422and3221}
Mass of a chiral multiplet in representation $R$ of $SO(10)$ (left) under the subgroups $SU(4)\times SU(2)_L\times SU(2)_R$ (middle) and $SU(3)_C\times SU(2)_L\times SU(2)_R\times U(1)_{3 (B-L)}$ (right).}
\end{table}
\endgroup

\begingroup
\begin{table}
\begin{tabular}{l|l}
$SO(10)$ & $SU(4)\times SU(2)_L\times SU(2)_R$\\
\hline
\rotatebox[origin=c]{90}{$\vec{126}\oplus\overline{\vec{126}}$} & $\begin{array}{lll}
 m_{(\vec{6}, \vec{1}, \vec{1})} & = & \sqrt{m_\vec{1}^2+12 m_{\vec{54},A}^2+12 m_{\vec{54},B}^2-2 \sqrt{6} \left(m_{\vec{54},A}-m_{\vec{54},B}\right) \sqrt{m_\vec{1}^2+6 \left(m_{\vec{54},A}+m_{\vec{54},B}\right)^2}} \\
 m_{(\vec{6}, \vec{1}, \vec{1})'} & = & \sqrt{m_\vec{1}^2+12 m_{\vec{54},A}^2+12 m_{\vec{54},B}^2+2 \sqrt{6} \left(m_{\vec{54},A}-m_{\vec{54},B}\right) \sqrt{m_\vec{1}^2+6 \left(m_{\vec{54},A}+m_{\vec{54},B}\right)^2}} \\
 m_{(\vec{10}, \vec{1}, \vec{3})} & = & m_\vec{1}-m_{\vec{210}_{(\vec{1},\vec{1},\vec{1})}} \\
 m_{(\vec{10}, \vec{3}, \vec{1})} & = & m_\vec{1}+m_{\vec{210}_{(\vec{1},\vec{1},\vec{1})}} \\
 m_{(\vec{15}, \vec{2}, \vec{2})} & = & \sqrt{m_\vec{1}^2+3 m_{\vec{54},A}^2+3 m_{\vec{54},B}^2+\sqrt{3} \left(-m_{\vec{54},A}+m_{\vec{54},B}\right) \sqrt{2 m_\vec{1}^2+3 \left(m_{\vec{54},A}+m_{\vec{54},B}\right)^2}} \\
 m_{(\vec{15}, \vec{2}, \vec{2})'} & = & \sqrt{m_\vec{1}^2+3 m_{\vec{54},A}^2+3 m_{\vec{54},B}^2+\sqrt{3} \left(m_{\vec{54},A}-m_{\vec{54},B}\right) \sqrt{2 m_\vec{1}^2+3 \left(m_{\vec{54},A}+m_{\vec{54},B}\right)^2}} \\
\end{array}$ \\
\hline
\end{tabular}
\caption{\label{tab:126_422}
Mass of a chiral $\vec{126}\oplus\overline{\vec{126}}$ under the subgroup $SU(4)\times SU(2)_L\times SU(2)_R$.}
\end{table}
\endgroup

\begingroup
\begin{table}
\begin{tabular}{l|l}
$SO(10)$ & $SU(3)_C\times SU(2)_L\times SU(2)_R\times U(1)_{3 (B-L)}$\\
\hline
\rotatebox[origin=c]{90}{$\vec{126}\oplus\overline{\vec{126}}$} & $\begin{array}{lll}
 m_{(\vec{3}, \vec{1}, \vec{1}, -2)} & = & \left[m_\vec{1}^2+m_{\vec{45}_{(\vec{15},\vec{1},\vec{1})}}^2+12 m_{\vec{54},A}^2+12 m_{\vec{54},B}^2 \right.\\
 & & \left. -2 \sqrt{\left(m_{\vec{45}_{(\vec{15},\vec{1},\vec{1})}}^2+6 \left(m_{\vec{54},A}-m_{\vec{54},B}\right)^2\right) \left(m_\vec{1}^2+6 \left(m_{\vec{54},A}+m_{\vec{54},B}\right)^2\right)}\right]^{1/2} \\
 m_{(\vec{3}, \vec{1}, \vec{1}, -2)'} & = & \left[m_\vec{1}^2+m_{\vec{45}_{(\vec{15},\vec{1},\vec{1})}}^2+12 m_{\vec{54},A}^2+12 m_{\vec{54},B}^2 \right.\\
 & & \left. +2 \sqrt{\left(m_{\vec{45}_{(\vec{15},\vec{1},\vec{1})}}^2+6 \left(m_{\vec{54},A}-m_{\vec{54},B}\right)^2\right) \left(m_\vec{1}^2+6 \left(m_{\vec{54},A}+m_{\vec{54},B}\right)^2\right)}\right]^{1/2} \\
 m_{(\vec{1}, \vec{1}, \vec{3}, -6)} & = & m_\vec{1}+3 m_{\vec{45}_{(\vec{15},\vec{1},\vec{1})}}-m_{\vec{210}_{(\vec{1},\vec{1},\vec{1})}}+3 m_{\vec{210}_{(\vec{15},\vec{1},\vec{1})}} \\
 m_{(\vec{3}, \vec{1}, \vec{3}, -2)} & = & m_\vec{1}+m_{\vec{45}_{(\vec{15},\vec{1},\vec{1})}}-m_{\vec{210}_{(\vec{1},\vec{1},\vec{1})}}+m_{\vec{210}_{(\vec{15},\vec{1},\vec{1})}} \\
 m_{(\vec{6}, \vec{1}, \vec{3}, -2)} & = & m_\vec{1}-m_{\vec{45}_{(\vec{15},\vec{1},\vec{1})}}-m_{\vec{210}_{(\vec{1},\vec{1},\vec{1})}}-m_{\vec{210}_{(\vec{15},\vec{1},\vec{1})}} \\
 m_{(\vec{1}, \vec{3}, \vec{1}, 6)} & = & m_\vec{1}-3 m_{\vec{45}_{(\vec{15},\vec{1},\vec{1})}}+m_{\vec{210}_{(\vec{1},\vec{1},\vec{1})}}+3 m_{\vec{210}_{(\vec{15},\vec{1},\vec{1})}} \\
 m_{(\vec{3}, \vec{3}, \vec{1}, -2)} & = & m_\vec{1}-m_{\vec{45}_{(\vec{15},\vec{1},\vec{1})}}+m_{\vec{210}_{(\vec{1},\vec{1},\vec{1})}}+m_{\vec{210}_{(\vec{15},\vec{1},\vec{1})}} \\
 m_{(\vec{6}, \vec{3}, \vec{1}, -2)} & = & m_\vec{1}+m_{\vec{45}_{(\vec{15},\vec{1},\vec{1})}}+m_{\vec{210}_{(\vec{1},\vec{1},\vec{1})}}-m_{\vec{210}_{(\vec{15},\vec{1},\vec{1})}} \\
 m_{(\vec{1}, \vec{2}, \vec{2}, 0)} & = & \left[3 m_{\vec{54},A}^2+3 m_{\vec{54},B}^2+\left(m_\vec{1}+2 m_{\vec{210}_{(\vec{15},\vec{1},\vec{1})}}\right)^2 \right.\\
 & & \left. +\sqrt{3} \left(-m_{\vec{54},A}+m_{\vec{54},B}\right) \sqrt{3 \left(m_{\vec{54},A}+m_{\vec{54},B}\right)^2+2 \left(m_\vec{1}+2 m_{\vec{210}_{(\vec{15},\vec{1},\vec{1})}}\right)^2}\right]^{1/2} \\
 m_{(\vec{8}, \vec{2}, \vec{2}, 0)} & = & \left[3 m_{\vec{54},A}^2+3 m_{\vec{54},B}^2 +\left(m_\vec{1}-m_{\vec{210}_{(\vec{15},\vec{1},\vec{1})}}\right)^2 \right.\\
 & & \left. +\sqrt{3} \left(-m_{\vec{54},A}+m_{\vec{54},B}\right) \sqrt{3 \left(m_{\vec{54},A}+m_{\vec{54},B}\right)^2+2 \left(m_\vec{1}-m_{\vec{210}_{(\vec{15},\vec{1},\vec{1})}}\right)^2}\right]^{1/2} \\
 m_{(\vec{3}, \vec{2}, \vec{2}, 4)} & = & \left[4 m_{\vec{45}_{(\vec{15},\vec{1},\vec{1})}}^2+3 m_{\vec{54},A}^2+3 m_{\vec{54},B}^2+\left(m_\vec{1}+m_{\vec{210}_{(\vec{15},\vec{1},\vec{1})}}\right)^2 \right.\\
 & & \left. -\sqrt{\left(8 m_{\vec{45}_{(\vec{15},\vec{1},\vec{1})}}^2+3 \left(m_{\vec{54},A}-m_{\vec{54},B}\right)^2\right) \left(3 \left(m_{\vec{54},A}+m_{\vec{54},B}\right)^2+2 \left(m_\vec{1}+m_{\vec{210}_{(\vec{15},\vec{1},\vec{1})}}\right)^2\right)}\right]^{1/2} \\
 m_{(\vec{1}, \vec{2}, \vec{2}, 0)'} & = & \left[3 m_{\vec{54},A}^2+3 m_{\vec{54},B}^2+\left(m_\vec{1}+2 m_{\vec{210}_{(\vec{15},\vec{1},\vec{1})}}\right)^2 \right.\\
 & & \left. +\sqrt{3} \left(m_{\vec{54},A}-m_{\vec{54},B}\right) \sqrt{3 \left(m_{\vec{54},A}+m_{\vec{54},B}\right)^2+2 \left(m_\vec{1}+2 m_{\vec{210}_{(\vec{15},\vec{1},\vec{1})}}\right)^2}\right]^{1/2} \\
 m_{(\vec{8}, \vec{2}, \vec{2}, 0)'} & = & \left[3 m_{\vec{54},A}^2+3 m_{\vec{54},B}^2 +\left(m_\vec{1}-m_{\vec{210}_{(\vec{15},\vec{1},\vec{1})}}\right)^2 \right.\\
 & & \left. +\sqrt{3} \left(m_{\vec{54},A}-m_{\vec{54},B}\right) \sqrt{3 \left(m_{\vec{54},A}+m_{\vec{54},B}\right)^2+2 \left(m_\vec{1}-m_{\vec{210}_{(\vec{15},\vec{1},\vec{1})}}\right)^2}\right]^{1/2} \\
 m_{(\vec{3}, \vec{2}, \vec{2}, 4)'} & = & \left[4 m_{\vec{45}_{(\vec{15},\vec{1},\vec{1})}}^2+3 m_{\vec{54},A}^2+3 m_{\vec{54},B}^2+\left(m_\vec{1}+m_{\vec{210}_{(\vec{15},\vec{1},\vec{1})}}\right)^2 \right.\\
 & & \left. +\sqrt{\left(8 m_{\vec{45}_{(\vec{15},\vec{1},\vec{1})}}^2+3 \left(m_{\vec{54},A}-m_{\vec{54},B}\right)^2\right) \left(3 \left(m_{\vec{54},A}+m_{\vec{54},B}\right)^2+2 \left(m_\vec{1}+m_{\vec{210}_{(\vec{15},\vec{1},\vec{1})}}\right)^2\right)}\right]^{1/2} \\
\end{array}$ \\
\hline
\end{tabular}
\caption{\label{tab:126_3221}
Mass of a chiral $\vec{126}\oplus\overline{\vec{126}}$ under the subgroup $SU(3)_C\times SU(2)_L\times SU(2)_R\times U(1)_{3 (B-L)}$.}
\end{table}
\endgroup

\clearpage

\subsection{$G_{421}$ and $G_{3211}$}

The same procedure can be applied to the $SO(10)$ subgroups $G_{421}$ and $G_{3211}$, which follow from the previous ones by breaking $SU(2)_R\to U(1)_R$. This introduces additional VEVs and makes the expressions very lengthy. Once again we pick $U(1)$ normalizations that avoid fractions, namely $U(1)_{3 (B-L)}$ and $U(1)_{2 R}$.

For the group $SU(3)_C\times SU(2)_L\times U(1)_{2 R}\times U(1)_{3 (B-L)}$, one ends up with $3\times 3$ matrices for some of the representations, which are tedious to diagonalize analytically. Instead, we give the matrices here. For the $\vec{210}$, one finds three representations with quantum numbers $(\vec{1}, \vec{1}, 0, 0)$, which share a symmetric mass matrix with entries
\begin{align}
\mathcal{M}_{11} & = m_\vec{1}-\frac{12 m_{\vec{45}_{(\vec{15},\vec{1},\vec{1})}}}{7}+\frac{12
   m_{\vec{210}_{(\vec{15},\vec{1},\vec{1})}}}{7}+\frac{24 m_\vec{54}}{7} \,,\\
\mathcal{M}_{12} & = \frac{2
   m_{\vec{45}_{(\vec{1},\vec{1},\vec{3})}}}{\sqrt{13}}-\frac{5 m_{\vec{45}_{(\vec{15},\vec{1},\vec{1})}}}{7
   \sqrt{13}}+\frac{40 m_{\vec{210}_{(\vec{15},\vec{1},\vec{1})}}}{7 \sqrt{13}}-\frac{200 m_\vec{54}}{7
   \sqrt{13}}\,,\\
\mathcal{M}_{13} & = 5 \sqrt{\frac{2}{91}} m_{\vec{45}_{(\vec{1},\vec{1},\vec{3})}}+\sqrt{\frac{2}{91}}
   m_{\vec{45}_{(\vec{15},\vec{1},\vec{1})}}-8 \sqrt{\frac{2}{91}} m_{\vec{210}_{(\vec{15},\vec{1},\vec{1})}}+40
   \sqrt{\frac{2}{91}} m_\vec{54}\,,\\
\mathcal{M}_{22} & = m_\vec{1}+\frac{14 m_{\vec{210}_{(\vec{1},\vec{1},\vec{1})}}}{39}+\frac{40
   m_{\vec{45}_{(\vec{1},\vec{1},\vec{3})}}}{39}+\frac{496 m_{\vec{45}_{(\vec{15},\vec{1},\vec{1})}}}{273}+\frac{264
   m_{\vec{210}_{(\vec{15},\vec{1},\vec{1})}}}{91}-\frac{140 m_{\vec{210}_{(\vec{15},\vec{1},\vec{3})}}}{39}+\frac{296
   m_\vec{54}}{273} \,,\\
\mathcal{M}_{23} & = \frac{5}{39} \sqrt{14} m_{\vec{210}_{(\vec{1},\vec{1},\vec{1})}}+\frac{22}{39}
   \sqrt{\frac{2}{7}} m_{\vec{45}_{(\vec{1},\vec{1},\vec{3})}}+\frac{10}{39} \sqrt{\frac{2}{7}}
   m_{\vec{45}_{(\vec{15},\vec{1},\vec{1})}}\\
   &\quad +\frac{20}{13} \sqrt{\frac{2}{7}}
   m_{\vec{210}_{(\vec{15},\vec{1},\vec{1})}}-\frac{11}{39} \sqrt{14}
   m_{\vec{210}_{(\vec{15},\vec{1},\vec{3})}}+\frac{50}{39} \sqrt{\frac{2}{7}} m_\vec{54}\,,\\
\mathcal{M}_{33} & = m_\vec{1}+\frac{25
   m_{\vec{210}_{(\vec{1},\vec{1},\vec{1})}}}{39}-\frac{40 m_{\vec{45}_{(\vec{1},\vec{1},\vec{3})}}}{39}+\frac{74
   m_{\vec{45}_{(\vec{15},\vec{1},\vec{1})}}}{39}+\frac{44 m_{\vec{210}_{(\vec{15},\vec{1},\vec{1})}}}{13}+\frac{140
   m_{\vec{210}_{(\vec{15},\vec{1},\vec{3})}}}{39}+\frac{58 m_\vec{54}}{39} \,,
\end{align}
with singular values denoted by $\mathcal{M}_s^A$, $\mathcal{M}_s^B$, $\mathcal{M}_s^C$ that can be calculated straightforwardly from the given matrix.
Similarly, the representation $\vec{126}\oplus\overline{\vec{126}}$ contains three copies of $(\vec{3}, \vec{1}, 0, -2)$ under $SU(3)_C\times SU(2)_L\times U(1)_{2 R}\times U(1)_{3 (B-L)}$, which have a mass matrix with entries
\begin{align}
\mathcal{M}_{11} & = m_\vec{1}-\frac{m_{\vec{210}_{(\vec{1},\vec{1},\vec{1})}}}{3}+m_{\vec{45}_{(\vec{15},\vec{1},\vec{1})}}+\frac{m_{\vec{210}_{(\vec{15},\vec{1},\vec{1})}}}{3}+\frac{8 m_{\vec{210}_{(\vec{15},\vec{1},\vec{3})}}}{3} \,,\\
\mathcal{M}_{12} & =  -\frac{1}{3} \sqrt{2} m_{\vec{210}_{(\vec{1},\vec{1},\vec{1})}}+\frac{1}{3} \sqrt{2} m_{\vec{210}_{(\vec{15},\vec{1},\vec{1})}}+\frac{2}{3}\sqrt{2} m_{\vec{210}_{(\vec{15},\vec{1},\vec{3})}}\,,\\
\mathcal{M}_{13} & = -4 m_{\vec{54},A} \,,\\
\mathcal{M}_{21} & =  -\frac{1}{3} \sqrt{2} m_{\vec{210}_{(\vec{1},\vec{1},\vec{1})}}+\frac{1}{3} \sqrt{2} m_{\vec{210}_{(\vec{15},\vec{1},\vec{1})}}+\frac{2}{3} \sqrt{2} m_{\vec{210}_{(\vec{15},\vec{1},\vec{3})}}\,,\\
\mathcal{M}_{22} & =  m_\vec{1}-\frac{2 m_{\vec{210}_{(\vec{1},\vec{1},\vec{1})}}}{3}+m_{\vec{45}_{(\vec{15},\vec{1},\vec{1})}}+\frac{2 m_{\vec{210}_{(\vec{15},\vec{1},\vec{1})}}}{3}-\frac{8 m_{\vec{210}_{(\vec{15},\vec{1},\vec{3})}}}{3}\,,\\
\mathcal{M}_{23} & =  2 \sqrt{2} m_{\vec{54},A} \,,\\
\mathcal{M}_{31} & = 4 m_{\vec{54},B} \,,\\
\mathcal{M}_{32} & = -2 \sqrt{2} m_{\vec{54},B}  \,,\\
\mathcal{M}_{33} & = m_\vec{1}-m_{\vec{45}_{(\vec{15},\vec{1},\vec{1})}} \,,
\end{align}
and singular values that we denote as $\mathcal{M}_t^A$, $\mathcal{M}_t^B$, and $\mathcal{M}_t^C$.

Tab.~\ref{tab:421and3211_10_54_210prime} lists the chiral representations $\vec{10}$, $\vec{54}$, and $\vec{210}'$  together with their tree-level mass splittings due to VEVs down to the subgroups $G_{421}$ and $G_{3211}$.
Tab.~\ref{tab:421and3211_45_120} gives the same for $\vec{45}$ and $\vec{120}$; Tab.~\ref{tab:421and3211_210} for $\vec{210}$.
The considerably lengthier representation $\vec{126}\oplus\overline{\vec{126}}$ is listed in Tab.~\ref{tab:126_421} for $G_{421}$ and in Tab.~\ref{tab:126_3211} for $G_{3211}$.

\begingroup
\begin{table}
\begin{tabular}{l|l|l}
$SO(10)$ & $SU(4)\times SU(2)_L\times U(1)_{2 R}$ & $SU(3)_C\times SU(2)_L\times U(1)_{2 R}\times U(1)_{3 (B-L)}$\\
\hline
\rotatebox[origin=c]{90}{$\vec{10}$} & $\begin{array}{lll}
 m_{(\vec{1}, \vec{2}, 1)} & = & m_\vec{1}+3 m_\vec{54} \\
 m_{(\vec{6}, \vec{1}, 0)} & = & m_\vec{1}-2 m_\vec{54} \\
\end{array}$ & $\begin{array}{lll}
 m_{(\vec{1}, \vec{2}, 1, 0)} & = & m_\vec{1}+3 m_\vec{54} \\
 m_{(\vec{3}, \vec{1}, 0, -2)} & = & m_\vec{1}-2 m_\vec{54} \\
\end{array}$ \\
\hline
\rotatebox[origin=c]{90}{$\vec{54}$} & $\begin{array}{lll}
 m_{(\vec{1}, \vec{1}, 0)} & = & m_\vec{1}+2 m_\vec{54} \\
 m_{(\vec{1}, \vec{3}, 0)} & = & m_\vec{1}+6 m_\vec{54} \\
 m_{(\vec{1}, \vec{3}, 2)} & = & m_\vec{1}+6 m_\vec{54} \\
 m_{(\vec{6}, \vec{2}, 1)} & = & m_\vec{1}+m_\vec{54} \\
 m_{(\vec{20'}, \vec{1}, 0)} & = & m_\vec{1}-4 m_\vec{54} \\
\end{array}$ & $\begin{array}{lll}
 m_{(\vec{1}, \vec{1}, 0, 0)} & = & m_\vec{1}+2 m_\vec{54} \\
 m_{(\vec{1}, \vec{3}, 0, 0)} & = & m_\vec{1}+6 m_\vec{54} \\
 m_{(\vec{1}, \vec{3}, 2, 0)} & = & m_\vec{1}+6 m_\vec{54} \\
 m_{(\vec{3}, \vec{2}, -1, -2)} & = & m_\vec{1}+m_\vec{54} \\
 m_{(\vec{3}, \vec{2}, 1, -2)} & = & m_\vec{1}+m_\vec{54} \\
 m_{(\vec{6}, \vec{1}, 0, 4)} & = & m_\vec{1}-4 m_\vec{54} \\
 m_{(\vec{8}, \vec{1}, 0, 0)} & = & m_\vec{1}-4 m_\vec{54} \\
\end{array}$ \\
\hline
\rotatebox[origin=c]{90}{$\vec{210}'$} & $\begin{array}{lll}
 m_{(\vec{1}, \vec{2}, 1)} & = & m_\vec{1}+12 m_\vec{54} \\
 m_{(\vec{1}, \vec{4}, 1)} & = & m_\vec{1}+27 m_\vec{54} \\
 m_{(\vec{1}, \vec{4}, 3)} & = & m_\vec{1}+27 m_\vec{54} \\
 m_{(\vec{6}, \vec{1}, 0)} & = & m_\vec{1}+2 m_\vec{54} \\
 m_{(\vec{6}, \vec{3}, 0)} & = & m_\vec{1}+12 m_\vec{54} \\
 m_{(\vec{6}, \vec{3}, 2)} & = & m_\vec{1}+12 m_\vec{54} \\
 m_{(\vec{20'}, \vec{2}, 1)} & = & m_\vec{1}-3 m_\vec{54} \\
 m_{(\vec{50}, \vec{1}, 0)} & = & m_\vec{1}-18 m_\vec{54} \\
\end{array}$ & $\begin{array}{lll}
 m_{(\vec{1}, \vec{2}, 1, 0)} & = & m_\vec{1}+12 m_\vec{54} \\
 m_{(\vec{1}, \vec{4}, 1, 0)} & = & m_\vec{1}+27 m_\vec{54} \\
 m_{(\vec{1}, \vec{4}, 3, 0)} & = & m_\vec{1}+27 m_\vec{54} \\
 m_{(\vec{3}, \vec{1}, 0, -2)} & = & m_\vec{1}+2 m_\vec{54} \\
 m_{(\vec{3}, \vec{3}, 0, -2)} & = & m_\vec{1}+12 m_\vec{54} \\
 m_{(\vec{3}, \vec{3}, -2, -2)} & = & m_\vec{1}+12 m_\vec{54} \\
 m_{(\vec{3}, \vec{3}, 2, -2)} & = & m_\vec{1}+12 m_\vec{54} \\
 m_{(\vec{6}, \vec{2}, -1, 4)} & = & m_\vec{1}-3 m_\vec{54} \\
 m_{(\vec{6}, \vec{2}, 1, 4)} & = & m_\vec{1}-3 m_\vec{54} \\
 m_{(\vec{8}, \vec{2}, 1, 0)} & = & m_\vec{1}-3 m_\vec{54} \\
 m_{(\vec{10}, \vec{1}, 0, -6)} & = & m_\vec{1}-18 m_\vec{54} \\
 m_{(\vec{15}, \vec{1}, 0, -2)} & = & m_\vec{1}-18 m_\vec{54} \\
\end{array}$ \\
\hline
\end{tabular}
\caption{\label{tab:421and3211_10_54_210prime}
Mass of a chiral multiplet in representation $R$ of $SO(10)$ (left) under the subgroups $SU(4)\times SU(2)_L\times U(1)_{2 R}$ (middle) and $SU(3)_C\times SU(2)_L\times U(1)_{2 R}\times U(1)_{3 (B-L)}$ (right).}
\end{table}
\endgroup

\begingroup
\squeezetable
\begin{table}
\begin{tabular}{l|l}
$SO(10)$ & $SU(4)\times SU(2)_L\times U(1)_{2 R}$\\
\hline
\rotatebox[origin=c]{90}{$\vec{126}\oplus\overline{\vec{126}}$} & $\begin{array}{lll}
 m_{(\vec{6}, \vec{1}, 0)} & = & \sqrt{m_\vec{1}^2+12 m_{\vec{54},A}^2+12 m_{\vec{54},B}^2-2 \sqrt{6} \left(m_{\vec{54},A}-m_{\vec{54},B}\right) \sqrt{m_\vec{1}^2+6 \left(m_{\vec{54},A}+m_{\vec{54},B}\right)^2}} \\
 m_{(\vec{6}, \vec{1}, 0)'} & = & \sqrt{m_\vec{1}^2+12 m_{\vec{54},A}^2+12 m_{\vec{54},B}^2+2 \sqrt{6} \left(m_{\vec{54},A}-m_{\vec{54},B}\right) \sqrt{m_\vec{1}^2+6 \left(m_{\vec{54},A}+m_{\vec{54},B}\right)^2}} \\
 m_{(\vec{10}, \vec{1}, -2)} & = & m_\vec{1}-2 m_{\vec{45}_{(\vec{1},\vec{1},\vec{3})}}-m_{\vec{210}_{(\vec{1},\vec{1},\vec{1})}} \\
 m_{(\vec{10}, \vec{1}, 0)} & = & m_\vec{1}-m_{\vec{210}_{(\vec{1},\vec{1},\vec{1})}} \\
 m_{(\vec{10}, \vec{1}, 2)} & = & m_\vec{1}+2 m_{\vec{45}_{(\vec{1},\vec{1},\vec{3})}}-m_{\vec{210}_{(\vec{1},\vec{1},\vec{1})}} \\
 m_{(\vec{10}, \vec{3}, 0)} & = & m_\vec{1}+m_{\vec{210}_{(\vec{1},\vec{1},\vec{1})}} \\
 m_{(\vec{15}, \vec{2}, 1)} & = & \sqrt{m_\vec{1}^2+m_{\vec{45}_{(\vec{1},\vec{1},\vec{3})}}^2+3 m_{\vec{54},A}^2+3 m_{\vec{54},B}^2-\sqrt{\left(2 m_{\vec{45}_{(\vec{1},\vec{1},\vec{3})}}^2+3 \left(m_{\vec{54},A}-m_{\vec{54},B}\right)^2\right) \left(2 m_\vec{1}^2+3 \left(m_{\vec{54},A}+m_{\vec{54},B}\right)^2\right)}} \\
 m_{(\vec{15}, \vec{2}, 1)'} & = & \sqrt{m_\vec{1}^2+m_{\vec{45}_{(\vec{1},\vec{1},\vec{3})}}^2+3 m_{\vec{54},A}^2+3 m_{\vec{54},B}^2+\sqrt{\left(2 m_{\vec{45}_{(\vec{1},\vec{1},\vec{3})}}^2+3 \left(m_{\vec{54},A}-m_{\vec{54},B}\right)^2\right) \left(2 m_\vec{1}^2+3 \left(m_{\vec{54},A}+m_{\vec{54},B}\right)^2\right)}} \\
\end{array}$ \\
\hline
\end{tabular}
\caption{\label{tab:126_421}
Mass of a chiral $\vec{126}\oplus\overline{\vec{126}}$ under the subgroup $SU(4)\times SU(2)_L\times U(1)_{2 R}$.}
\end{table}
\endgroup

\begin{turnpage}
\begingroup
\begin{table}
\begin{tabular}{l|l|l}
$SO(10)$ & $SU(4)\times SU(2)_L\times U(1)_{2 R}$ & $SU(3)_C\times SU(2)_L\times U(1)_{2 R}\times U(1)_{3 (B-L)}$\\
\hline
\rotatebox[origin=c]{90}{$\vec{45}$} & $\begin{array}{lll}
 m_{(\vec{1}, \vec{1}, 0)} & = & m_\vec{1}-6 m_\vec{54}-m_{\vec{210}_{(\vec{1},\vec{1},\vec{1})}} \\
 m_{(\vec{1}, \vec{1}, 2)} & = & m_\vec{1}-6 m_\vec{54}-m_{\vec{210}_{(\vec{1},\vec{1},\vec{1})}} \\
 m_{(\vec{1}, \vec{3}, 0)} & = & m_\vec{1}-6 m_\vec{54}+m_{\vec{210}_{(\vec{1},\vec{1},\vec{1})}} \\
 m_{(\vec{6}, \vec{2}, 1)} & = & m_\vec{1}-m_\vec{54} \\
 m_{(\vec{15}, \vec{1}, 0)} & = & m_\vec{1}+4 m_\vec{54} \\
\end{array}$ & $\begin{array}{lll}
 m_{(\vec{1}, \vec{1}, 0, 0)} & = & m_\vec{1}-m_\vec{54}-\frac{m_{\vec{210}_{(\vec{1},\vec{1},\vec{1})}}}{2}-m_{\vec{210}_{(\vec{15},\vec{1},\vec{1})}} \\
 & & -\frac{1}{2} \sqrt{\left(10 m_\vec{54}+m_{\vec{210}_{(\vec{1},\vec{1},\vec{1})}}-2 m_{\vec{210}_{(\vec{15},\vec{1},\vec{1})}}\right)^2+24 m_{\vec{210}_{(\vec{15},\vec{1},\vec{3})}}^2} \\
 m_{(\vec{1}, \vec{1}, 2, 0)} & = & m_\vec{1}-6 m_\vec{54}-m_{\vec{210}_{(\vec{1},\vec{1},\vec{1})}} \\
 m_{(\vec{1}, \vec{3}, 0, 0)} & = & m_\vec{1}-6 m_\vec{54}+m_{\vec{210}_{(\vec{1},\vec{1},\vec{1})}} \\
 m_{(\vec{3}, \vec{2}, -1, -2)} & = & m_\vec{1}-m_\vec{54}+m_{\vec{210}_{(\vec{15},\vec{1},\vec{3})}} \\
 m_{(\vec{3}, \vec{2}, 1, -2)} & = & m_\vec{1}-m_\vec{54}-m_{\vec{210}_{(\vec{15},\vec{1},\vec{3})}} \\
 m_{(\vec{1}, \vec{1}, 0, 0)'} & = & m_\vec{1}-m_\vec{54}-\frac{m_{\vec{210}_{(\vec{1},\vec{1},\vec{1})}}}{2}-m_{\vec{210}_{(\vec{15},\vec{1},\vec{1})}} \\
 & & +\frac{1}{2} \sqrt{\left(10 m_\vec{54}+m_{\vec{210}_{(\vec{1},\vec{1},\vec{1})}}-2 m_{\vec{210}_{(\vec{15},\vec{1},\vec{1})}}\right)^2+24 m_{\vec{210}_{(\vec{15},\vec{1},\vec{3})}}^2} \\
 m_{(\vec{3}, \vec{1}, 0, 4)} & = & m_\vec{1}+4 m_\vec{54}-m_{\vec{210}_{(\vec{15},\vec{1},\vec{1})}} \\
 m_{(\vec{8}, \vec{1}, 0, 0)} & = & m_\vec{1}+4 m_\vec{54}+m_{\vec{210}_{(\vec{15},\vec{1},\vec{1})}} \\
\end{array}$ \\
\hline
\rotatebox[origin=c]{90}{$\vec{120}$} & $\begin{array}{lll}
 m_{(\vec{1}, \vec{2}, 1)} & = & m_\vec{1}-9 m_\vec{54} \\
 m_{(\vec{6}, \vec{1}, 0)} & = & m_\vec{1}-4 m_\vec{54}-m_{\vec{210}_{(\vec{1},\vec{1},\vec{1})}} \\
 m_{(\vec{6}, \vec{1}, 2)} & = & m_\vec{1}-4 m_\vec{54}-m_{\vec{210}_{(\vec{1},\vec{1},\vec{1})}} \\
 m_{(\vec{6}, \vec{3}, 0)} & = & m_\vec{1}-4 m_\vec{54}+m_{\vec{210}_{(\vec{1},\vec{1},\vec{1})}} \\
 m_{(\vec{10}, \vec{1}, 0)} & = & m_\vec{1}+6 m_\vec{54} \\
 m_{(\vec{15}, \vec{2}, 1)} & = & m_\vec{1}+m_\vec{54} \\
\end{array}$ & $\begin{array}{lll}
 m_{(\vec{1}, \vec{2}, 1, 0)} & = & -5 m_\vec{54}+m_{\vec{210}_{(\vec{15},\vec{1},\vec{1})}} \\
 & & +\frac{1}{2} \sqrt{4 \left(-m_\vec{1}+4 m_\vec{54}+m_{\vec{210}_{(\vec{15},\vec{1},\vec{1})}}\right)^2-3 m_{\vec{210}_{(\vec{15},\vec{1},\vec{3})}}^2} \\
 m_{(\vec{3}, \vec{1}, 0, -2)} & = & m_\vec{1}+m_\vec{54}-\frac{m_{\vec{210}_{(\vec{1},\vec{1},\vec{1})}}}{2}-\frac{m_{\vec{210}_{(\vec{15},\vec{1},\vec{1})}}}{2} \\
 & & -\frac{1}{2} \sqrt{\left(10 m_\vec{54}+m_{\vec{210}_{(\vec{1},\vec{1},\vec{1})}}-m_{\vec{210}_{(\vec{15},\vec{1},\vec{1})}}\right)^2+4 m_{\vec{210}_{(\vec{15},\vec{1},\vec{3})}}^2} \\
 m_{(\vec{3}, \vec{1}, -2, -2)} & = & m_\vec{1}-4 m_\vec{54}-m_{\vec{210}_{(\vec{1},\vec{1},\vec{1})}}+m_{\vec{210}_{(\vec{15},\vec{1},\vec{3})}} \\
 m_{(\vec{3}, \vec{1}, 2, -2)} & = & m_\vec{1}-4 m_\vec{54}-m_{\vec{210}_{(\vec{1},\vec{1},\vec{1})}}-m_{\vec{210}_{(\vec{15},\vec{1},\vec{3})}} \\
 m_{(\vec{3}, \vec{3}, 0, -2)} & = & m_\vec{1}-4 m_\vec{54}+m_{\vec{210}_{(\vec{1},\vec{1},\vec{1})}} \\
 m_{(\vec{1}, \vec{1}, 0, -6)} & = & m_\vec{1}+6 m_\vec{54}-3 m_{\vec{210}_{(\vec{15},\vec{1},\vec{1})}} \\
 m_{(\vec{3}, \vec{1}, 0, -2)'} & = & m_\vec{1}+m_\vec{54}-\frac{m_{\vec{210}_{(\vec{1},\vec{1},\vec{1})}}}{2}-\frac{m_{\vec{210}_{(\vec{15},\vec{1},\vec{1})}}}{2} \\
 & & +\frac{1}{2} \sqrt{\left(10 m_\vec{54}+m_{\vec{210}_{(\vec{1},\vec{1},\vec{1})}}-m_{\vec{210}_{(\vec{15},\vec{1},\vec{1})}}\right)^2+4 m_{\vec{210}_{(\vec{15},\vec{1},\vec{3})}}^2} \\
 m_{(\vec{6}, \vec{1}, 0, -2)} & = & m_\vec{1}+6 m_\vec{54}+m_{\vec{210}_{(\vec{15},\vec{1},\vec{1})}} \\
 m_{(\vec{1}, \vec{2}, 1, 0)'} & = & 5 m_\vec{54}-m_{\vec{210}_{(\vec{15},\vec{1},\vec{1})}} \\
 & & +\frac{1}{2} \sqrt{4 \left(-m_\vec{1}+4 m_\vec{54}+m_{\vec{210}_{(\vec{15},\vec{1},\vec{1})}}\right)^2-3 m_{\vec{210}_{(\vec{15},\vec{1},\vec{3})}}^2} \\
 m_{(\vec{3}, \vec{2}, -1, 4)} & = & m_\vec{1}+m_\vec{54}-m_{\vec{210}_{(\vec{15},\vec{1},\vec{1})}}-m_{\vec{210}_{(\vec{15},\vec{1},\vec{3})}} \\
 m_{(\vec{3}, \vec{2}, 1, 4)} & = & m_\vec{1}+m_\vec{54}-m_{\vec{210}_{(\vec{15},\vec{1},\vec{1})}}+m_{\vec{210}_{(\vec{15},\vec{1},\vec{3})}} \\
 m_{(\vec{8}, \vec{2}, 1, 0)} & = & m_\vec{1}+m_\vec{54}+m_{\vec{210}_{(\vec{15},\vec{1},\vec{1})}} \\
\end{array}$ \\
\hline
\end{tabular}
\caption{\label{tab:421and3211_45_120}
Mass of a chiral multiplet in representation $R$ of $SO(10)$ (left) under the subgroups $G_{421}$ (middle) and $G_{3211}$ (right).}
\end{table}
\endgroup
\end{turnpage}

\begin{turnpage}
\begingroup
\squeezetable
\begin{table}
\begin{tabular}{l|l|l}
$SO(10)$ & $SU(4)\times SU(2)_L\times U(1)_{2 R}$ & $SU(3)_C\times SU(2)_L\times U(1)_{2 R}\times U(1)_{3 (B-L)}$\\
\hline
\rotatebox[origin=c]{90}{$\vec{210}$} & $\begin{array}{lll}
 m_{(\vec{1}, \vec{1}, 0)} & = & m_\vec{1}+12 m_\vec{54} \\
 m_{(\vec{6}, \vec{2}, 1)} & = & m_\vec{1}+7 m_\vec{54} \\
 m_{(\vec{10}, \vec{2}, 1)} & = & m_\vec{1}-3 m_\vec{54}-m_{\vec{45}_{(\vec{1},\vec{1},\vec{3})}} \\
 m_{(\vec{10}, \vec{2}, -1)} & = & m_\vec{1}-3 m_\vec{54}+m_{\vec{45}_{(\vec{1},\vec{1},\vec{3})}} \\
 m_{(\vec{15}, \vec{1}, 0)'} & = & m_\vec{1}-3 m_\vec{54}+\frac{m_{\vec{210}_{(\vec{1},\vec{1},\vec{1})}}}{2} \\
 & & -\frac{\sqrt{8 m_{\vec{45}_{(\vec{1},\vec{1},\vec{3})}}^2+\left(10 m_\vec{54}+m_{\vec{210}_{(\vec{1},\vec{1},\vec{1})}}\right)^2}}{2}  \\
 m_{(\vec{15}, \vec{1}, 0)} & = & m_\vec{1}-3 m_\vec{54}+\frac{m_{\vec{210}_{(\vec{1},\vec{1},\vec{1})}}}{2} \\
 & & +\frac{\sqrt{8 m_{\vec{45}_{(\vec{1},\vec{1},\vec{3})}}^2+\left(10 m_\vec{54}+m_{\vec{210}_{(\vec{1},\vec{1},\vec{1})}}\right)^2}}{2}  \\
 m_{(\vec{15}, \vec{1}, 2)} & = & m_\vec{1}+2 m_\vec{54}+m_{\vec{210}_{(\vec{1},\vec{1},\vec{1})}} \\
 m_{(\vec{15}, \vec{3}, 0)} & = & m_\vec{1}+2 m_\vec{54}-m_{\vec{210}_{(\vec{1},\vec{1},\vec{1})}} \\
\end{array}$ & $\begin{array}{lll}
 m_{(\vec{1}, \vec{1}, 0, 0)} & = & \mathcal{M}_s^A \\
 m_{(\vec{3}, \vec{2}, -1, -2)} & = & 5 m_\vec{54}-\frac{m_{\vec{45}_{(\vec{1},\vec{1},\vec{3})}}}{2}-m_{\vec{210}_{(\vec{15},\vec{1},\vec{1})}} \\
 & & +\frac{1}{2} \left[\left(2 m_\vec{1}+4 m_\vec{54}+m_{\vec{45}_{(\vec{1},\vec{1},\vec{3})}}+2 m_{\vec{210}_{(\vec{15},\vec{1},\vec{1})}}\right)^2 -4 m_{\vec{210}_{(\vec{15},\vec{1},\vec{3})}}^2-8 m_{\vec{45}_{(\vec{15},\vec{1},\vec{1})}}^2\right.\\
 & & \left.-4 \left(2 m_\vec{1}+4 m_\vec{54}+m_{\vec{45}_{(\vec{1},\vec{1},\vec{3})}}-4 m_{\vec{45}_{(\vec{15},\vec{1},\vec{1})}}+2 m_{\vec{210}_{(\vec{15},\vec{1},\vec{1})}}\right) m_{\vec{210}_{(\vec{15},\vec{1},\vec{3})}}\right]^{1/2} \\
 m_{(\vec{3}, \vec{2}, 1, -2)} & = & 5 m_\vec{54}+\frac{m_{\vec{45}_{(\vec{1},\vec{1},\vec{3})}}}{2}-m_{\vec{210}_{(\vec{15},\vec{1},\vec{1})}} \\
 & & +\frac{1}{2} \left[\left(2 m_\vec{1}+4 m_\vec{54}-m_{\vec{45}_{(\vec{1},\vec{1},\vec{3})}}+2 m_{\vec{210}_{(\vec{15},\vec{1},\vec{1})}}\right)^2-4 m_{\vec{210}_{(\vec{15},\vec{1},\vec{3})}}^2-8 m_{\vec{45}_{(\vec{15},\vec{1},\vec{1})}}^2\right.\\
 & & \left.+4 \left(2 m_\vec{1}+4 m_\vec{54}-m_{\vec{45}_{(\vec{1},\vec{1},\vec{3})}}-4 m_{\vec{45}_{(\vec{15},\vec{1},\vec{1})}}+2 m_{\vec{210}_{(\vec{15},\vec{1},\vec{1})}}\right) m_{\vec{210}_{(\vec{15},\vec{1},\vec{3})}}\right]^{1/2} \\
 m_{(\vec{1}, \vec{2}, 1, -6)} & = & m_\vec{1}-3 m_\vec{54}-m_{\vec{45}_{(\vec{1},\vec{1},\vec{3})}}+6 m_{\vec{210}_{(\vec{15},\vec{1},\vec{1})}}+3 m_{\vec{210}_{(\vec{15},\vec{1},\vec{3})}} \\
 m_{(\vec{3}, \vec{2}, 1, -2)'} & = & -5 m_\vec{54}-\frac{m_{\vec{45}_{(\vec{1},\vec{1},\vec{3})}}}{2}+m_{\vec{210}_{(\vec{15},\vec{1},\vec{1})}} \\
 & & +\frac{1}{2} \left[\left(2 m_\vec{1}+4 m_\vec{54}-m_{\vec{45}_{(\vec{1},\vec{1},\vec{3})}}+2 m_{\vec{210}_{(\vec{15},\vec{1},\vec{1})}}\right)^2-4 m_{\vec{210}_{(\vec{15},\vec{1},\vec{3})}}^2-8 m_{\vec{45}_{(\vec{15},\vec{1},\vec{1})}}^2\right.\\
 & & \left.+4 \left(2 m_\vec{1}+4 m_\vec{54}-m_{\vec{45}_{(\vec{1},\vec{1},\vec{3})}}-4 m_{\vec{45}_{(\vec{15},\vec{1},\vec{1})}}+2 m_{\vec{210}_{(\vec{15},\vec{1},\vec{1})}}\right) m_{\vec{210}_{(\vec{15},\vec{1},\vec{3})}}\right]^{1/2} \\
 m_{(\vec{6}, \vec{2}, -1, -2)} & = & m_\vec{1}-3 m_\vec{54}-m_{\vec{45}_{(\vec{1},\vec{1},\vec{3})}}-2 m_{\vec{210}_{(\vec{15},\vec{1},\vec{1})}}-m_{\vec{210}_{(\vec{15},\vec{1},\vec{3})}} \\
 m_{(\vec{1}, \vec{2}, -1, -6)} & = & m_\vec{1}-3 m_\vec{54}+m_{\vec{45}_{(\vec{1},\vec{1},\vec{3})}}+6 m_{\vec{210}_{(\vec{15},\vec{1},\vec{1})}}-3 m_{\vec{210}_{(\vec{15},\vec{1},\vec{3})}} \\
 m_{(\vec{3}, \vec{2}, -1, -2)'} & = & -5 m_\vec{54}+\frac{m_{\vec{45}_{(\vec{1},\vec{1},\vec{3})}}}{2}+m_{\vec{210}_{(\vec{15},\vec{1},\vec{1})}} \\
 & & +\frac{1}{2} \left[\left(2 m_\vec{1}+4 m_\vec{54}+m_{\vec{45}_{(\vec{1},\vec{1},\vec{3})}}+2 m_{\vec{210}_{(\vec{15},\vec{1},\vec{1})}}\right)^2-4 m_{\vec{210}_{(\vec{15},\vec{1},\vec{3})}}^2-8 m_{\vec{45}_{(\vec{15},\vec{1},\vec{1})}}^2\right.\\
 & & \left.-4 \left(2 m_\vec{1}+4 m_\vec{54}+m_{\vec{45}_{(\vec{1},\vec{1},\vec{3})}}-4 m_{\vec{45}_{(\vec{15},\vec{1},\vec{1})}}+2 m_{\vec{210}_{(\vec{15},\vec{1},\vec{1})}}\right) m_{\vec{210}_{(\vec{15},\vec{1},\vec{3})}}\right]^{1/2} \\
 m_{(\vec{6}, \vec{2}, 1, -2)} & = & m_\vec{1}-3 m_\vec{54}+m_{\vec{45}_{(\vec{1},\vec{1},\vec{3})}}-2 m_{\vec{210}_{(\vec{15},\vec{1},\vec{1})}}+m_{\vec{210}_{(\vec{15},\vec{1},\vec{3})}} \\
 m_{(\vec{1}, \vec{1}, 0, 0)'} & = & \mathcal{M}_s^B \\
 m_{(\vec{3}, \vec{1}, 0, 4)'} & = & m_\vec{1}-3 m_\vec{54}+\frac{m_{\vec{45}_{(\vec{15},\vec{1},\vec{1})}}}{2}+\frac{m_{\vec{210}_{(\vec{1},\vec{1},\vec{1})}}}{2}+2 m_{\vec{210}_{(\vec{15},\vec{1},\vec{1})}} \\
 & & -\frac{1}{2} \left[\left(10 m_\vec{54}+m_{\vec{45}_{(\vec{15},\vec{1},\vec{1})}}+m_{\vec{210}_{(\vec{1},\vec{1},\vec{1})}}\right)^2+8 \left(m_{\vec{45}_{(\vec{1},\vec{1},\vec{3})}}-m_{\vec{210}_{(\vec{15},\vec{1},\vec{3})}}\right)^2\right]^{1/2} \\
 m_{(\vec{8}, \vec{1}, 0, 0)'} & = & m_\vec{1}-3 m_\vec{54}-\frac{m_{\vec{45}_{(\vec{15},\vec{1},\vec{1})}}}{2}+\frac{m_{\vec{210}_{(\vec{1},\vec{1},\vec{1})}}}{2}-2 m_{\vec{210}_{(\vec{15},\vec{1},\vec{1})}} \\
 & & -\frac{1}{2} \left[\left(10 m_\vec{54}-m_{\vec{45}_{(\vec{15},\vec{1},\vec{1})}}+m_{\vec{210}_{(\vec{1},\vec{1},\vec{1})}}\right)^2+8 \left(m_{\vec{45}_{(\vec{1},\vec{1},\vec{3})}}+m_{\vec{210}_{(\vec{15},\vec{1},\vec{3})}}\right)^2\right]^{1/2} \\
 m_{(\vec{1}, \vec{1}, 0, 0)''} & = & \mathcal{M}_s^C \\
 m_{(\vec{3}, \vec{1}, 0, 4)} & = & m_\vec{1}-3 m_\vec{54}+\frac{m_{\vec{45}_{(\vec{15},\vec{1},\vec{1})}}}{2}+\frac{m_{\vec{210}_{(\vec{1},\vec{1},\vec{1})}}}{2}+2 m_{\vec{210}_{(\vec{15},\vec{1},\vec{1})}} \\
 & & +\frac{1}{2} \left[\left(10 m_\vec{54}+m_{\vec{45}_{(\vec{15},\vec{1},\vec{1})}}+m_{\vec{210}_{(\vec{1},\vec{1},\vec{1})}}\right)^2+8 \left(m_{\vec{45}_{(\vec{1},\vec{1},\vec{3})}}-m_{\vec{210}_{(\vec{15},\vec{1},\vec{3})}}\right)^2\right]^{1/2} \\
 m_{(\vec{8}, \vec{1}, 0, 0)} & = & m_\vec{1}-3 m_\vec{54}-\frac{m_{\vec{45}_{(\vec{15},\vec{1},\vec{1})}}}{2}+\frac{m_{\vec{210}_{(\vec{1},\vec{1},\vec{1})}}}{2}-2 m_{\vec{210}_{(\vec{15},\vec{1},\vec{1})}} \\
 & & +\frac{1}{2} \left[\left(10 m_\vec{54}-m_{\vec{45}_{(\vec{15},\vec{1},\vec{1})}}+m_{\vec{210}_{(\vec{1},\vec{1},\vec{1})}}\right)^2+8 \left(m_{\vec{45}_{(\vec{1},\vec{1},\vec{3})}}+m_{\vec{210}_{(\vec{15},\vec{1},\vec{3})}}\right)^2\right]^{1/2} \\
 m_{(\vec{1}, \vec{1}, 2, 0)} & = & m_\vec{1}+2 m_\vec{54}+2 m_{\vec{45}_{(\vec{15},\vec{1},\vec{1})}}+m_{\vec{210}_{(\vec{1},\vec{1},\vec{1})}}+4 m_{\vec{210}_{(\vec{15},\vec{1},\vec{1})}} \\
 m_{(\vec{3}, \vec{1}, 2, 4)} & = & m_\vec{1}+2 m_\vec{54}+m_{\vec{45}_{(\vec{15},\vec{1},\vec{1})}}+m_{\vec{210}_{(\vec{1},\vec{1},\vec{1})}}+2 m_{\vec{210}_{(\vec{15},\vec{1},\vec{1})}} -4 m_{\vec{210}_{(\vec{15},\vec{1},\vec{3})}} \\
 m_{(\vec{3}, \vec{1}, -2, 4)} & = & m_\vec{1}+2 m_\vec{54}+m_{\vec{45}_{(\vec{15},\vec{1},\vec{1})}}+m_{\vec{210}_{(\vec{1},\vec{1},\vec{1})}}+2 m_{\vec{210}_{(\vec{15},\vec{1},\vec{1})}} +4 m_{\vec{210}_{(\vec{15},\vec{1},\vec{3})}} \\
 m_{(\vec{8}, \vec{1}, 2, 0)} & = & m_\vec{1}+2 m_\vec{54}-m_{\vec{45}_{(\vec{15},\vec{1},\vec{1})}}+m_{\vec{210}_{(\vec{1},\vec{1},\vec{1})}}-2 m_{\vec{210}_{(\vec{15},\vec{1},\vec{1})}} \\
 m_{(\vec{1}, \vec{3}, 0, 0)} & = & m_\vec{1}+2 m_\vec{54}-2 m_{\vec{45}_{(\vec{15},\vec{1},\vec{1})}}-m_{\vec{210}_{(\vec{1},\vec{1},\vec{1})}}+4 m_{\vec{210}_{(\vec{15},\vec{1},\vec{1})}} \\
 m_{(\vec{3}, \vec{3}, 0, 4)} & = & m_\vec{1}+2 m_\vec{54}-m_{\vec{45}_{(\vec{15},\vec{1},\vec{1})}}-m_{\vec{210}_{(\vec{1},\vec{1},\vec{1})}}+2 m_{\vec{210}_{(\vec{15},\vec{1},\vec{1})}} \\
 m_{(\vec{8}, \vec{3}, 0, 0)} & = & m_\vec{1}+2 m_\vec{54}+m_{\vec{45}_{(\vec{15},\vec{1},\vec{1})}}-m_{\vec{210}_{(\vec{1},\vec{1},\vec{1})}}-2 m_{\vec{210}_{(\vec{15},\vec{1},\vec{1})}} \\
\end{array}$ \\
\hline
\end{tabular}
\caption{\label{tab:421and3211_210}
Mass of a chiral multiplet in representation $R$ of $SO(10)$ (left) under the subgroups $G_{421}$ (middle) and $G_{3211}$ (right).}
\end{table}
\endgroup
\end{turnpage}

\begin{turnpage}
\begingroup
\squeezetable
\begin{table}
\begin{tabular}{l|l}
$SO(10)$ & $SU(3)_C\times SU(2)_L\times U(1)_{2 R}\times U(1)_{3 (B-L)}$\\
\hline
\rotatebox[origin=c]{90}{$\vec{126}\oplus\overline{\vec{126}}$} & $\begin{array}{lll}
 m_{(\vec{3}, \vec{1}, 0, -2)} & = & \mathcal{M}_t^A \\
 m_{(\vec{3}, \vec{1}, 0, -2)'} & = & \mathcal{M}_t^B \\
 m_{(\vec{1}, \vec{1}, -2, -6)} & = & m_\vec{1}-2 m_{\vec{45}_{(\vec{1},\vec{1},\vec{3})}}+3 m_{\vec{45}_{(\vec{15},\vec{1},\vec{1})}}-m_{\vec{210}_{(\vec{1},\vec{1},\vec{1})}}+3 m_{\vec{210}_{(\vec{15},\vec{1},\vec{1})}}+6 m_{\vec{210}_{(\vec{15},\vec{1},\vec{3})}} \\
 m_{(\vec{3}, \vec{1}, -2, -2)} & = & m_\vec{1}-2 m_{\vec{45}_{(\vec{1},\vec{1},\vec{3})}}+m_{\vec{45}_{(\vec{15},\vec{1},\vec{1})}}-m_{\vec{210}_{(\vec{1},\vec{1},\vec{1})}}+m_{\vec{210}_{(\vec{15},\vec{1},\vec{1})}}+2 m_{\vec{210}_{(\vec{15},\vec{1},\vec{3})}} \\
 m_{(\vec{6}, \vec{1}, 2, -2)} & = & m_\vec{1}-2 m_{\vec{45}_{(\vec{1},\vec{1},\vec{3})}}-m_{\vec{45}_{(\vec{15},\vec{1},\vec{1})}}-m_{\vec{210}_{(\vec{1},\vec{1},\vec{1})}}-m_{\vec{210}_{(\vec{15},\vec{1},\vec{1})}}-2 m_{\vec{210}_{(\vec{15},\vec{1},\vec{3})}} \\
 m_{(\vec{1}, \vec{1}, 0, -6)} & = & m_\vec{1}+3 m_{\vec{45}_{(\vec{15},\vec{1},\vec{1})}}-m_{\vec{210}_{(\vec{1},\vec{1},\vec{1})}}+3 m_{\vec{210}_{(\vec{15},\vec{1},\vec{1})}} \\
 m_{(\vec{3}, \vec{1}, 0, -2)''} & = & \mathcal{M}_t^C \\
 m_{(\vec{6}, \vec{1}, 0, -2)} & = & m_\vec{1}-m_{\vec{45}_{(\vec{15},\vec{1},\vec{1})}}-m_{\vec{210}_{(\vec{1},\vec{1},\vec{1})}}-m_{\vec{210}_{(\vec{15},\vec{1},\vec{1})}} \\
 m_{(\vec{1}, \vec{1}, 2, -6)} & = & m_\vec{1}+2 m_{\vec{45}_{(\vec{1},\vec{1},\vec{3})}}+3 m_{\vec{45}_{(\vec{15},\vec{1},\vec{1})}}-m_{\vec{210}_{(\vec{1},\vec{1},\vec{1})}}+3 m_{\vec{210}_{(\vec{15},\vec{1},\vec{1})}}-6 m_{\vec{210}_{(\vec{15},\vec{1},\vec{3})}} \\
 m_{(\vec{3}, \vec{1}, 2, -2)} & = & m_\vec{1}+2 m_{\vec{45}_{(\vec{1},\vec{1},\vec{3})}}+m_{\vec{45}_{(\vec{15},\vec{1},\vec{1})}}-m_{\vec{210}_{(\vec{1},\vec{1},\vec{1})}}+m_{\vec{210}_{(\vec{15},\vec{1},\vec{1})}}-2 m_{\vec{210}_{(\vec{15},\vec{1},\vec{3})}} \\
 m_{(\vec{6}, \vec{1}, -2, -2)} & = & m_\vec{1}+2 m_{\vec{45}_{(\vec{1},\vec{1},\vec{3})}}-m_{\vec{45}_{(\vec{15},\vec{1},\vec{1})}}-m_{\vec{210}_{(\vec{1},\vec{1},\vec{1})}}-m_{\vec{210}_{(\vec{15},\vec{1},\vec{1})}}+2 m_{\vec{210}_{(\vec{15},\vec{1},\vec{3})}} \\
 m_{(\vec{1}, \vec{3}, 0, 6)} & = & m_\vec{1}-3 m_{\vec{45}_{(\vec{15},\vec{1},\vec{1})}}+m_{\vec{210}_{(\vec{1},\vec{1},\vec{1})}}+3 m_{\vec{210}_{(\vec{15},\vec{1},\vec{1})}} \\
 m_{(\vec{3}, \vec{3}, 0, -2)} & = & m_\vec{1}-m_{\vec{45}_{(\vec{15},\vec{1},\vec{1})}}+m_{\vec{210}_{(\vec{1},\vec{1},\vec{1})}}+m_{\vec{210}_{(\vec{15},\vec{1},\vec{1})}} \\
 m_{(\vec{6}, \vec{3}, 0, -2)} & = & m_\vec{1}+m_{\vec{45}_{(\vec{15},\vec{1},\vec{1})}}+m_{\vec{210}_{(\vec{1},\vec{1},\vec{1})}}-m_{\vec{210}_{(\vec{15},\vec{1},\vec{1})}} \\
 m_{(\vec{1}, \vec{2}, 1, 0)} & = & \left[3 m_{\vec{54},A}^2+3 m_{\vec{54},B}^2+\left(m_\vec{1}+2 m_{\vec{210}_{(\vec{15},\vec{1},\vec{1})}}\right)^2+\left(m_{\vec{45}_{(\vec{1},\vec{1},\vec{3})}}-2 m_{\vec{210}_{(\vec{15},\vec{1},\vec{3})}}\right)^2\right.\\
 & & \left.-\sqrt{\left(3 \left(m_{\vec{54},A}+m_{\vec{54},B}\right)^2+2 \left(m_\vec{1}+2 m_{\vec{210}_{(\vec{15},\vec{1},\vec{1})}}\right)^2\right) \left(3 \left(m_{\vec{54},A}-m_{\vec{54},B}\right)^2+2 \left(m_{\vec{45}_{(\vec{1},\vec{1},\vec{3})}}-2 m_{\vec{210}_{(\vec{15},\vec{1},\vec{3})}}\right)^2\right)}\right]^{1/2} \\
 m_{(\vec{3}, \vec{2}, -1, 4)} & = & \left[\left(m_{\vec{45}_{(\vec{1},\vec{1},\vec{3})}}+2 m_{\vec{45}_{(\vec{15},\vec{1},\vec{1})}}\right)^2+3 m_{\vec{54},A}^2+3 m_{\vec{54},B}^2+\left(m_\vec{1}+m_{\vec{210}_{(\vec{15},\vec{1},\vec{1})}}\right)^2-2 \left(2 m_\vec{1}+m_{\vec{45}_{(\vec{1},\vec{1},\vec{3})}}+2 \left(m_{\vec{45}_{(\vec{15},\vec{1},\vec{1})}}+m_{\vec{210}_{(\vec{15},\vec{1},\vec{1})}}\right)\right) m_{\vec{210}_{(\vec{15},\vec{1},\vec{3})}}+5 m_{\vec{210}_{(\vec{15},\vec{1},\vec{3})}}^2\right.\\
 & & \left.-\sqrt{\left(3 \left(m_{\vec{54},A}+m_{\vec{54},B}\right)^2+2 \left(m_\vec{1}+m_{\vec{210}_{(\vec{15},\vec{1},\vec{1})}}-2 m_{\vec{210}_{(\vec{15},\vec{1},\vec{3})}}\right)^2\right) \left(3 \left(m_{\vec{54},A}-m_{\vec{54},B}\right)^2+2 \left(m_{\vec{45}_{(\vec{1},\vec{1},\vec{3})}}+2 m_{\vec{45}_{(\vec{15},\vec{1},\vec{1})}}-m_{\vec{210}_{(\vec{15},\vec{1},\vec{3})}}\right)^2\right)}\right]^{1/2} \\
 m_{(\vec{3}, \vec{2}, 1, 4)} & = & \left[\left(m_{\vec{45}_{(\vec{1},\vec{1},\vec{3})}}-2 m_{\vec{45}_{(\vec{15},\vec{1},\vec{1})}}\right)^2+3 m_{\vec{54},A}^2+3 m_{\vec{54},B}^2+\left(m_\vec{1}+m_{\vec{210}_{(\vec{15},\vec{1},\vec{1})}}\right)^2+2 \left(2 m_\vec{1}-m_{\vec{45}_{(\vec{1},\vec{1},\vec{3})}}+2 \left(m_{\vec{45}_{(\vec{15},\vec{1},\vec{1})}}+m_{\vec{210}_{(\vec{15},\vec{1},\vec{1})}}\right)\right) m_{\vec{210}_{(\vec{15},\vec{1},\vec{3})}}+5 m_{\vec{210}_{(\vec{15},\vec{1},\vec{3})}}^2\right.\\
 & & \left.-\sqrt{\left(3 \left(m_{\vec{54},A}-m_{\vec{54},B}\right)^2+2 \left(-m_{\vec{45}_{(\vec{1},\vec{1},\vec{3})}}+2 m_{\vec{45}_{(\vec{15},\vec{1},\vec{1})}}+m_{\vec{210}_{(\vec{15},\vec{1},\vec{3})}}\right)^2\right) \left(3 \left(m_{\vec{54},A}+m_{\vec{54},B}\right)^2+2 \left(m_\vec{1}+m_{\vec{210}_{(\vec{15},\vec{1},\vec{1})}}+2 m_{\vec{210}_{(\vec{15},\vec{1},\vec{3})}}\right)^2\right)}\right]^{1/2} \\
 m_{(\vec{8}, \vec{2}, 1, 0)} & = & \left[3 m_{\vec{54},A}^2+3 m_{\vec{54},B}^2+\left(m_\vec{1}-m_{\vec{210}_{(\vec{15},\vec{1},\vec{1})}}\right)^2+\left(m_{\vec{45}_{(\vec{1},\vec{1},\vec{3})}}+m_{\vec{210}_{(\vec{15},\vec{1},\vec{3})}}\right)^2\right.\\
 & & \left.-\sqrt{\left(3 \left(m_{\vec{54},A}+m_{\vec{54},B}\right)^2+2 \left(m_\vec{1}-m_{\vec{210}_{(\vec{15},\vec{1},\vec{1})}}\right)^2\right) \left(3 \left(m_{\vec{54},A}-m_{\vec{54},B}\right)^2+2 \left(m_{\vec{45}_{(\vec{1},\vec{1},\vec{3})}}+m_{\vec{210}_{(\vec{15},\vec{1},\vec{3})}}\right)^2\right)}\right]^{1/2} \\
 m_{(\vec{1}, \vec{2}, 1, 0)'} & = & \left[3 m_{\vec{54},A}^2+3 m_{\vec{54},B}^2+\left(m_\vec{1}+2 m_{\vec{210}_{(\vec{15},\vec{1},\vec{1})}}\right)^2+\left(m_{\vec{45}_{(\vec{1},\vec{1},\vec{3})}}-2 m_{\vec{210}_{(\vec{15},\vec{1},\vec{3})}}\right)^2\right.\\
 & & \left.+\sqrt{\left(3 \left(m_{\vec{54},A}+m_{\vec{54},B}\right)^2+2 \left(m_\vec{1}+2 m_{\vec{210}_{(\vec{15},\vec{1},\vec{1})}}\right)^2\right) \left(3 \left(m_{\vec{54},A}-m_{\vec{54},B}\right)^2+2 \left(m_{\vec{45}_{(\vec{1},\vec{1},\vec{3})}}-2 m_{\vec{210}_{(\vec{15},\vec{1},\vec{3})}}\right)^2\right)}\right]^{1/2} \\
 m_{(\vec{3}, \vec{2}, -1, 4)'} & = & \left[\left(m_{\vec{45}_{(\vec{1},\vec{1},\vec{3})}}+2 m_{\vec{45}_{(\vec{15},\vec{1},\vec{1})}}\right)^2+3 m_{\vec{54},A}^2+3 m_{\vec{54},B}^2+\left(m_\vec{1}+m_{\vec{210}_{(\vec{15},\vec{1},\vec{1})}}\right)^2-2 \left(2 m_\vec{1}+m_{\vec{45}_{(\vec{1},\vec{1},\vec{3})}}+2 \left(m_{\vec{45}_{(\vec{15},\vec{1},\vec{1})}}+m_{\vec{210}_{(\vec{15},\vec{1},\vec{1})}}\right)\right) m_{\vec{210}_{(\vec{15},\vec{1},\vec{3})}}+5 m_{\vec{210}_{(\vec{15},\vec{1},\vec{3})}}^2\right.\\
 & & \left.+\sqrt{\left(3 \left(m_{\vec{54},A}+m_{\vec{54},B}\right)^2+2 \left(m_\vec{1}+m_{\vec{210}_{(\vec{15},\vec{1},\vec{1})}}-2 m_{\vec{210}_{(\vec{15},\vec{1},\vec{3})}}\right)^2\right) \left(3 \left(m_{\vec{54},A}-m_{\vec{54},B}\right)^2+2 \left(m_{\vec{45}_{(\vec{1},\vec{1},\vec{3})}}+2 m_{\vec{45}_{(\vec{15},\vec{1},\vec{1})}}-m_{\vec{210}_{(\vec{15},\vec{1},\vec{3})}}\right)^2\right)}\right]^{1/2} \\
 m_{(\vec{3}, \vec{2}, 1, 4)'} & = & \left[\left(m_{\vec{45}_{(\vec{1},\vec{1},\vec{3})}}-2 m_{\vec{45}_{(\vec{15},\vec{1},\vec{1})}}\right)^2+3 m_{\vec{54},A}^2+3 m_{\vec{54},B}^2+\left(m_\vec{1}+m_{\vec{210}_{(\vec{15},\vec{1},\vec{1})}}\right)^2+2 \left(2 m_\vec{1}-m_{\vec{45}_{(\vec{1},\vec{1},\vec{3})}}+2 \left(m_{\vec{45}_{(\vec{15},\vec{1},\vec{1})}}+m_{\vec{210}_{(\vec{15},\vec{1},\vec{1})}}\right)\right) m_{\vec{210}_{(\vec{15},\vec{1},\vec{3})}}+5 m_{\vec{210}_{(\vec{15},\vec{1},\vec{3})}}^2\right.\\
 & & \left.+\sqrt{\left(3 \left(m_{\vec{54},A}-m_{\vec{54},B}\right)^2+2 \left(-m_{\vec{45}_{(\vec{1},\vec{1},\vec{3})}}+2 m_{\vec{45}_{(\vec{15},\vec{1},\vec{1})}}+m_{\vec{210}_{(\vec{15},\vec{1},\vec{3})}}\right)^2\right) \left(3 \left(m_{\vec{54},A}+m_{\vec{54},B}\right)^2+2 \left(m_\vec{1}+m_{\vec{210}_{(\vec{15},\vec{1},\vec{1})}}+2 m_{\vec{210}_{(\vec{15},\vec{1},\vec{3})}}\right)^2\right)}\right]^{1/2} \\
 m_{(\vec{8}, \vec{2}, 1, 0)'} & = & \left[3 m_{\vec{54},A}^2+3 m_{\vec{54},B}^2+\left(m_\vec{1}-m_{\vec{210}_{(\vec{15},\vec{1},\vec{1})}}\right)^2+\left(m_{\vec{45}_{(\vec{1},\vec{1},\vec{3})}}+m_{\vec{210}_{(\vec{15},\vec{1},\vec{3})}}\right)^2\right.\\
 & & \left.+\sqrt{\left(3 \left(m_{\vec{54},A}+m_{\vec{54},B}\right)^2+2 \left(m_\vec{1}-m_{\vec{210}_{(\vec{15},\vec{1},\vec{1})}}\right)^2\right) \left(3 \left(m_{\vec{54},A}-m_{\vec{54},B}\right)^2+2 \left(m_{\vec{45}_{(\vec{1},\vec{1},\vec{3})}}+m_{\vec{210}_{(\vec{15},\vec{1},\vec{3})}}\right)^2\right)}\right]^{1/2} \\
\end{array}$ \\
\hline
\end{tabular}
\caption{\label{tab:126_3211}
Mass of a chiral $\vec{126}\oplus\overline{\vec{126}}$ under the subgroup $SU(3)_C\times SU(2)_L\times U(1)_{2 R}\times U(1)_{3 (B-L)}$.}
\end{table}
\endgroup
\end{turnpage}

\clearpage

\twocolumngrid

\bibliographystyle{utcaps_mod}
\bibliography{BIB}

\end{document}